\documentclass[a4paper,10pt]{article}
\pdfoutput=1
\usepackage{jheppub}
\usepackage{amsmath}
\usepackage{amssymb}
\usepackage{multirow}
\usepackage{multicol}
\usepackage{tabularx}
\newcommand{\ep}{\epsilon}
\newcommand{\be}{\begin{equation}}
\newcommand{\ee}{\end{equation}}
\newcommand{\gI}{{\sf I}}
\newcommand{\gJ}{{\sf J}}
\newcommand{\gK}{{\sf K}}

\newcommand{\dOm}{[{\rm d} \Omega_{\perp}]}
\newcommand{\TTPaff}{Institute for Theoretical Particle Physics,
  KIT, 76128 Karlsruhe, Germany}
\newcommand{\IAPaff}{Institute for Astroparticle Physics,
  KIT, 76344 Eggenstein-Leopoldshafen, Germany}
\newcommand{\EDBaff}{Higgs Centre for Theoretical Physics,School of Physics and Astronomy,
  The University of Edinburgh, Edinburgh EH9 3FD, Scotland, UK}

\allowdisplaybreaks[4]
\preprint{
  \begin {flushright}
    TTP26-036, P3H-26-077 
  \end{flushright}
}
\title{NNLO $N$-jettiness soft function for processes with massless and massive partons}
\author[a]{Prem Agarwal,}
\author[a]{Kirill Melnikov,}
\author[a,b]{Ivan Pedron,}
\author[c]{and Andrey Pikelner}
\affiliation[a]{\TTPaff}
\affiliation[b]{\IAPaff}
\affiliation[c]{\EDBaff}
\emailAdd{prem.agarwal@kit.edu}
\emailAdd{kirill.melnikov@kit.edu}
\emailAdd{ivan.pedron@kit.edu}
\emailAdd{a.pikelner@ed.ac.uk}
\abstract{
  We compute the NNLO QCD contribution to the $N$-jettiness soft function for
  processes with two massive and an arbitrary number of massless final-state hard
  partons. The cancellation of all divergencies is demonstrated analytically. The
  finite remainder of the soft function is calculated numerically without an
  infrared regulator. The calculation employs the recently proposed method that
  uses the so-called inclusive $N$-jettiness variable to simplify the computation
  of the non-iterative contribution to the soft function.
}
\begin{document}
\maketitle
%%%
\section{Introduction}
Hadron collisions with large momentum transfers occur at short distances. They
can be studied using perturbative Quantum Chromodynamics which provides a
theoretical framework for describing them through a systematic expansion in the
strong coupling constant. To reach higher accuracy, inclusion of higher orders
in such an expansion is necessary and, unfortunately, the required calculations
rapidly become very complicated. Although there are many reasons for this
increase in complexity, one of the major ones is the fact that conventional QCD
perturbation theory is organized in terms of subprocesses with \emph{fixed
partonic} multiplicities. This leads to soft and collinear infrared divergences
which only disappear when contributions of processes with different
multiplicities are combined. The quest for organizing these cancellations in an
efficient manner for next-to-next-to-leading (NNLO) and
next-to-next-to-next-to-leading (N3LO) QCD computations is an important research
direction in contemporary theoretical collider physics \cite{Caola:2022ayt}.
Currently, the development of two main methods is pursued; they are referred to
as subtraction and slicing. In recent years many realizations of these methods
were proposed and applied to a large number of physical processes at the LHC and
elsewhere
\cite{Czakon:2010td,Czakon:2011ve,Czakon:2014oma,Devoto:2023rpv,Caola:2017dug,Gehrmann-DeRidder:2005btv,Magnea:2018hab,DelDuca:2016ily,Herzog:2018ily,Catani:2007vq,Boughezal:2015aha,Gaunt:2015pea}.

In this paper we deal with a particular variant of the slicing method. In
general, slicing methods are based on the idea that, at each order in
perturbation theory, one can isolate the most singular part of a subprocess with
the highest partonic multiplicity and treat it using simplified expressions for
matrix elements and multi-particle phase spaces. The simplifications are applied
separately to the description of soft and collinear radiation, which naturally
specifies the so-called soft, beam and jet functions. Once, these functions are
computed at a particular perturbative order for a suitable slicing (or
resolution) parameter, slicing computations at this order can be performed, at
least as a matter of principle.

In this paper we will consider the $N$-jettiness observable
\cite{Stewart:2010tn} as a resolution parameter since it is suitable for the
description of processes with final-state jets
\cite{Boughezal:2015dva,Boughezal:2015aha,Gaunt:2015pea}. Process-independent
beam and jet functions for this resolution variable are available through N3LO
in perturbative QCD \cite{Ebert:2020unb,
Baranowski:2022vcn,Behring:2019quf,Bruser:2018rad,Banerjee:2018ozf}.

The situation with the soft functions is more complicated. Physically, soft
functions correspond to cross sections of a particular process computed in an
approximation where unresolved radiation, both real and virtual, is soft. While
the universality of the soft limits of scattering amplitudes ensures the
possibility to provide process-independent description of soft functions, they
still depend on particularities of a specific process such as multiplicities of
final-state particles, as well as their masses, momenta and color charges and on
the resolution variable.

These dependencies make it difficult to compute $N$-jettiness soft functions in
high orders of perturbative QCD; because of that, progress was both slow and
case-specific. Indeed, the NLO soft functions were computed in
Ref.~\cite{Stewart:2010tn}. This was followed by NNLO computations for the
zero-jettiness \cite{Monni:2011gb,Kelley:2011ng}, one-jettiness
\cite{Boughezal:2015eha,Campbell:2017hsw} and two-jettiness \cite{Jin:2019dho}
soft functions for massless emitters. The $N$-jettiness soft function for
arbitrary number of massless emitters was computed in
Refs.~\cite{Bell:2023yso,Agarwal:2024gws,Buonocore:2026yai}. The zero-jettiness
soft function for massless emitters at N3LO was obtained in
Refs.~\cite{Baranowski:2024vxg,Buonocore:2026FutureJettiness}.

For processes with massive particles, much less is known about $N$-jettiness
soft functions. In fact, the NNLO QCD zero-jettiness soft function with
\emph{one} massive emitter either in the initial or in the final state was
computed in Ref.~\cite{Li:2018tsq}. For two back-to-back massive emitters
produced in the collisions of massless partons, the zero-jettiness soft function
was obtained very recently in Ref.~\cite{Bell:2026pou}. No results for
$N$-jettiness soft functions for processes with heavy quarks and \emph{any}
number of jets are currently available.\footnote{ The situation with other
resolution variables is similar in a sense that a large number of NNLO soft
functions for heavy quark production without additional jets is available
\cite{Liu:2025ldi,Devoto:2025eyc,Liu:2024hfa,Catani:2023tby,Angeles-Martinez:2018mqh,Wang:2018vgu,vonManteuffel:2014mva}.}

In this paper, we remedy this situation by presenting the computation of the
NNLO QCD $N$-jettiness soft function for processes with two massive and
\emph{arbitrary} number of massless emitters in the final state. This result is
obtained using the computational method proposed recently in
Ref.~\cite{Buonocore:2026yai} which dramatically simplifies the complexity of
the required calculations. The availability of the soft function computed in
this paper implies that the calculation of the NNLO QCD corrections for
processes $pp \to Q_1 \bar Q_2 + N~{\rm jets}+X$ using the $N$-jettiness slicing
method becomes possible barring the availability of the corresponding two-loop
amplitudes.\footnote{For recent progress in computing virtual amplitudes for
such processes, see Refs.~\cite{Badger:2024fgb,Badger:2024dxo,Badger:2025ilt,Becchetti:2025oyb,Mandal:2022vju,Agarwal:2024jyq,Wang:2024pmv,Becchetti:2026yxl}.}

Since soft radiation is sensitive to the color flow in the whole process, the
soft function has a complex dependence on the color charges of hard partons.
This dependence involves terms with two color charges that we refer to as
dipoles, terms with three color charges that we refer to as tripoles and terms
with four color charges that we refer to as quadrupoles. The tripoles and
quadrupoles appear at NNLO for the first time, whereas dipoles are present
starting at NLO. Moreover, since at NNLO the quadrupole terms arise entirely
from the iteration of the NLO result, they are the easiest contribution to deal
with. Dipoles and tripoles require a different treatment and we deal with them
separately in what follows.

The rest of the paper is organized as follows. In Section~\ref{sec:definitions} we
define the soft function that we wish to calculate. In Section~\ref{sec:nlo-soft-function} we
describe its computation at next-to-leading order where we first derive the
master formula and then apply it to the massless-massless, massive-massless and
massive-massive dipoles. We continue with the NNLO computation. We describe the
dipole contribution to the soft function in Section~\ref{sec:nnlo-dipole} and provide a
detailed discussion of the triple-color correlated contributions to the soft
function in Section~\ref{sec:nnlo-tripole}. In both cases, we start with the derivation of
the master formula for a generic mass assignment and then consider specific
cases. We present numerical results in Section~\ref{sec:results} and conclude in
Section~\ref{sec:conclusion}. Some technical details are relegated to several appendices.

\section{Definitions}
\label{sec:definitions}
We consider a process where two massless QCD partons collide, producing $N$ jets
and a pair of massive quarks. We wish to compute the NNLO QCD soft function for
this process using the $N$-jettiness \cite{Stewart:2010tn} as the resolution
variable. Since emissions of gluons from heavy quarks do not lead to collinear
singularities, the $N$-jettiness function in this case is independent of the
momenta of massive quarks.

In principle, the $N$-jettiness variable depends on momenta of the reconstructed
jets and, therefore, on the adopted jet algorithm. Fortunately, for the
soft-function computation, jet momenta coincide with momenta of massless hard
partons. Hence, assuming that $N_s$ soft partons with momenta
$k_1, \dots, k_{N_s}$ have been emitted, the $N$-jettiness function reads
\begin{equation} \mathcal{T}_{N}(k_1, \dots, k_{N_s}) = \sum_{i=1}^{N_s} \min \left \{ \frac{ p_1 \cdot k_i}{Q_1}, \frac{p_2 \cdot k_i}{Q_2}, \dots,\frac{p_{N+2} \cdot k_i}{Q_{N+2}} \right \},
  \label{eq:TN-std-def}
\end{equation}
where momenta $p_{1,2}$ refer to two incoming partons, momenta
$p_{3,\dots, N+2}$ to $N$ outgoing hard massless partons and $Q_{i=1,\dots,N+2}$
are arbitrary normalization factors.

To apply the method of Ref.~\cite{Buonocore:2025ucn}, we also require the
so-called \emph{inclusive} $N$-jettiness variable
\begin{equation}
  {T}_{N}(q) = \min \left \{ \frac{p_1 \cdot q}{Q_1}, \frac{p_2 \cdot q}{Q_2}, \dots,\frac{p_{N+2} \cdot q}{Q_{N+2}}  \right \},
  \label{eq:TN-inc-def}
\end{equation}
where
\begin{equation}
  q = \sum_{i=1}^{N_s} k_i.
\end{equation}
As explained in Ref.~\cite{Buonocore:2026yai}, computing the soft function with
the inclusive $N$-jettiness variable is simpler but, since $T_N$ and
$\mathcal{T}_N$ become identical in the strongly-ordered soft and certain
collinear limits, the inclusive $N$-jettiness $T_N$ can be used to construct
suitable subtraction terms for calculating the $N$-jettiness soft function,
leading to significant simplifications.

The $N$-jettiness soft function corresponds to integrals of soft contributions
to scattering cross sections, both real and virtual, computed for
$\tau = {\cal T}_N$, in the limit of vanishingly small $\tau$. The bare soft
function can be written as an expansion in the bare coupling constant
\begin{equation}
  S(\tau) = \delta(\tau) +  S_{1} (\tau) +  S_{2} (\tau) + \dots,
  \label{eq:Stau-bare}
\end{equation}
where the term $S^{(i)}$ corresponds to the ${\cal O}(\alpha_s^{i})$
contribution. Although the soft function is originally defined for the fixed
value of the $N$-jettiness variable $\tau$, it is actually convenient to work
with its Laplace transform. The Laplace transform is defined as follows
\begin{equation}
  S (u) = \int \limits_0^\infty d\tau \: S(\tau) \; e^{-\tau u}.
  \label{eq:Su-bare}
\end{equation}
The convenience of the Laplace transform stems from the fact that it makes all
$1/\ep$ singularities explicit, without the need to introduce plus-distributions
in $\tau$, and that the renormalization of the Laplace image of the soft
function is simpler than the renormalization in the $\tau$-space.

Indeed, in the Laplace space, the soft function renormalization is given by
\begin{equation}
  S = Z \tilde S Z^+,
  \label{eq:sf-ren-ZSZ}
\end{equation}
where $\tilde S$ is the $\overline{\rm MS}$-renormalized soft function, and
$Z$ is a matrix in color space. We present this matrix through NNLO in
perturbative QCD in Appendix~\ref{sec:app-ren}. It is important that
this matrix is complex-valued; for this reason $Z^+$ and $Z$ are different and
do not commute with each other.

We write the renormalized soft function as follows
\begin{equation}
  \tilde S = 1 + \tilde S_1 +
  \left ( \frac{1}{2} {\tilde S}_1^2 + \tilde S_2 \right )+\mathcal{O}(\alpha_s^3),
  \label{eq:sf-nnlo-ren-exp}
\end{equation}
where the structure of the ${\cal O}(\alpha_s^2)$ contribution emphasizes the
fact that the soft function exponentiates \cite{Frenkel:1984pz}. Comparing
Eqs~\eqref{eq:Stau-bare}, \eqref{eq:sf-ren-ZSZ} and \eqref{eq:sf-nnlo-ren-exp}, we easily find
\begin{equation}
  \tilde S_1 = S_1 - Z_1 - Z_1^+,
\label{eq:S1_nlo_ren}
\end{equation}
where $Z_1,Z_1^+$ are the ${\cal O}(\alpha_s)$ contributions to the
renormalization matrix $Z$.

To write the non-iterative ${\cal O}(\alpha_s^2)$ contribution to the
renormalized soft function, we separate dipole $\vec T_\gI \cdot \vec T_\gJ$
terms and triple-color $f_{abc} T^a_\gI T^{b}_{\gJ} T^c_{\sf K}$ terms
\begin{equation}
  \tilde S_2 = {\tilde S}_2^{\rm dip} + {\tilde S}_2^{\rm tri}.
\end{equation}
The dipole term reads
\begin{equation}
  {\tilde S}_2^{\rm dip} = S_{RR,T^2}+S_{RV,T^2} -Z_{2,\rm dip} - Z_{2,\rm dip}^+ - \frac{\beta_0 a_s}{\ep} S_1,
  \label{eq:S2-dip-ren}
\end{equation}
where $Z_{2,\rm dip}$ are the relevant parts of the renormalization matrices
defined in Appendix~\ref{sec:app-ren},
$\beta_0 = \frac{11}{6} C_A - \frac{2}{3} T_R n_f$ is the one-loop QCD
$\beta$-function and the subscript $T^2$ indicates that only dipole terms should
be taken from the double-real (RR) and real-virtual (RV) contributions to the
eikonal function.

The triple-color-correlated term  reads
\begin{equation}
  {\tilde S}_2^{\rm tri} = \frac{1}{2} [Z_1,Z_1^+] + \frac{1}{2} [S_1,Z_1 - Z_1^+]
  + S_{RV,T^3} - Z_{2,\rm tri} - Z_{2, \rm tri}^+.
  \label{eq:S2-trip-ren}
\end{equation}
We note that the triple-color-correlated contributions appear in the
real-virtual contribution, and also arise from the iterations of lower-order
terms. Furthermore, if massive emitters are considered, there are
triple-correlated terms in the renormalization matrices $Z$, see
Appendix~\ref{sec:app-ren}.

Our goal in this paper is to compute $\tilde S_1$, $\tilde S_2^{\rm dip}$ and
$\tilde S_{2}^{\rm tri}$ for an arbitrary process at a hadron collider with up
to two heavy quarks. We begin with the calculation of $\tilde S_1$ which we
describe in the next section. To the best of our knowledge, even the NLO results
for the $N$-jettiness soft function are not available in the literature in full
generality.

\section{The \texorpdfstring{$N$}{N}-jettiness soft function at NLO}
\label{sec:nlo-soft-function}

At next-to-leading order, the bare soft function is given by an integral of the
square of the full soft current of the process which involves both massless and
massive emitters.\footnote{Throughout the paper we employ gothic letters
$\gI,\gJ,\gK$ etc. to refer to a generic color-charged particle, capital Latin
letters $I,J,K$ etc. to refer to massive quarks, and small Latin letters $i,j,k$
etc. to refer to massless partons.} Using color conservation, we can write
\begin{equation}
  S_1(\tau) = \sum_{{\sf I}{\sf J}} \;
  {\bf T}_{\sf I} \cdot {\bf T}_{\sf J} \; W_{{\sf I} {\sf J} }(\tau),
  \label{eq:S1-nlo-TT-W}
\end{equation}
where
\begin{equation}
  W_{{\sf I} {\sf J} }(\tau)
  = -g_s^2
  \int [{\rm d} k_1] \delta(\tau - {\cal T}_{N} ) \;
  \tilde S_{{\sf I} {\sf J} }(k_1).
\end{equation}
In the above equation $g_s$ is the bare strong coupling constant,
\begin{equation}
  [ {\rm d} k_1] = \frac{{\rm d}^d k_1}{(2 \pi)^{d-1}} \; \delta_+(k_1^2),
\end{equation}
is the phase-space element of the soft gluon. The eikonal function reads 
\begin{equation}
  \tilde S_{\gI \gJ}(k_1) = \frac{1}{2} \Big[ 2 S_{\gI \gJ}(k_1) - S_{\gI \gI}(k_1) - S_{\gJ \gJ }(k_1) \Big],
  \label{eq:tldS-nonit-def}
\end{equation}
with
\begin{equation}
  S_{ \gI \gJ  }(k_1) = \frac{p_\gI p_\gJ}{(p_\gI \cdot k_1) ( p_\gJ \cdot k_1)}.
\end{equation}
We note that all hard emitters are on-shell, i.e. $p_\gI^2 = m_\gI^2$ where
$m_\gI^2$ is either zero, for a massless hard parton, or $m_I^2$, for a heavy
quark $I$.

The sum in Eq.~\eqref{eq:S1-nlo-TT-W} runs over all $(\gI, \gJ)$ dipoles, where emitters
$\gI$ and $\gJ$ are either massless or massive. We denote massless emitters with
the lower-case letters, i.e. ($i,j$), and massive emitters with the upper-case
letters, i.e. ($I,J$). Then, separating contributions of massless-massless,
massive-massive and massless-massive dipoles, we write
\begin{equation}
  S_1(\tau) = \sum_{ij} \;
  {\bf T}_{i} \cdot {\bf T}_{j} \; W_{ij }(\tau)
  + \sum_{IJ} \;
  {\bf T}_{I} \cdot {\bf T}_{J} \; W_{IJ }(\tau)
  + \sum_{Ij} \;
  {\bf T}_{I} \cdot {\bf T}_{j} \; W_{Ij }(\tau)
  + \sum_{iJ} \;
  {\bf T}_{i} \cdot {\bf T}_{J} \; W_{iJ }(\tau).
  \label{eq:S1-sum-00-M0-MM}
\end{equation}
We note that, thanks to the use of color conservation to define eikonal
functions $\tilde S_{ \gI \gJ}$, there are no diagonal contributions $\gI=\gJ$
even for the massive-massive dipoles. We write the renormalized soft function as
\begin{equation}
  \tilde S_1 = a_s\sum_{{\sf I}{\sf J}} \;
  {\bf T}_{\sf I} \cdot {\bf T}_{\sf J} \; \tilde{W}_{{\sf I} {\sf J} }^{(1)},
  \label{eq:S1-nlo-TT-rem}
\end{equation}
where obtaining $\tilde{W}_{{\sf I} {\sf J} }^{(1)}$ follows from Eq. \eqref{eq:S1_nlo_ren}.
In what follows, we first describe how the
calculation is set up in the general case and then specialize to the three cases
of different mass assignments.

\subsection{Derivation of the master formula}
\label{sec:nlo-master-formula}

We consider the following quantity
\begin{equation}
  W_{\gI \gJ} = - g_s^2 \int [{\rm d} k_1] \; \tilde S_{\gI \gJ}(k_1)  \; \delta(\tau - \mathcal{T}_{N}).
  \label{eq:WIJ-nlo-int-def}
\end{equation}
The function $\tilde S_{\gI \gJ}$ depends on the four-momenta of hard emitters
$p_\gI$ and $p_\gJ$. For each such momentum we introduce its ``light-cone''
version that we denote as $\tilde p_\gI$ or $\tilde p_\gJ$.
We then write
\begin{align}
  && p_\gI = \tilde p_\gI + \kappa_{\gI \gJ} \; \tilde p_\gJ,
  && p_\gJ = \tilde p_\gJ + \kappa_{\gJ \gI} \; \tilde p_\gI,
  &&
\end{align}
with ${\tilde p}_\gI^2= {\tilde p}_\gJ^2 = 0$ and
\begin{equation}
  \kappa_{\gI \gJ} = \frac{m_\gI^2}{2 \tilde p_\gI \tilde p_\gJ}.
\end{equation}
We note that the inverse mapping $p_{\gI, \gJ} \to \tilde p_{\gI, \gJ}$ as well
as some useful kinematic relation for momenta $\tilde p_\gI, \tilde p_\gJ$ are
described in Appendix~\ref{sec:app-kin}.

We employ momenta $\tilde p_\gI$ and $\tilde p_\gJ$ as basis vectors for the
Sudakov decomposition of the soft-gluon momentum and write
\begin{equation}
  k_1 = \alpha \tilde p_\gI + \beta \tilde p_\gJ + k_{1,\perp}.
\end{equation}
Ignoring the on-shell condition of the gluon for the moment, and allowing for
the positive-definite invariant mass $k_1^2 > 0$, we write
\begin{equation}
  \textrm{d}^d k_1 = \tilde p_\gI \tilde p_\gJ \; \textrm{d}\alpha \;\textrm{d}\beta \;
  \textrm{d}^{d-2} k_{1,\perp} = \Omega^{(d-2)} \: \frac{\tilde p_\gI \tilde p_\gJ}{2}\;
  \textrm{d}\alpha \;\textrm{d}\beta \; (k^2_{1,\perp})^{-\ep} \; \textrm{d} k_{1,\perp}^{2}  \dOm,
\end{equation}
where $\dOm =\textrm{d} \Omega^{(d-2)}/\Omega^{(d-2)}$ is the normalized measure of the
$(d-2)$-dimensional space, orthogonal to vectors $\tilde p_\gI, \tilde p_\gJ$.
Since,
\begin{equation}
  k_{1,\perp}^2 = 2 \tilde p_\gI \tilde p_\gJ \alpha \beta - k_1^2,
\end{equation}
it follows that
\begin{equation}
  \textrm{d} k_{1,\perp}^2 = \textrm{d} k_1^2 .
\end{equation}
Since the square of the transverse momentum is positive, for fixed values of the
Sudakov parameters the invariant mass of the parton $k_1$ is restricted from
above. To make this explicit, we write
\begin{equation}
  k_1^2 = (2 \tilde p_\gI \tilde p_\gJ ) \; \alpha \beta \; x ,
\end{equation}
with $ 0 < x < 1$. Hence, for the general case $k_1^2 > 0 $, we find
\begin{equation}
  \textrm{d}^d k_1 = \Omega^{(d-2)} \frac{\tilde p_\gI \tilde p_\gJ }{2} \; (2 \tilde p_\gI \tilde p_\gJ \alpha\beta)^{1-\ep}\;
  \textrm{d}\alpha \;\textrm{d}\beta \; \textrm{d} x \; (1-x)^{-\ep} \;   \dOm.
  \label{eq:ps-measure-x-arb}
\end{equation}
A similar computation for the real $(k_1^2 = 0)$ gluon gives
\begin{equation}
  [\textrm{d} k_1] = \frac{\Omega^{(d-2)}}{2(2\pi)^{d-1}} \;( \tilde p_\gI \tilde p_\gJ ) \; (2 \tilde p_\gI \tilde p_\gJ  )^{-\ep} \;
  \textrm{d}\alpha \;\textrm{d}\beta \; \textrm{d} x \; (\alpha \beta)^{-\ep} \; \delta(x) \; \dOm,
\end{equation}
where we keep the delta-function $\delta(x)$ to emphasize that, with the above
parametrization, $x = 0$ corresponds to the on-shell condition.

We can now combine $g_s^2$ that appears in Eq.~\eqref{eq:WIJ-nlo-int-def} with the
phase-space parametrization described above. This naturally leads to the
following quantity
\begin{equation}
  [\alpha_s] = g_s^2 \frac{\Omega^{(d-2)}}{2 (2 \pi)^{d-1}} =  a_s(\mu) \;
  \mu^{2 \ep}\frac{ e^{\gamma_E \epsilon}}{\Gamma(1-\ep)}
  \left ( 1 + {\cal O}(a_s) \right ),
\end{equation}
that we will use throughout this paper. We note that in the above equation
$a_s(\mu) = \alpha_s(\mu)/(2\pi)$ and $\alpha_s(\mu)$ is the QCD coupling
constant renormalized in the $\overline{\rm MS}$ scheme.
\\

We proceed with the calculation of $W_{\gI \gJ}$. To this end, we introduce the
massless version of the dipole $\tilde S_{\gI \gJ}$,
\begin{equation}
  \mathcal{S}_{\gI \gJ} = \frac{1}{\tilde p_\gI \tilde p_\gJ} \frac{1}{\alpha \beta},
\end{equation}
and denote
\begin{equation}
  {\cal R}_{\gI \gJ}(\alpha, \beta) = \frac{\tilde S_{\gI \gJ}}{\mathcal{S}_{\gI \gJ}}.
\end{equation}
Using these notations and the phase-space parametrization described above, we find
\begin{equation}
  W_{\gI \gJ}(\tau)
  = -[\alpha_s] \; (2 \tilde p_\gI \tilde p_\gJ)^{-\ep} \int \limits_{0}^{\infty} d \alpha \; d\beta \; (\alpha \beta)^{-1-\ep} \; dx \; \;\delta(x) \;
  \dOm  \; {\cal R}_{\gI \gJ}(\alpha, \beta)  \;  \delta(\tau - \mathcal{T}_N).
  \label{eq:WIJ-nlo-int-albe}
\end{equation}

The $N$-jettiness function $\mathcal{T}_N$ depends on all scalar products of the
soft-gluon momentum $k_1$ with momenta of the hard massless partons, see
Eq.~\eqref{eq:TN-std-def}. To illustrate the dependence of such scalar products on the
integration parameters $\alpha, \beta$ and the angles in $(d-2)$-dimensional
transverse space, we consider a hard massless parton $k$, parametrize its
momentum as
\begin{equation}
  p_k = \alpha_k \tilde p_\gI + \beta_k \tilde p_\gJ + p_{k,\perp},
\end{equation}
and find
\begin{equation}
  p_k\cdot k_1 = \tilde p_\gI \tilde p_\gJ \left( \beta_k \alpha + \alpha_k \beta  - 2 \sqrt{\alpha_k \beta_k \alpha \beta} \sqrt{1-x} \cos \varphi \right).
  \label{eq:sp-sudakov-def}
\end{equation}
In Eq.~\eqref{eq:sp-sudakov-def}, we retained the dependence of the scalar
product on the gluon off-shellness $x$. For the single-real emission case that
we discuss in this section, we have to set $x \to 0$ in
Eq.~\eqref{eq:sp-sudakov-def}, but the general formula for $x \ne 0$ will be
useful for computing the double real-emission contribution to the soft function
at NNLO.

It follows from Eq.~\eqref{eq:sp-sudakov-def} and from the definition of the
$N$-jettiness function that the dependence of $T_N$ on $\alpha, \beta$ is
uniform
\begin{equation}
    T_{N,\gI \gJ }(\lambda \alpha ,\lambda \beta,x)= \lambda T_{N,\gI \gJ }(\alpha, \beta,x).
\end{equation}
The subscript $\gI \gJ$ introduced for the $N$-jettiness function in the above
equation, indicates that we have used ${\tilde p}_\gI$ and ${\tilde p}_\gJ$ as
the basis vectors for the Sudakov decomposition of the soft momentum $k_1$.

We use the homogeneity of $T_N$ to simplify the calculation of $W_{\gI \gJ}$ for
arbitrary number of hard emitters. To this end, we insert the partition of unity
\begin{equation}
  1 = \theta(\alpha-\beta) + \theta(\beta-\alpha),
  \label{eq:one-al-be-thetas}
\end{equation}
into Eq.~\eqref{eq:WIJ-nlo-int-albe}, and write $\beta = \alpha \xi$ in the first and
$\alpha=\beta \xi$ in the second term. The $N$-jettiness $\delta$-function
becomes
\begin{equation}
  \delta(\tau - \alpha {\cal T}_{N,\gI \gJ }(1,\xi,x) ),
\end{equation}
in the first case and
\begin{equation}
  \delta(\tau - \beta {\cal T}_{N,\gI \gJ }(\xi,1, x) ),
\end{equation}
in the second one. We then integrate over $\alpha$ or $\beta$ to remove the
corresponding $\delta$-function, set $x$ to zero, use the fact that
${\cal R}_{\gI \gJ}(\lambda \alpha, \lambda \beta) = {\cal R}_{\gI \gJ} (\alpha, \beta) $
and finally find
\begin{equation}
  \begin{split}
    W_{\gI \gJ } (\tau)
    & = -\frac{[\alpha_s]}{\tau^{1+2\ep}} \left(\frac{\tilde p_\gI
      \tilde p_\gJ}{2}\right)^{\ep} \;
      \int \limits_{0}^{1} d \xi \; \xi^{-1-\ep} \; \dOm \;
    \\
    & \times
      \left [
      \left ( \frac{{\cal T}_{N,\gI \gJ }(1,\xi,0) }{(\tilde p_\gI \tilde p_\gJ)} \right )^{2 \ep} {\cal R}_{\gI \gJ}(1,\xi) +
      \left ( \frac{ {\cal T}_{N,\gI \gJ }(\xi,1,0)}{(\tilde p_\gI  \tilde p_\gJ )} \right )^{2\ep}
      {\cal R}_{\gI \gJ}(\xi,1)
      \right ].
  \end{split}
\end{equation}

To make singularities in $\tau$ explicit, we compute the Laplace transform of
$W_{\gI \gJ}$ with respect to $\tau$ using Eq.~\eqref{eq:Su-bare} and obtain
\begin{equation}
  \begin{split}
    W_{\gI \gJ }
    & = \frac{N_{1}(u)}{\ep}
      \left( \frac{\tilde p_\gI \tilde p_\gJ }{2}\right)^{\ep} \int \limits_0^1  d\xi \; \xi^{-1-\ep} \; [d \Omega^{(d-2)}]
    \\
    & \times \left [
      \left ( \frac{{\cal T}_{N,\gI \gJ }(1,\xi,0) }{(\tilde p_\gI \tilde p_\gJ )} \right )^{2 \ep} {\cal R}_{\gI \gJ}(1,\xi) +
      \left ( \frac{{\cal  T}_{N,\gI \gJ }(\xi,1,0)}{(\tilde p_\gI \tilde p_\gJ)} \right )^{2\ep}
      {\cal R}_{\gI \gJ}(\xi,1)  \right ].
  \end{split}
  \label{eq:WijLap}         %Former eq3.28(1)
\end{equation}
The function $N_n(u)$ is defined as follows
\begin{equation}
  N_n(u) = \frac{[\alpha_s]^{n} {\bar u}^{2n \ep} e^{-2 n \gamma_E \ep} \Gamma(1-2n \ep)}{2n},
\end{equation}
with $\bar u = u e^{\gamma_E}$. It provides a short-hand notation for the
general Laplace transform multiplied with an appropriate power of $[\alpha_s]$
\begin{equation}
  [\alpha_s]^n \int_{0}^{\infty}
  {\rm d} \tau \tau^{-1 - 2n \ep}
  e^{-u \tau} = -\frac{N_n(u)}{\ep}.
\end{equation}

The remaining integration over $\xi$ in Eq.~\eqref{eq:WijLap} can produce a $1/\ep$
singularity, but it is easy to extract it. Indeed, we consider the $\xi \to 0$
asymptotic of the $N$-jettiness function and write it as
\begin{equation}
  \frac{{\cal T}_{N,\gI \gJ }(1,\xi,0) }{(\tilde p_\gI \tilde p_\gJ )} \sim   C_{\gI} f_\gI(\xi),
  \quad
  {\rm at~} \xi \to 0.
\end{equation}
Here, $C_{\gI} = 1/Q_\gI$, and $f_\gI(\xi) = \xi$ if $\gI = i$, and
$C_{\gI} f_\gI(\xi)$ is a complicated, $\xi$-independent constant if $i = I$. In
what follows, we use the correct asymptotic of the $N$-jettiness function for
$\gI = i$ to regulate the $\xi = 0$ singularity. On the contrary, for $\gI = I$
we choose the function $f_I(\xi)$ and the normalization factor $C_I$ based on
considerations of convenience since in this case, the function $R_{\gI \gJ}$
regulates the potential $\xi = 0$ singularity completely.\footnote{ We choose
$f_I = 1$ in the majority of cases. However, when considering
triple-color-correlated contributions at NNLO, it turns out to be useful to
choose $f_I(\xi) = \sqrt{\xi}$ for a massive emitter. See Section~\ref{sec:trip-MM0}
for further discussion.} Hence, we obtain
\begin{equation}
  \begin{split}
    W_{\gI \gJ }
    & = \frac{N_{1}(u)}{\ep}
      \left( \frac{\tilde p_\gI \tilde p_\gJ }{2}\right)^{\ep} \int \limits_0^1  d\xi \; \xi^{-1-\ep} \; \dOm
    \\
    & \times \left [
      C_\gI^{2 \ep} f_{\gI}^{2\ep}
      \left ( \frac{{\cal T}_{N,\gI \gJ }(1,\xi,0) }{(\tilde p_\gI \tilde p_\gJ )  C_\gI f_\gI(\xi) } \right )^{2 \ep} {\cal R}_{\gI \gJ}(1,\xi)
      +
      C_\gJ^{2 \ep} f_{\gJ}^{2\ep}
      \left ( \frac{{\cal  T}_{N,\gI \gJ }(\xi,1,0)}{(\tilde p_\gI \tilde p_\gJ) C_{\gJ} f_{\gJ}(\xi)  } \right )^{2\ep}
      {\cal R}_{\gI \gJ}(\xi,1)  \right ].
  \end{split}
  \label{eq:WijLapCfact}      %Former eq3.27
\end{equation}
To separate the $N$-jettiness-dependent and $N$-jettiness-independent
contributions to $W_{\gI \gJ}$, it is convenient to write
\begin{align}
  \label{eq:TNoneExpNorm}           %Former eq3.28(2)
  && \left ( \frac{{\cal T}_{N,\gI \gJ }(1,\xi,0) }{(\tilde p_\gI \tilde p_\gJ )  C_\gI f_\gI(\xi) } \right )^{2 \ep}
     =  1 + 2 \ep g^{(2)}_{\gI \gJ  }(\xi),
  &&
     \left ( \frac{{\cal  T}_{N,\gI \gJ }(\xi,1,0)}{(\tilde p_\gI \tilde p_\gJ) C_{\gJ} f_{\gJ}(\xi)  } \right )^{2\ep}
     =  1 + 2 \ep g^{(2)}_{\gJ\,\gI }(\xi),
  &&
\end{align}
where the functions $g^{(2)}(\xi)$ can be further expanded in powers of $\ep$.

A critical feature of the functions $g^{(2)}$ and ${\cal R}$ that appear in
Eqs~(\ref{eq:TNoneExpNorm},\ref{eq:WijLapCfact}) is that their \emph{products}
vanish at $\xi = 0$, regulating the $\xi = 0$ divergence. Hence, since the
functions $g^{(2)}$ always appear with a prefactor $\ep$, they only contribute
to $W_{\gI \gJ}$ at ${\cal O}(\ep^0)$. Therefore, with the ${\cal O}(\ep^0)$
accuracy, we find
\begin{equation}
  \begin{split}
    W_{\gI \gJ }
    & = \frac{N_1(u)}{\ep}
      \; \left( \frac{\tilde p_\gI  \tilde p_\gJ}{2}\right)^{\ep}
      \Bigg \{
      \int \limits_{0}^{1}
      {\rm d} \xi \; \xi^{-1-\ep} \dOm
      \left (   C_\gI^{2 \ep} f_\gI^{2\ep} R_{\gI \gJ}(1,\xi)
      + C_\gJ^{2 \ep} f_\gJ^{2 \ep}  R_{\gI \gJ}(\xi,1)
      \right )
    \\
    &    +
      2 \ep
      \int \limits_0^1  \frac{ d\xi}{\xi}
      \int \limits_{0}^{2 \pi}
      \frac{{\rm d} \varphi}{2 \pi}  \left [
      \ln \left ( \frac{{\cal T}_{N,\gI \gJ }(1,\xi,0) }{(\tilde p_\gI \tilde p_\gJ )  C_\gI f_\gI(\xi) } \right ) {\cal R}_{\gI \gJ}(1,\xi)
      +
      \ln \left ( \frac{{\cal  T}_{N,\gI \gJ }(\xi,1,0)}{(\tilde p_\gI \tilde p_\gJ) C_{\gJ} f_{\gJ}(\xi)  } \right )
      {\cal R}_{\gI \gJ}(\xi,1)
      \right ]
      \Bigg  \}.
  \end{split}
  \label{eq:WIJ-nlo-master}
\end{equation}
We apply the above master formula to the three different cases of mass
assignments, as described below. Since there is no difference between inclusive
and conventional $N$-jettiness functions at NLO, the above master formula
applies to both cases.

\subsection{Massless-massless dipoles}

In the massless-massless case
$\gI = i, \gJ = j$ and
\begin{align}
  && \tilde  p_\gI = p_i,
  &&  \tilde p_\gJ = p_j,
  && {\cal R}_{ij}(x,y) = 1,
  && f_{\gI}(\xi) = f_{\gJ}(\xi) =\xi,
  && C_\gI = Q_i^{-1},
  && C_\gJ = Q_j^{-1}.
  &&
\end{align}
Using these expression in the master formula for $W_{ij}$ in Eq.~\eqref{eq:WIJ-nlo-master}
and combining these results with the contribution of the renormalization
constants for the massless emitters from Appendix~\ref{sec:app-ren}, we
obtain
\begin{equation}
  \begin{split}
    \tilde{W}_{ij}^{(1)}
     =  \frac{1}{2} L_{ij}^2+
      \frac{1}{2} \ln^2 \frac{Q_i}{Q_j}
      +  \frac{\pi^2}{4}
      + \int \limits_0^1  \frac{d\xi}{\xi} \;
      \int \limits_{0}^{2\pi} \frac{d\varphi}{2 \pi} \;
      \ln \left ( \frac{{\cal T}_{N,ij} (1,\xi,0) {\cal T}_{N,ij}(\xi,1,0)   Q_i Q_j}
      {\xi^2 (p_i p_j)^2}
      \right ).
  \end{split}
  \label{eq:dip-nlo-00-final}
\end{equation}
We note that the quantity $L_{ij}$ is defined
in Eq.~(\ref{eqA.4}).

\subsection{Massive-massless dipoles}

We continue with the case of the massive-massless dipole and identify $\gI = I,$
and $\gJ = j$. The case $\gI = i,$ and $\gJ = J$ is completely analogous, and
both results can be combined into a single contribution. We use
\begin{align}
  && p_I =
     \tilde p_I + \kappa_{Ij} p_j,
  &&
     \tilde p_j = p_j,
  &&
     {\cal R}_{I j}(\alpha, \beta)  =
     \left (\frac{\beta }{\beta + \alpha \; \kappa_{Ij}}
     \right )^2
  &&
\end{align}
and
\begin{align}
  && C_\gJ = Q_j^{-1},
  &&  C_\gI = Q_j^{-1},
  &&f_\gI = 1,
  && f_\gJ = \xi.
  &&
\end{align}
With these ingredients in Eq.~\eqref{eq:WIJ-nlo-master},  we easily derive
\begin{equation}
\begin{split}
W_{Ij} & =N_1(u)
\left (\frac{p_I p_j}{2} \right )^\ep
\Bigg  ( Q_j^{-2\ep}
\left (
\frac{1}{\ep^2} - \frac{1}{\ep}
\left( \ln \kappa_{Ij} + 1 \right )
+ \frac{1}{2} \ln^2 \kappa_{Ij} + \ln \kappa_{Ij} + \frac{\pi^2}{6}
\right )
\\
& +
2 \int \limits_{0}^{1}
\frac{ {\rm d} \xi }{\xi} \int \limits_{0}^{2 \pi} \frac{{\rm d} \varphi}{2 \pi}
\left [{\cal R}_{Ij}(1,\xi) \ln
\left ( \frac{T_{N,Ij}(1,\xi,0) Q_j}{(p_I p_j)} \right )
+ {\cal R}_{Ij}(\xi,1) \ln
\left ( \frac{T_{N,Ij}(\xi,1,0) Q_j}{\xi (p_I p_j)} \right )
\right ]
\Bigg ).
\end{split}
\label{eq:dip-nlo-M0-bare}
\end{equation}
Finally, combining this result with the renormalization constants in
Appendix~\ref{sec:app-ren}, we obtain the
massive-massless contribution to the soft function. It reads
\begin{equation}
\begin{split}
    \tilde{W}_{Ij}^{(1)} &=
    \frac{1}{4} L^2_{Ij} - \frac{L_{Ij}}{2} + \frac{5\pi^2}{24}
\\
& +  \int \limits_{0}^{1} \frac{ {\rm d} \xi }{\xi}
  \int \limits_{0}^{2 \pi} \frac{{\rm d} \varphi}{2 \pi}
\bigg [ {\cal R}_{Ij}(1,\xi) \ln
\left ( \frac{T_{N,Ij}(1,\xi,0) Q_j}{(p_I p_j)} \right ) + {\cal R}_{Ij}(\xi,1) \ln
\left ( \frac{T_{N,Ij}(\xi,1,0) Q_j}{\xi (p_I p_j)} \right )
\bigg  ] .
\end{split}
\label{eq:dip-nlo-M0-final}
\end{equation}
The quantity $L_{Ij}$ can be found in Eq.~\eqref{eqA.4}. Finally, we note that the term $\tilde{W}_{iJ}^{(1)}$ is identical to $\tilde{W}_{Ij}^{(1)}$.
\subsection{Massive-massive dipoles}

The last contribution we need is that of a massive-massive dipole. Hence,  we identify  $\gI = I$,  $\gJ = J$, and write
\begin{align}
  && p_I = \tilde p_I + \kappa_{IJ} \tilde p_J,
  &&  p_J = \tilde p_J + \kappa_{JI} \tilde p_I.
  &&
\end{align}
In the massive-massive  case, the function ${\cal R}_{I J}(\alpha,\beta)$  reads
\begin{equation}
  \begin{split}
    {\cal R}_{I J}(\alpha, \beta)
    &= \frac{\alpha \beta }{(\beta + \alpha \kappa_{IJ})(\beta \kappa_{JI} + \alpha )}
      \left ( \frac{2}{1+v_{IJ}} - \kappa_{IJ} \frac{\beta \kappa_{JI} + \alpha }{\beta+ \kappa_{IJ} \alpha } - \kappa_{JI} \frac{\beta+ \kappa_{IJ} \alpha }{\beta \kappa_{JI} + \alpha}
      \right ) \\
    & =  \left( \frac{2 \alpha \beta v_{IJ}}{(1+v_{IJ})(\alpha \kappa_{IJ} + \beta) (\alpha + \beta \kappa_{JI} )}\right)^2,
  \end{split}
\end{equation}
where in the last step  we used  the relation (see
Appendix~\ref{sec:app-kin} for details)
\begin{equation}
  \kappa_{IJ} \kappa_{JI} = \frac{1-v_{IJ}}{1 + v_{IJ}},
\label{eq:kappaIJ-def}
\end{equation}
with 
\begin{equation}
  v_{IJ} = \sqrt{1-\frac{m_I^2 m_J^2}{(p_I p_J)^2}}.
\label{eq:vIJ-def}
\end{equation}
Furthermore, for the massive-massive case we also choose
\begin{align}
  && C_{\gI} = C_{\gJ} = Q^{-1},
  && f_\gI(\xi) = f_\gJ(\xi) = 1,
  &&
\end{align}
where $Q$ is an arbitrary dimension-full quantity.
Using the above ingredients, we find
\begin{equation}
  \begin{split}
    & W_{IJ}  =N_1(u)
             \left (\frac{p_I p_J (1+v_{IJ})}{4} \right )^\ep
             \Bigg  ( Q^{-2\ep}
             \bigg [
             \frac{1}{\ep}
             \bigg[\frac{1}{v_{IJ}} L_{v_{IJ}}-2 \bigg]
             +\frac{2}{v_{IJ}}\left({\rm Li}_2(-\kappa_{IJ})+ {\rm Li}_2(-\kappa_{JI})\right) \\
           & + \frac{1}{v_{IJ}}\frac{\pi^2}{3} +\frac{1}{2v_{IJ}}L_{v_{IJ}}^2- L_{v_{IJ}} -2\ln[(1+\kappa_{IJ})(1+\kappa_{JI})]- \frac{1}{v_{IJ}}\ln\kappa_{IJ}\ln\kappa_{JI}
             \bigg]
    \\
           & +
             2 \int \limits_{0}^{1}
             \frac{ {\rm d} \xi }{\xi} \int \limits_{0}^{2 \pi} \frac{{\rm d} \varphi}{2 \pi}
             \left [{\cal R}_{IJ}(1,\xi) \ln
             \left (  \frac{T_{N,IJ}(1,\xi,0) \; Q}{\tilde p_I \tilde p_J} \right )
             + {\cal R}_{IJ}(\xi,1) \ln
             \left ( \frac{T_{N,IJ}(\xi,1,0) \; Q}{\tilde p_I \tilde p_J} \right )
             \right ]
             \Bigg ),
  \end{split}
\end{equation}
where
\begin{equation}
  L_{v_{IJ}}= \ln \left( \frac{1-v_{IJ}}{1+v_{IJ}}\right).
\end{equation}
It is straightforward to obtain the final result for the massive-massive dipole
by combining $W_{IJ}$ with the renormalization constants in Eq.~\eqref{eqA11}.
We find
\begin{equation}
  \begin{split}
    &  \tilde{W}_{IJ}^{(1)} =  \frac{1}{v_{IJ}} \Bigg\{ -\left(\frac{1}{2}L_{v_{IJ}} + v_{IJ} \right) L_{IJ} + \frac{1}{4}L^2_{v_{IJ}} + \frac{v_{IJ}}{2}L_{v_{IJ}} + \frac{\pi^2}{6}
    \\
    &     - v_{IJ} \ln[(1+\kappa_{IJ})(1+\kappa_{JI})]+
      {\rm Li}_2(-\kappa_{IJ})+ {\rm Li}_2(-\kappa_{JI}) - \frac{1}{2} \ln\kappa_{IJ}\ln\kappa_{JI}
    \\
    &+ v_{IJ}
      \int \limits_{0}^{1}
      \frac{ {\rm d} \xi }{\xi} \int \limits_{0}^{2 \pi} \frac{{\rm d} \varphi}{2 \pi}
      \left [{\cal R}_{IJ}(1,\xi) \ln
      \left (  \frac{T_{N,IJ}(1,\xi,0) Q}{\tilde p_I \tilde p_J} \right )
      + {\cal R}_{IJ}(\xi,1) \ln
      \left ( \frac{T_{N,IJ}(\xi,1,0) Q}{\tilde p_I   \tilde p_J} \right )
      \right ]
      \Bigg \},
  \end{split}
  \label{eq:dip-nlo-MM-final}
\end{equation}
where
\begin{equation}
  L_{IJ} = \ln \left(\frac{\bar u^2 \mu^2 \; \tilde p_I \tilde p_J}{2 Q^2 }  \right).
\end{equation}

\section{NNLO \texorpdfstring{$N$}{N}-jettiness soft function: the dipole contribution  }
\label{sec:nnlo-dipole}

We continue with the discussion of the soft function at NNLO in QCD, focusing on
the dipole contribution. As shown in Eq.~\eqref{eq:S2-dip-ren}, it contains
real-virtual and double-real contributions, as well as terms related to the
renormalization of the soft function and the coupling constant. We will start
with discussing the first two contributions and then combine them with the
remaining ones.

\subsection{General considerations}

\subsubsection{The dipole part of the real-virtual  contribution}

The dipole part of the real-virtual contribution is written as follows
\begin{equation}
  S_{RV,T^2} =
  \sum \limits_{\gI \gJ} {\bf  T}_{\gI}
  \;
  \cdot {\bf T}_{\gJ} \; W_{RV,\gI \gJ}^{\rm dip},
\end{equation}
where the quantity $W_{RV}^{\rm dip}$ is determined by the \emph{dipole} part of
the soft one-loop current. It was computed in full generality in
Ref.~\cite{Czakon:2018iev}.\footnote{The one-loop soft current for massless
emitters was computed in Ref.~\cite{Catani:2000pi}. } We write
\begin{equation}
  W_{RV,\gI \gJ}^{\rm dip}(\tau) = [\alpha_s] \; 2^{-\ep}
  g_s^2 \int [{\rm d} k_1 ] \; F_{RV,\gI \gJ}^{\rm dip} \;
  \delta(\tau - {\cal T}_N ),
\end{equation}
where  the dependence of $F_{RV,\gI \gJ}^{\rm dip}$ on $p_\gI,p_\gJ$ and $k_1$ is suppressed for brevity.
The one-gluon phase space in the above formula is identical to the NLO case
discussed in the previous section. There we have seen that the scaling
properties of the integrand with respect to the Sudakov parameters $\alpha$ and
$\beta$ play an important role in the extraction of divergences. The scaling of
$T_N$ has already been discussed in Section~\ref{sec:nlo-soft-function}, and the scaling
property of $F^{\rm dip}_{RV,\gI \gJ}$ is described by the following equation
\begin{equation}
  F^{\rm dip}_{RV, \gI \gJ}(\lambda \alpha, \lambda \beta) = \lambda^{-2 -2 \ep} F^{\rm dip}_{RV,\gI \gJ}(\alpha, \beta).
  \label{eq:FdipRVIJ-scaling}
\end{equation}
We normalize $F_{\rm RV}^{\rm dip}$ with respect to the massless dipole writing
\begin{equation}
  F^{\rm dip}_{RV,\gI \gJ}(\alpha, \beta)
  = \left ( \frac{1}{( \tilde p_\gI \tilde p_\gJ ) \alpha \beta}
  \right )^{1+\ep} {\tilde F}^{\rm dip}_{RV,\gI \gJ}(\alpha, \beta).
  \label{eq:FdipRVIJ-tilde}
\end{equation}
It follows from Eqs~(\ref{eq:FdipRVIJ-scaling},\ref{eq:FdipRVIJ-tilde}) that the function
${\tilde F}^{\rm dip}_{\rm RV, \gI \gJ}$ is invariant under the simultaneous
rescaling of the Sudakov parameters $\alpha$ and $\beta$.

Hence, following the steps described in the NLO case, and performing the Laplace
transform, we obtain
\begin{equation}
  \begin{split}
    W^{\rm dip}_{RV,\gI \gJ}  =
    & -\frac{N_2(u)}{\ep}
      \left ( \frac{  \tilde p_{\gI} \tilde p_{\gJ} }{2}
      \right )^{2 \ep}
      \int \limits_{0}^{1}
      {\rm d} \xi \; \xi^{-1-2 \ep} \;
      \dOm \;
      \Bigg [ C_\gI^{4 \ep} f^{4\ep}_\gI(\xi)
      \left ( \frac{ T_{N,\gI \gJ}(1,\xi,0)}{\tilde p_\gI \tilde p_\gJ C_\gI f_\gI(\xi) }  \right )^{4 \ep}
      \tilde F_{RV, \gI \gJ}^{\rm dip}(1,\xi)
    \\
    & +
      C_\gJ^{4 \ep} f^{4\ep}_\gJ(\xi)  \left ( \frac{ T_{N,\gI \gJ}(\xi,1,0)}{ \tilde p_\gI \tilde p_\gJ C_\gJ f_\gJ(\xi) }  \right )^{4 \ep}
      \tilde F_{RV, \gI \gJ}^{\rm dip}(\xi,1)
      \Bigg  ].
  \end{split}
  \label{eq:dip-nnlo-RV-bare}
\end{equation}
We can further split this contribution into the $N$-jettiness dependent and
$N$-jettiness independent parts following Eq.~\eqref{eq:TNoneExpNorm},
\begin{align}
  \label{eq:g4-def-x0}
  &&
     \left ( \frac{{\cal T}_{N,\gI \gJ }(1,\xi,0) }{(\tilde p_\gI \tilde p_\gJ )  C_\gI f_\gI(\xi) } \right )^{4 \ep}
     =  1 + 4 \ep g^{(4)}_{\gI \gJ  }(\xi,0),
  &&
     \left ( \frac{{\cal  T}_{N,\gI \gJ }(\xi,1,0)}{(\tilde p_\gI \tilde p_\gJ) C_{\gJ} f_{\gJ}(\xi)  } \right )^{4\ep}
     =  1 +4 \ep g^{(4)}_{\gJ\,\gI }(\xi,0)
     .
  &&
\end{align}
Using the above equation, we obtain the expression for $W_{\gI \gJ}^{\rm RV}$,
suitable for combining it with other contributions relevant for computing the
soft function at NNLO. It reads
\begin{equation}
  \begin{split}
    W^{\rm RV}_{\gI \gJ}  =
    & -\frac{N_2(u)}{\ep}
      \left ( \frac{  \tilde p_{\gI} \tilde p_{\gJ} }{2}
      \right )^{2 \ep}
      \int \limits_{0}^{1}
      {\rm d} \xi \; \xi^{-1-2 \ep}
      [ {\rm d} \Omega ^{(d-2)}]
      \Bigg [ C_\gI^{4 \ep} f^{4\ep}_\gI(\xi)
      (1 + 4 \ep g^{(4)}_{\gI \gJ}(\xi,0) )
      \tilde F_{RV,\gI \gJ}^{\rm dip}(1,\xi)
    \\
    & +
      C_\gJ^{4 \ep} f^{4\ep}_\gJ(\xi)
      ( 1 + 4 \ep g^{(4)}_{\gJ \gI}(\xi,0)
      )
      \tilde F_{RV,\gI \gJ}^{\rm dip}(\xi,1)
      \Bigg  ].
  \end{split}
  \label{eq:WIJ-RV-final}
\end{equation}

\subsubsection{The double-real dipole contribution}

The non-iterative double-real-emission contribution  reads
\begin{equation}
  S_{RR,T^2} = \sum_{\gI \gJ} {\bf T}_{\gI} \cdot {\bf T}_{\gJ } \; W_{RR,\gI \gJ}^{\rm dip},
\end{equation}
where
\begin{equation}
  W_{RR,\gI \gJ}^{\rm dip} =
  g_s^4 \int [\textrm{d} k_1] [\textrm{d}k_2] \; \tilde S_{\gI \gJ }(k_1,k_2) \; \delta(\tau - \mathcal{T}_N),
\end{equation}
The two partons $1$ and $2$ can be either two gluons or a $q \bar q$ pair. The
relevant symmetry factor (for the two gluons) and the remaining color factors
(i.e $C_A, T_R$) are accounted for in the function $\tilde S_{\gI \gJ}(k_1,k_2)$.

To compute $W_{RR,\gI \gJ}^{\rm dip}$, we follow Ref.~\cite{Buonocore:2026yai}
and write
\begin{equation}
  W_{RR,\gI \gJ}^{\rm dip} =  W_{RR,\gI \gJ}^{\rm dip,inc}
  + \Delta W_{RR,\gI \gJ}^{\rm dip}.
  \label{eq:WIJ-split-inc-delta}
\end{equation}
The first term on the right hand side is defined as follows
\begin{equation}
  W_{RR,\gI \gJ}^{\rm dip, inc} =
  g_s^4 \int [\textrm{d}k_1] [\textrm{d}k_2] \; \tilde S_{\gI \gJ }(k_1,k_2) \; \delta(\tau - T_N(k_{12})),
\end{equation}
where $k_{12} = k_1 + k_2$ and $T_N$ is the
\emph{inclusive} $N$-jettiness variable
that depends on the sum of the soft-parton momenta, see Eq.~\eqref{eq:TN-inc-def}. The
second term on the right-hand side of Eq.~\eqref{eq:WIJ-split-inc-delta} is the difference of
the dipole contribution computed with the regular and inclusive $N$-jettiness
variables. It reads
\begin{equation}
  \Delta  W_{RR,\gI \gJ}^{\rm dip} =
  g_s^4 \int [\textrm{d}k_1] [\textrm{d}k_2] \; \tilde S_{\gI \gJ }(k_1,k_2) \;
  \big [
  \delta(\tau - {\cal T}_N(k_1,k_2) ) - \delta(\tau - T_N(k_{12})) \big ],
  \label{eq:deltaWIJ-def}
\end{equation}
As explained in Ref.~\cite{Buonocore:2026yai}, this term is finite and can be
integrated numerically in four space-time dimensions without further ado.

The key simplification in computing the double-real dipole contribution arises
because inclusive jettiness observable that appears in the function
$W_{RR,\gI \gJ}^{\rm dip, inc}$ depends on the \emph{sum} of two soft momenta.
Because of this, integration over $k_1$ and $k_2$ at fixed $k_{12}$ becomes
observable-independent. Hence, we write
\begin{equation}
  \int [\textrm{d}k_1][\textrm{d}k_2]
  \; \tilde{S}_{\gI \gJ}(k_1,k_2) \; \delta(\tau - T_N(k_{12}))
  =  \int {\rm d}^d q \; \tilde{S}^{\rm inc}_{\gI \gJ}(q) \; \delta(\tau - T_N(q)),
\end{equation}
where  $\tilde S_{\gI \gJ}^{\rm inc}(q)$ is analogous to  the double-real contribution to the fully-differential soft function introduced in
Ref.~\cite{Li:2011zp}. It reads
\begin{equation}
  {\tilde  S}_{\gI \gJ}^{\rm inc}(q) = \int [\textrm{d}k_1]\; [\textrm{d}k_2] \; \delta^{(d)} \left(q-k_1 - k_2 \right) \tilde S_{\gI \gJ}(k_1, k_2).
  \label{eq:WIJinc-def}
\end{equation}

It is convenient to normalize the function
${\tilde S}_{\gI \gJ}^{\rm inc}$ in the following way
\begin{equation}
  {\tilde S}_{\gI \gJ}^{\rm inc}(q)
  = I_0(q^2) \; F_{\gI \gJ}(q,p_\gI,p_\gJ),
\end{equation}
where $I_0(q^2)$ is the phase space of two partons which evaluates to
\begin{equation}
  I_0 = \int [\textrm{d} k_1] [\textrm{d} k_2] \;\delta^{(d)}(q -k_1-k_2 ) = \frac{\Omega^{(d-2)}}{( 2(2\pi)^{d-1})^2}  \frac{\Gamma(1-\ep)^2}{\Gamma(2-2\ep)} \; (q^2)^{-\ep}.
  \label{eq:ps-norm-I0}
\end{equation}
Using the above definitions, we can write
\begin{equation}
  \begin{split}
    W_{RR, \gI \gJ}^{\rm dip, inc}
    &= g_s^4 \int [\textrm{d}k_1][\textrm{d}k_2] \;
      \tilde{S}_{\gI \gJ}(k_1,k_2) \; \delta(\tau - T_N(k_{12}))
      =  g_s^4 \int {\rm d}^d q  \; \tilde{S}^{\rm inc}_{\gI \gJ}(q) \; \delta(\tau - T_N(q))\\
    &= [\alpha_s]^2 \;  \frac{\Gamma(1-\ep)^2}{\Gamma(2-2\ep)} \; \int \frac{ {\rm d}^d q}{ \Omega^{(d-2)}} \;
      (q^2)^{-\ep} F_{\gI \gJ}(q,p_\gI,p_\gJ)\; \delta(\tau - T_N(q)).
  \end{split}
\end{equation}

To proceed further, we require a suitable parametrization of the integration
measure ${\rm d}^d q$. Such a parametrization was provided in
Section~\ref{sec:nlo-soft-function} where formulas with $q^2 \ne 0$ (i.e. $ x \ne 0$) should be
used, c.f. Eq.~\eqref{eq:ps-measure-x-arb}. Similar to the NLO case, we compute
$W_{RR, \gI \gJ}^{\rm dip, inc}$ by integrating over Sudakov parameters
$\alpha, \beta$ to remove the delta-function that involves the inclusive
$N$-jettiness function $T_N$. To do that, we need to know the properties of the
function $F_{\gI \gJ}$ under a simultaneous rescaling of $\alpha$ and $\beta$.
Since the function $F_{\gI \gJ}$ is an integral of the eikonal function
$\tilde S_{\gI \gJ}$ one can convince oneself that the function
$\tilde F_{\gI \gJ}$ defined as follows
\begin{equation}
  F_{\gI \gJ} (q,p_\gI,p_\gJ)
  = \frac{1}{ \alpha^2 \beta^2 (\tilde p_{\gI}  \tilde p_{\gJ})^2 x} \tilde  F_{\gI \gJ}(\alpha, \beta, x),
  \label{eq:FIJ-to-tld}
\end{equation}
is a uniform function of degree zero, i.e.
\begin{equation}
  \tilde  F_{\gI \gJ}(\lambda \alpha, \lambda \beta, x)
  = \tilde F_{\gI \gJ}(\alpha,  \beta, x).
\end{equation}
Using this representation, we integrate over $\alpha$ and $\beta$ to obtain
\begin{equation}
  \begin{split}
    & W_{RR,\gI \gJ}^{\rm dip, inc}  =  -\frac{N_2}{\ep}
      \frac{\Gamma^2(1-\ep)}{\Gamma(2-2\ep)}
      \left ( \frac{  \tilde p_{\gI} \tilde p_{\gJ} }{2}
      \right )^{2\ep}
      \int \limits_{0}^{1} {\rm d} x (1-x)^{-\ep}
      x^{-\ep-1} \;  {\rm d} \xi \;  \xi^{-2 \ep-1} \; \dOm \;
    \\
    & \times
      \Bigg [ C_\gI^{4 \ep} f^{4\ep}_\gI(\xi)
      \left ( \frac{ T_{N,\gI \gJ}(1,\xi,x)}{\tilde p_\gI \tilde p_\gJ C_\gI f_\gI(\xi) }  \right )^{4 \ep}
      \tilde F_{\gI \gJ}(1,\xi, x)
      +
      C_\gJ^{4 \ep} f^{4\ep}_\gJ(\xi)  \left ( \frac{ T_{N,\gI \gJ}(\xi,1,x)}{ \tilde p_\gI \tilde p_\gJ C_\gJ f_\gJ(\xi) }  \right )^{4 \ep}
      \tilde F_{\gI \gJ}(\xi,1, x)
      \Bigg  ].
  \end{split}
  \label{eq:WIJ-nnlo-bare}
\end{equation}
We emphasize that all singularities of the integrand in Eq.~\eqref{eq:WIJ-nnlo-bare} have
been made explicit. Indeed, although the functions $\tilde F_{\gI \gJ}$ may
contain \emph{logarithmic} singularities in $x$ and $\xi$, which will have to be
properly treated at the singular points of the integrand, they do not possess
power singularities in these variables.

Since the NNLO double-real contribution involves two singular limits
($\xi \to 0$, $x \to 0$), it requires a more sophisticated treatment than the
NLO case. However, its basic elements remain unchanged. Hence, following earlier
discussion, we separate the double-real contribution into the
$N$-jettiness-dependent and $N$-jettiness-independent parts using the following
equations
\begin{equation}
  \left ( \frac{ T_{N,\gI \gJ}(1,\xi,x)}{\tilde p_\gI \tilde p_\gJ C_\gI f_\gI(\xi) }  \right )^{4 \ep} = 1 + 4\ep \; g_{\gI \gJ}^{(4)}(\xi ,x) ,\quad \left ( \frac{ T_{N,\gI \gJ}(\xi,1,x)}{ \tilde p_\gI \tilde p_\gJ C_\gJ f_\gJ(\xi) }  \right )^{4 \ep} = 1 + 4\ep \; g_{ \gJ \gI}^{(4)}(\xi ,x),
  \label{eq:g4-def-x}
\end{equation}
and note that the product of functions $g_{\gI \gJ}^{(4)}$ and
$g_{\gJ \gI}^{(4)}$ with functions $\tilde F_{\gI \gJ}(1,\xi,x)$ and
$\tilde F_{\gI \gJ}(\xi,1,x)$ respectively, vanishes at $\xi=0$. Furthermore,
the functions $g_{\gI \gJ}^{(4)}$ and $g_{\gJ \gI}^{(4)}$ have smooth $x \to 0$
limit.

We make use of these observations to facilitate the extraction of singularities
from $W_{\gI \gJ}^{\rm dip, inc}$. To this end, we write
\begin{equation}
  g_{\gI \gJ}^{(4)}(\xi,x) =
  g_{\gI \gJ}^{(4)}(\xi,0) + \Delta g_{\gI \gJ}^{(4)}(\xi,x),
\end{equation}
where
\begin{equation}
  \Delta g_{\gI \gJ}^{(4)}(\xi,x) =  g_{\gI \gJ}^{(4)}(\xi,x) - g_{\gI \gJ}^{(4)}(\xi,0),
\end{equation}
and do the same for the second  function $g_{\gJ \gI}^{(4)}$.  Following this separation, we define
\begin{equation}
  W^{\rm dip, inc}_{RR,\gI \gJ} =  W^{\rm dip, inc}_{RR,\gI \gJ}|_{g(x = 0)}
  + \Delta W^{\rm dip, inc}_{RR,\gI \gJ},
  \label{eq:WIJ-nnlo-reg-subtr}
\end{equation}
with
\begin{equation}
  \begin{split}
    & W_{RR,\gI \gJ}^{\rm dip, inc} |_{g(x = 0)}
      =  -\frac{N_2}{\ep}
      \left ( \frac{ \tilde p_{\gI} \tilde p_{\gJ} }{2}
      \right )^{2\ep}
      \int \limits_{0}^{1} \;  {\rm d} \xi \;  \xi^{-2 \ep-1} \dOm  \;\
    \\
    & \times
      \Bigg [ C_\gI^{4 \ep} f^{4\ep}_\gI(\xi) F_{\gI \gJ}^{RR}(1,\xi)
      \left(1 + 4\ep \; g_{\gI \gJ}^{(4)}(\xi ,0) \right)
      +
      C_\gJ^{4 \ep} f^{4\ep}_\gJ(\xi)
      F_{\gI \gJ}^{RR}(\xi,1) \left(1 + 4\ep \; g_{\gJ \gI}^{(4)}(\xi ,0) \right)
      \Bigg  ].
  \end{split}
  \label{eq:WIJ-gx0-bare}
\end{equation}
and
\begin{equation}
  \begin{split}
    \Delta W_{RR, \gI \gJ}^{\rm dip, inc} =
    & -4 N_2 \;
      \frac{\Gamma(1-\ep)^2}{\Gamma(2-2\ep)}
      \left ( \frac{ \tilde p_{\gI} \tilde p_{\gJ} }{2}
      \right )^{2\ep}
      \int \limits_{0}^{1} {\rm d} x (1-x)^{-\ep}
      x^{-\ep-1} \;  {\rm d} \xi \;  \xi^{-2 \ep-1} \;  \dOm \;\
    \\
    & \times
      \Bigg [ C_\gI^{4 \ep} f^{4\ep}_\gI(\xi)
      \;\Delta  g_{\gI \gJ}^{(4)}(\xi ,x)
      \tilde F_{\gI \gJ}(1,\xi, x)
      +
      C_\gJ^{4 \ep} f^{4\ep}_\gJ(\xi)   \; \Delta  g_{\gJ \gI}^{(4)}(\xi ,x)
      \tilde F_{\gI \gJ}(\xi,1, x)
      \Bigg  ].
  \end{split}
  \label{eq:deltaWIJ-reg-x0}
\end{equation}
The functions $F_{\gI \gJ}^{RR}(1,\xi)$ and $F_{\gI \gJ}^{RR}(\xi,1)$ in
Eq.~\eqref{eq:WIJ-gx0-bare} are defined as follows
\begin{equation}
  \begin{split}
    & F_{\gI \gJ}^{RR}(\alpha,\beta)
      = \frac{\Gamma(1-\ep)^2}{\Gamma(2-2\ep)} \int \limits_{0}^{1} {\rm d} x \;
      x^{-\ep-1} (1-x)^{-\ep} \tilde F_{\gI \gJ}(\alpha,\beta,x).
  \end{split}
  \label{eq:FIJRR-def}
\end{equation}

Since products of the functions $\Delta g_{\gI \gJ}^{(4)}(\xi,x)$ and
$\tilde F_{\gI \gJ}$ vanish at $\xi = 0$ and $x = 0$, integrations in
$\Delta W_{RR,\gI \gJ}^{\rm dip,inc}$ over $x$, $\xi$ and the azimuthal angle
can be performed for the vanishing infrared regulator. Furthermore, since
functions $\tilde F_{\gI \gJ}(1,\xi,x)$ and $\tilde F_{\gI \gJ}(\xi,1,x)$ do not
contain any $\ep$-divergencies at non-vanishing $x$ and $\xi$, their expansion
in $\ep$ is straightforward. Hence, writing
\begin{align}
   \label{eq:tldFIJ-ep-exp}
  &&
     \tilde F_{\gI \gJ}(1,\xi,x) = \tilde F_{\gI \gJ}^{(0)}(1,\xi,x)  + {\cal O}(\ep),
  &&
     \tilde F_{\gI \gJ}(\xi,1,x) = \tilde F_{\gI \gJ}^{(0)}(\xi,1,x)  + {\cal O}(\ep),
  &&
\end{align}
we obtain
\begin{equation}
  \begin{split}
    \Delta W_{RR,\gI \gJ}^{\rm dip, inc} =  -a_s^2 \;
    \int \limits_{0}^{1} \frac{ {\rm d} x}{x}   \frac{{\rm d} \xi}{\xi}  \;  \int \limits_{0}^{2 \pi} \frac{{\rm d} \varphi}{2 \pi}
    \Bigg [
    & \tilde F^{(0)}_{\gI \gJ}(1,\xi, x) \;  \ln \left ( \frac{T_{N, \gI \gJ}(1,\xi,x)}{T_{N,\gI \gJ}(1,\xi,0)}
      \right )
    \\
    &  +
      \tilde F^{(0)}_{\gI \gJ}(\xi,1, x)
      \ln
      \left ( \frac{T_{N,\gI \gJ}(\xi,1,x)}{T_{N, \gI \gJ}(\xi,1,0)}
      \right )
      \Bigg  ].
  \end{split}
  \label{eq:deltaWIJ-ep-exp}
\end{equation}
We note that we have used $N_2|_{\ep \to 0} = a_s^2/4$ and reconstructed
functions $g^{(4)}_{\gI \gJ}$ and $g^{(4)}_{\gJ \gI}$ from their definitions in
Eq.~\eqref{eq:g4-def-x}. The function $\Delta W_{RR,\gI \gJ}^{\rm dip, inc}$ will
have to be computed for the different cases of mass assignments of hard
emitters.

The master formula for the double-real dipole contribution is given by the
combination of Eqs~\eqref{eq:deltaWIJ-def}, \eqref{eq:WIJ-nnlo-reg-subtr}, \eqref{eq:WIJ-gx0-bare} and
\eqref{eq:deltaWIJ-ep-exp}. We write this combination explicitly in the following section.
We then apply the derived master formula to three different cases of the mass
assignments for both the two-gluon and the quark-antiquark emissions.

\subsubsection{The master formula  for the dipole contribution}
\label{sec:nnlo-rr-dipole-master-formula}

We write the renormalized dipole contributions,
\begin{equation}
  {\tilde S}_2^{\rm dip}
  = \sum_{\gI \gJ} {\bf T}_{\gI} \cdot {\bf T}_{\gJ } \; \tilde{W}_{\gI \gJ}^{(2)},
\end{equation}
as a sum of three terms
\begin{equation}
  \tilde{W}_{\gI \gJ}^{(2)}
  = A_{\gI \gJ}^{(1)}
  + A_{\gI \gJ}^{(2)}
  + A_{\gI \gJ}^{(3)},
\end{equation}
where
\begin{equation}
  \begin{split}
    A_{\gI \gJ}^{(1)}
    & =  \Delta W_{RR,\gI \gJ}^{\rm dip}
      -a_s^2 \;
      \int \limits_{0}^{1} \frac{ {\rm d} x}{x}   \frac{{\rm d} \xi}{\xi}    \int \limits_{0}^{2 \pi} \frac{{\rm d} \varphi}{2 \pi}
      \Bigg [
      \tilde F^{(0)}_{\gI \gJ}(1,\xi, x) \;  \ln \left ( \frac{T_{N, \gI \gJ}(1,\xi,x)}{T_{N, \gI \gJ}(1,\xi,0)}
      \right ) \\
    & + \tilde F^{(0)}_{\gI \gJ}(\xi,1, x)
      \ln
      \left ( \frac{T_{N, \gI \gJ}(\xi,1,x)}{T_{N, \gI \gJ}(\xi,1,0)}
      \right )
      \Bigg  ]
    \\
    & + a_s^2\beta_0
      \int \limits_{0}^{1} \frac{{\rm d} \xi}{\xi}
      \int \limits_{0}^{2 \pi} \frac{{\rm d} \varphi}{2 \pi}
      \Bigg [
      R_{\gI \gJ}(1,\xi)
      \; \ln \left(\frac{C_\gI f_\gI(\xi) ( \bar \mu u)^2 T_{N,\gI \gJ}(1,\xi,0)}{2 \xi}\right)
      \ln \left(\frac{T_{N, \gI \gJ}(1,\xi,0)}{\tilde p_\gI \tilde p_\gJ C_\gI f_\gI(\xi) }\right)\\
    & + R_{\gI \gJ}(\xi,1)
      \ln \left(\frac{C_\gJ f_\gJ(\xi) ( \bar \mu u)^2 T_{N,\gI \gJ}(\xi,1,0)}{2 \xi}\right)
      \ln \left(\frac{T_{N, \gI \gJ}(\xi,1,0)}{\tilde p_\gI \tilde p_\gJ C_\gJ f_\gJ(\xi) }\right)
      \Bigg ],
  \end{split}
\end{equation}
is the finite term that only depends on the real-emission contributions and
involves the $N$-jettiness functions $T_N$ and ${\cal T}_N$.

The second contribution $A_{\gI \gJ}^{(2)}$ also depends on the $N$-jettiness
variable. However, it becomes finite only after the double-real, the
real-virtual and the coupling constant renormalization terms are combined. It
reads
\begin{equation}
  \begin{split}
    A_{\gI \gJ}^{(2)} =
    -a_s^2\;
    \int \limits_{0}^{1} \frac{{\rm d} \xi}{\xi}
    \int \limits_{0}^{2 \pi} \frac{{\rm d} \varphi}{2 \pi}
    \Bigg [
    K_{\gI \gJ}(\xi)
    \ln \left(\frac{T_{N, \gI \gJ}(1,\xi,0)}{\tilde p_\gI \tilde p_\gJ C_\gI f_\gI(\xi) }\right)
    + \;
    K_{\gJ \gI}(\xi)
    \ln \left(\frac{T_{N, \gI \gJ}(\xi,1,0)}{\tilde p_\gI \tilde p_\gJ C_\gJ f_\gJ(\xi) }\right)
    \Bigg ],
  \end{split}
  \label{eq:RR-A2-master}
\end{equation}
with
\begin{equation}
  \begin{split}
    & K_{\gI \gJ}(\xi) =
      \lim_{\ep \to 0} \left ( \tilde F_{\gI \gJ}^{RV}(1,\xi)
      + F_{\gI \gJ}^{RR}(1,\xi)
      +
      \frac{\beta_0}{\ep} \; R_{\gI \gJ}(1,\xi)
      \right ),
    \\
    & K_{\gJ \gI}(\xi) =
      \lim_{\ep \to 0}
      \left ( \tilde F_{\gI \gJ}^{RV}(\xi,1)
      + F_{\gI \gJ}^{RR}(\xi,1)
      +
      \frac{\beta_0}{\ep} \; R_{\gI \gJ}(\xi, 1)
      \right ).
  \end{split}
  \label{eq:RR-KIJ-master}
\end{equation}

The last  term, $A_{\gI \gJ}^{(3)}$ is independent of the $N$-jettiness function. It reads
\begin{equation}
  \begin{split}
    A_{\gI \gJ}^{(3)} =
    &   - \frac{N_2}{\ep}\;
      \left ( \frac{ \tilde p_{\gI} \tilde p_{\gJ} }{2}
      \right )^{2\ep}
      \int \limits_{0}^{1}
      {\rm d} \xi \xi^{-1-2\ep}
      \Bigg [
      C_\gI^{4 \ep} f^{4\ep}_\gI(\xi)
      \;
      \tilde K_{\gI \gJ}(\xi)
      +
      C_\gJ^{4 \ep} f^{4\ep}_\gJ(\xi)
      \;
      \tilde K_{\gJ \gI}(\xi)
      \Bigg ]
      -Z_{2,\gI \gJ} - Z_{2,\gI \gJ}^+,
  \end{split}
  \label{eq:RR-A3-master}
\end{equation}
where we have used the independence of the integrand on the azimuthal angle to
integrate over $\dOm$. Furthermore, the renormalization constants $Z_2$ can be
found in Appendix~\ref{sec:app-ren}, and the two auxiliary functions
read
\begin{equation}
  \begin{split}
    & \tilde K_{\gI \gJ}(\xi) = \tilde F_{\gI \gJ}^{RV}(1,\xi)
      + F_{\gI \gJ}^{RR}(1,\xi)
      +
      \frac{\beta_0}{\ep} \frac{a_s N_1}{N_2}\left(\frac{\tilde p_{\gI} \tilde p_{\gJ}}{2}\right)^{-\ep}
      \xi^{\ep} \; C_\gI^{-2 \ep} f^{-2\ep}_\gI(\xi) \; R_{\gI \gJ}(1,\xi),
    \\
    & \tilde K_{\gJ \gI}(\xi) = \tilde F_{\gI \gJ}^{RV}(\xi,1)
      + F_{\gI \gJ}^{RR}(\xi,1)
      +
      \frac{\beta_0}{\ep} \frac{a_s N_1}{N_2}\left(\frac{\tilde p_{\gI} \tilde p_{\gJ}}{2}\right)^{-\ep} \xi^{\ep} \; C_\gJ^{-2 \ep} f^{-2\ep}_\gJ(\xi) \; R_{\gI \gJ}(\xi, 1)
      .
      \label{eq:RR-tldKIJ-master}
  \end{split}
\end{equation}
The term $A^{(3)}_{\gI \gJ}$ is finite in the $\ep \to 0$ limit as well. In the
next sections we apply the general formulas presented here to the three cases of
mass assignments.

\subsection{Two massless emitters}

The case of two massless emitters has been extensively studied in the literature
\cite{Bell:2023yso,Agarwal:2024gws,Buonocore:2026yai}. Nevertheless, we present
it here for completeness, to illustrate how the general formulas of the previous
section apply to this well-known case.

For the massless emitters,
$\gI = i$, $\gJ = j$, $C_\gI = Q_i^{-1}$, $C_\gJ = Q_j^{-1}$,
$f_{\gI, \gJ}(\xi) = \xi $ and $\tilde p_{\gI} = p_i$, $\tilde p_{\gJ} = p_j$.
In addition, we require the following functions
\begin{align}
  && \tilde F_{ij}^{RV}(\alpha, \beta) = \frac{C_A}{\ep^2} A_K(\ep),
  && R_{ij}(\alpha,\beta) = 1,
  &&
\end{align}
where
\begin{equation}
  A_K = \frac{\Gamma^3(1+\ep)\Gamma^5(1-\ep)}{\Gamma(1+2\ep) \Gamma^2(1-2\ep)}.
\end{equation}
The function $\tilde F_{ij}$ in this case is $\xi$-independent. It reads
\begin{equation}
  \tilde F_{ij} = -\frac{C_A}{4} \tilde F^{gg}_{ij} + \frac{T_R n_f}{2}
  \tilde F_{ij}^{q \bar q},
\end{equation}
where
\begin{equation}
  \begin{split}
    \tilde F^{gg}_{ij}
     =
      \frac{2(3-\ep)(-2+\ep)}{\ep (3-2\ep)} +
      \frac{8 (1-2\ep) x}{\ep}  \; & {}_2F_{1}(1,1,1-\ep,1-x) \\
    - \frac{4(1-2\ep)}{\ep} (1-x) \; & {}_2F_{1}(1,1,1-\ep,x),
  \end{split}
\end{equation}
and
\begin{equation}
  \tilde F_{ij}^{q\bar q}= -\frac{4(1-\ep)}{(3-2\ep)}.
\end{equation}

We use these definitions to compute the three quantities $A_{ij}^{(1,..,3)}$.
The first step is to expand the function $\tilde F_{ij}$ in $\ep$. We obtain
\begin{equation}
  \tilde F_{ij}^{(0)}
  = \left(\frac{11}{6} - \ln(1-x) + 2 \ln x \right)C_A
  -   \frac{2}{3} T_R n_f.
\end{equation}

Integrals over $x$ of the function $\tilde F_{ij}$ are needed to construct the
function $F_{RR}$ which, for the massless emitters, is $\xi$-independent. We
write
\begin{equation}
  F_{RR}(1,\xi) = F_{RR}(\xi,1)  =
  \frac{\Gamma(1-\ep)^2}{\Gamma(2-2 \ep)}
  \left (-
    \frac{C_A}{4}
    \gamma_g + \frac{T_R n_f}{2}
    \gamma_q
  \right ),
\end{equation}
where
\begin{equation}
  \begin{split}
    & \gamma_g  = \frac{4}{\ep^2}
      -\frac{2}{3 \ep} +  \left (
      \frac{2}{9} - \frac{4  \pi^2}{3}
      \right )
      + \ep \left (
      \frac{4}{27} + \frac{13 \pi^2}{9} - 28 \zeta_3 \right )
      +
      \ep^2\left (  \frac{8}{81} - \frac{\pi^2}{27}
      - \frac{61 \pi^4}{90} + \frac{124}{3} \zeta_3 \right ),
    \\
    & \gamma_q =
      \frac{4}{3 \ep} -
      \frac{4}{9}
      -\ep \left (
      \frac{8}{27}
      + \frac{6 \pi^2}{27}
      \right )
      +\ep^2
      \left (
      -\frac{16}{81}
      + \frac{2 \pi^2}{27}
      - \frac{8 \zeta_3}{3}
      \right ).
  \end{split}
\end{equation}

Functions $K_{ij}$ and $K_{ji}$, defined in Eq.~\eqref{eq:RR-KIJ-master}, are finite and
$\xi$-independent. They evaluate to
\begin{equation}
  K_{ij}(\xi) = K_{ji}(\xi) = C_A \left ( -\frac{67}{18}
    + \frac{\pi^2}{6}
  \right ) + \frac{10}{9} n_f T_R.
\end{equation}
We note that, up to a minus sign, this quantity is the  two-loop cusp anomalous dimension first computed in Ref.~\cite{Catani:1990rr}.

It remains to determine the quantity $A_{\gI \gJ}^{(3)}$ which requires
integration of $F_{RR}$ and $\tilde F^{RV}_{ij}$ over $\xi$. However, since
these functions do not depend on $\xi$, this integration is trivial. We obtain
\begin{equation}
  \begin{split}
    & A_{ij}^{(3)} =   \left(\frac{11}{36} C_A - \frac{ T_R n_f}{9}\right) L_{ij}^3 + \bigg[\left(\frac{67}{36}   - \frac{\pi^2}{12}   \right)C_A - \frac{5}{9} T_R n_f \bigg]\left(L_{ij}^2 + \ln^2 \left(\frac{Q_i}{Q_j}\right) \right)
      -\frac{5\pi^4}{72} C_A
    \\
    & + \bigg[\left(\frac{101}{27} + \frac{11 \pi^2}{36} + \frac{11}{12} \ln^2 \left(\frac{Q_i}{Q_j}\right) - \frac{7}{2} \zeta_3 \right) C_A - \left(\frac{28}{27} +\frac{\pi^2}{9} - \frac{1}{3} \ln^2 \left(\frac{Q_i}{Q_j}\right)\right)T_R n_f \bigg]L_{ij}
    \\
    &  + \left(\frac{335}{432} C_A - \frac{25}{108} T_R n_f\right) \pi^2 +\left(\frac{121}{36} C_A - \frac{11}{9} T_R n_f \right) \zeta_3 + \left(\frac{607}{162} C_A - \frac{82}{81}T_R n_f \right).
  \end{split}
\end{equation}

Finally, putting the different terms together, we derive  the final result for the case of two massless emitters
\begin{equation}
  \begin{split}
    & \tilde{W}_{ij}^{(2)} =   \left(\frac{11}{36} C_A - \frac{ T_R n_f}{9}\right) L_{ij}^3 + \bigg[\left(\frac{67}{36}   - \frac{\pi^2}{12}   \right)C_A - \frac{5}{9} T_R n_f \bigg]\left(L_{ij}^2 + \ln^2 \left(\frac{Q_i}{Q_j}\right) \right) \\
    & + \bigg[\left(\frac{101}{27} + \frac{11 \pi^2}{36} + \frac{11}{12} \ln^2 \left(\frac{Q_i}{Q_j}\right) - \frac{7}{2} \zeta_3 \right) C_A - \left(\frac{28}{27} +\frac{\pi^2}{9} - \frac{1}{3} \ln^2 \left(\frac{Q_i}{Q_j}\right)\right)T_R n_f \bigg]L_{ij}
    \\
    &  -\frac{5 \pi^4}{72} C_A +\left(\frac{335}{432} C_A - \frac{25}{108} T_R n_f\right) \pi^2 +\left(\frac{121}{36} C_A - \frac{11}{9} T_R n_f \right) \zeta_3 + \left(\frac{607}{162} C_A - \frac{82}{81}T_R n_f \right) \\
    &+ \int \limits_{0}^{1} {\rm d} x \frac{{\rm d} \xi}{\xi}  \;  \int \limits_{0}^{2 \pi} \frac{{\rm d} \varphi}{2 \pi} \Bigg \{ -\frac{\tilde F_{ij}^{(0)}(x)}{x} \ln \left(\frac{t_i(\xi,x) \; t_j (\xi,x)}{t_i(\xi,0) \; t_j (\xi,0)}\right)  + \bigg[\frac{11}{12} C_A - \frac{1}{3}T_R n_f\bigg]\\
    & \times \left(\ln^2 \left(t_i(\xi,0) \; t_j (\xi,0)\right)
      +  \ln^2 \left(\frac{t_i(\xi,0) }{t_j(\xi,0)  }\right) \right) + \ln \left(t_i(\xi,0) \; t_j (\xi,0)\right)\\
    &\times \bigg[\left(\frac{67}{18} -\frac{\pi^2}{6} + \frac{11}{6} L_{ij} + \frac{11}{6} \ln \xi \right) C_A - \left(\frac{10}{9} + \frac{2}{3} L_{ij} + \frac{2}{3} \ln \xi \right) T_R n_f\bigg]\\
    &+ \ln\left(\frac{t_i(\xi,0)}{t_j(\xi,0)}\right) \left( \frac{11}{6}C_A - \frac{2}{3} T_R n_f \right) \ln \left(\frac{Q_j}{Q_i}\right) \bigg  \}  +\Delta W_{RR,ij}^{\rm dip} ,
      \label{eq:dip-nnlo-00-final}
  \end{split}
\end{equation}
where we have used
\begin{align}
  && t_i(\xi,x) = \frac{Q_i T_{N,ij}(1,\xi,x)}{(p_i  p_j) \xi},
  && t_j(\xi,x) = \frac{Q_j T_{N,ij}(\xi,1,x)}{(p_i  p_j) \xi},
  &&
\end{align}
to  simplify the result.

%%%%%%%%%%%%%%%
\subsection{Massive and massless emitters}
We continue with the case of one massive and one massless emitter, and take
  $\gI = I,$ and $\gJ = j$. Then,
\begin{align}
  && p_I = \tilde p_I + \kappa_{Ij} p_j,
  && \tilde p_j = p_j,
  && \kappa_{Ij} = \frac{m_I^2}{2 p_I p_j},
  && {\cal R}_{I j}(\alpha, \beta)  =
     \left (\frac{\beta }{\beta + \alpha \; \kappa_{Ij}} \right )^2.
  &&
\end{align}
We also choose

\begin{align}
  && C_\gJ = C_\gI = Q_j^{-1},
  && f_\gI = 1,
  && f_\gJ = \xi.
  &&
\end{align}

We proceed with the discussion of functions that describe the double-real and
real-virtual contributions. We begin with the double-real emission and write
\begin{equation}
\tilde F_{I j}
= -\frac{C_A}{4} \tilde F_{I j}^{gg} + \frac{T_R n_f}{2}
\tilde F_{Ij}^{q \bar q}.
\end{equation}

To compute the two functions, we use angular integrals defined in
Appendix~\ref{sec:app-angular}. We find
\begin{equation}
\begin{split}
    \tilde F_{Ij}^{gg}(q,p_I,p_j) &  = -2m_I^2 \beta q^2 \frac{ p_{Ij} \; q_I - m_I^2 \; q_j}{q_I^2 \; (q_I^2 - m_I^2 q^2)} -\beta \frac{ 6 (1 + \ep - \ep^2) p_{Ij} \;q_I - \ep (11 - 7 \ep) m_I^2 \;q_j}{\ep (3 -
   2 \ep) q_I^2}   \\
   &+ 2 \beta  \left(\frac{2 q_I^2 - m_I^2 q^2}{q_I^2 - m_I^2 q^2 }\right) \frac{p_{Ij} q_I^3 -p_I^4 q^2 q_j }{\;q_I^4  } I_{1}^{(1)}\left[ \frac{m_I^2 q^2}{q_I^2}\right]
   \\
   & + 4 \beta q^2 \frac{p_{Ij}(m_I^2 q_j - 2 p_{Ij} q_I)}{q_I^3 \;q_j} I_{1,1}^{(1)}\left[\frac{p_{Ij} q^2}{q_I q_j}, \frac{m_I^2 q^2}{q_I^2}, 0\right]  \\
   & + 4 \beta \frac{(q_I p_{Ij} - m_I^2 q_j)(2q_I q_j - q^2 p_{Ij})}{q_I^3 q_j } I_{1,1}^{(1)}\left[2-\frac{p_{Ij} q^2}{q_I q_j}, \frac{m_I^2 q^2}{q_I^2}, 0\right],
   \label{eq:tld-FIj-M0}
\end{split}
\end{equation}
and
\begin{equation}
\tilde F_{Ij}^{qq}(q,p_I,p_j)
= -\frac{4(1-\ep)}{3-2\ep}
\frac{\beta}{2 q_I^2}
\left ( 2 q_I p_{Ij} -  m_I^2 q_j
\right ).
\end{equation}
In the above expressions, we have used  the following notations
\begin{align}
  &&
     p_{Ij} = p_I p_j,
  &&
     q_I = q \cdot p_I,
  &&
     q_j = q \cdot p_j,
  &&
\end{align}
as well as the Sudakov decomposition of
the vector $q$
\begin{equation}
q = \alpha \tilde p_I + \beta p_j + q_\perp.
\end{equation}
Computing the scalar products, we  find
\begin{align}
  &&
     p_{Ij} = \tilde p_I p_j,
  &&
     q^2 = 2 p_{Ij}  \alpha \beta x,
  &&
     q_I = p_{Ij}\left ( \beta + \alpha  \kappa_{Ij}  \right),
  &&
     q_j = p_{Ij} \alpha.
  &&
\end{align}
\\

We continue with the calculation of the three quantities $A^{(i)}_{Ij}$, with
$i = 1,2,3$, introduced in Section~\ref{sec:nnlo-rr-dipole-master-formula}. To compute $A^{(1)}_{Ij}$, we
require the function $\tilde F_{Ij}$ expanded in $\ep$. This requires
$\ep$-expansion of the function $I_{1,1}^{(1)}$ which turns out to be quite
involved; we explain how to derive it in Appendix~\ref{sec:app-ps-ints-simp}. For $gg$ and
$q \bar q$ contributions, we find
\begin{equation}
\begin{split}
    \tilde F_{Ij}^{gg,(0)}
    (1,\xi,x)=
    & -\frac{2}{(1+u)^2} \bigg [ \frac{(1+ \eta^2)}{\eta^3} \left(\frac{1-(1-2\eta^2)u}{2}\right) \ln \left(\frac{1-\eta}{1+\eta}\right)
    - 2 \ln\left( \frac{u}{x} \right) \\
    & + (1-u) \ln\left( \frac{u x}{(1+u-x)^2} \right) +\frac{1-u}{\eta^2} + u + \frac{8}{3}\bigg],
    \end{split}
\end{equation}
and
\begin{equation}
  \tilde F_{Ij}^{qq,(0)}(1,\xi,x)
  = -\frac{4}{3} \; \frac{1}{( 1+ u)^2}.
\end{equation}
where the following notations have been used
\begin{equation}
  \eta= \sqrt{1-\frac{4 x u}{(1+u)^2}}, \quad u= \frac{\kappa_{Ij}}{\xi}.
  \label{eq:eta-u-M0}
\end{equation}
Functions $\tilde F^{(0),gg}(\xi,1,x)$ and $\tilde F^{(0),qq}(\xi,1,x)$, that
are also needed for computing $A^{(1)}_{I j}$, can be obtained from the previous
results by using the following identities
\begin{align}
  &&
     \tilde F_{Ij}^{(0),gg}(\xi,1,x) = \tilde F_{Ij}^{(0),gg}(1,\xi^{-1},x),
  &&
     \tilde F_{Ij}^{(0),qq}(\xi,1,x) = \tilde F_{Ij}^{(0),qq}(1,\xi^{-1},x),
  &&
\end{align}
which follow from the homogeneity of the function $\tilde F_{Ij}$.

For the second term $A_{Ij}^{(2)}$ two quantities are needed -- the integral of
the double-real contribution over $x$ and the (rescaled) real-virtual function
$\tilde F^{\rm RV}_{Ij}$. We begin with the discussion of the double-real
integral.

At variance with the case of the massless emitters, integrating
$\tilde F_{Ij}^{ gg}(1,\xi,x)$ and $\tilde F_{Ij}^{gg}(\xi,1,x)$ over $x$ is
non-trivial as this quantity involves complicated special functions, and,
furthermore, these integrals diverge. To get around this problem, we write
\begin{equation}
  \tilde F_{Ij}^{gg}(q,p_I,p_j)
  = \tilde F_{Ij}^{gg, x \sim 0} (q,p_I,p_j)
  + \Delta \tilde F_{Ij}^{gg} (q,p_I,p_j),
\end{equation}
where $\tilde F_{Ij}^{ gg, x \sim 0}$ is obtained by extracting the $x \sim 0$
asymptotic from the function $\tilde F_{Ij}^{gg}$. Since the function
$\Delta \tilde F^{gg}_{Ij}$ is sufficiently smooth at $x=0$, one can integrate
it over $x$ by expanding the corresponding \emph{integrand} in powers of
$\ep$.

We note that the extraction of the $x \sim 0$ asymptotic of the function
$\tilde F_{Ij}^{gg}$ requires non-trivial manipulations of integrals $I_1^{(1)}$
and $I_{1,1}^{(1)}$; the corresponding discussion can be found in Appendix~\ref{sec:app-ps-ints-simp}. We find
\begin{equation}
  \begin{split}
    &  \tilde F_{Ij}^{gg,x \sim 0} (q,p_I,p_j)
      =  B_{Ij} \;\beta
      \Bigg \{
      2^{2+\ep}\frac{\Gamma(1+\ep) \Gamma(2-2\ep)}{\ep \;\Gamma(1-\ep)}
      \left ( \frac{m_I^2 \;q_j \beta \;  x }{q_I^2} \right )^{-\ep}
      \times
    \\
    & \left [
      1 + \left ( \frac{m_I^4 q_j^2}{4 p_{Ij}^2 q_I^2} \right )^{\ep}
      {}_2F_{1}(-2\ep,-\ep,1-\ep,1-\frac{2 p_{Ij} q_I}{m_I^2 q_j})
      \right  ] -\frac{6(1-\ep)(4-3\ep)}{\ep (3-2\ep)}
      \Bigg \},
  \end{split}
\end{equation}
where
\begin{equation}
 B_{Ij} =  \frac{2 q_I p_{Ij}-m_I^2 q_j}{2 q_I  }.
\end{equation}
This allows us to compute $F_{Ij}^{RR}$ in the following way
\begin{equation}
  F_{I j}^{RR}(\xi)
  = \frac{\Gamma(1-\ep)^2}{\Gamma(2-2\ep)} \int \limits_{0}^{1} {\rm d} x \;
  x^{-\ep-1} (1-x)^{-\ep} \tilde F_{I j}^{x \sim 0}(1,\xi,x) + \int \limits_{0}^{1} \frac{{\rm d} x}{x} \;\Delta \tilde F_{I j}(1,\xi,x).
\end{equation}
We also note that the function $\tilde F_{Ij}^{q \bar q}$ is $x$-independent, so
that $ \Delta \tilde F_{Ij}^{q \bar q} = 0 $.

According to Eq.~\eqref{eq:RR-KIJ-master}, the above integral needs to be combined with
the real-virtual contribution and the remnant of the coupling constant
renormalization. The real-virtual contribution is described by the function
$\tilde F^{\rm RV}_{Ij}$ which can be extracted from Ref.~\cite{Bierenbaum:2011gg}
\begin{equation}
  \begin{split}
    & \tilde F^{\rm RV}_{Ij} (\alpha, \beta) = \frac{2^{-\ep}C_A}{\ep^2} \; \frac{ \Gamma(1+2\ep)}{  \Gamma(1-\ep)} \beta^{1+\ep}  \Bigg\{ \cos(2\pi\ep) \Gamma(1-\ep)  \bigg[ \left( \frac{m_I^2 \;q_j}{q_I^2} \right)^{1+\ep} \\
    &+2 \left( \frac{m_{I}^2 \;q_j}{q_I^2} \right)^\ep  \frac{q_I p_{Ij} - m_I^2 q_j }{q_I^2 } \; {}_2F_{1}\left(1,-\ep,1-\ep,1-\frac{2 p_{Ij} q_I}{m_I^2 q_j}\right) \bigg]+ \Gamma(1-2\ep) \Gamma(1+\ep) \\
    & \times \bigg[ \frac{p_{Ij}}{q_I} \left( \frac{4 p_{Ij}^2}{m_I^2 q_j} \right)^\ep + \left( \frac{m_{I}^2\; q_j}{q_I^2} \right)^\ep  \frac{q_I p_{Ij} - m_I^2 q_j }{q_I^2 } \; {}_2F_{1}\left(-2\ep,-\ep,1-\ep,1-\frac{2 p_{Ij} q_I}{m_I^2 q_j}\right) \bigg] \Bigg\}.
  \end{split}
  \label{eq:tldFRV-M0}
\end{equation}

Combining the real-real, real-virtual and the coupling constant renormalization
contributions and expanding  $\ep$, we obtain finite
$\xi$-dependent functions $K_{Ij}$ and $K_{jI}$. They are given by
\begin{equation}
  \begin{split}
    K_{Ij}(\xi) =
    & \frac{1}{(1+u)^2}\bigg[-\frac{C_A}{2}\Big[\frac{67}{9}  +2 u \;{\rm Li}_2(-u) -u \ln^2 u - 2(1+u)\ln(1+u) \\
    &- 2 (1-u) \left( \ln(1+u) \ln u + \frac{u \ln u}{1+u} + \frac{\pi^2}{6} \right)\Big] +  \frac{10 \;T_R n_f}{9}\bigg],
    \\
    K_{jI}(\xi) =
    & K_{Ij}(\xi^{-1}),
  \end{split}
\end{equation}
where the variable $u$ can be found in Eq.~\eqref{eq:eta-u-M0}. Thanks to the
availability of the functions $K_{Ij}$ and $K_{jI}$, the computation of the
$A^{(2)}_{Ij}$ term defined in Eq.~\eqref{eq:RR-A2-master} becomes straightforward.

What remains is the computation of the third quantity $A^{(3)}_{Ij}$ defined in
Eq.~\eqref{eq:RR-A3-master}. Compared to the $A^{(2)}_{Ij}$ case, this computation is
more complicated because, in addition to the $x$-integration, also integration
over $\xi$ is needed. Since both of these integrations are singular, and since
dependencies of functions $\tilde F_{Ij}^{\rm RV}$ and $\tilde F_{Ij}^{RR}$ on
$x$ and $\xi$ are complex, these integrations are not straightforward. For this
reason, we briefly explain how to perform them.

First, we note that integrals of $R_{Ij}(1,\xi)$ and $R_{Ij}(\xi,1)$ can be
immediately expressed in terms of hypergeometric functions, so we do not need to
discuss them further. Second, functions $F^{\rm RV}_{Ij}(1, \xi)$ and
$F^{\rm RV}_{Ij}(\xi,1)$ have to be integrated over $\xi$ from zero to one with
the weights $\xi^{-1-2\ep}$ and $\xi^{-1+2\ep}$, respectively. We note that
integration of the first function is straightforward because
$F_{Ij}^{\rm RV}(1,\xi)$ vanishes at $\xi = 0$, and, therefore, we only need its
$\ep$ expansion.

The integration of $\tilde F^{\rm RV}_{Ij}(\xi,1)$ is more involved because this
function does not vanish at $\xi = 0$. Hence, we need to extract its $\xi \to 0$
asymptotic before integrating. To this end, we note that the only terms in
Eq.~\eqref{eq:tldFRV-M0} which require discussion are the two hypergeometric
functions. Both of these functions depend on
\begin{equation}
  1 - \frac{2p_{Ij} q_I}{m_I^2q_j} = -\frac{1}{\kappa_{Ij} \xi},
\end{equation}
where we have substituted $\alpha =\xi$ and $\beta = 1$. Hence, we need to
analyse the $\xi \to 0$ asymptotics of the two functions
\begin{align}
  && F_{21}\left (-2\ep,-\ep,1-\ep,-\frac{1}{\kappa_{Ij} \xi}
     \right )
  &&
     {\rm and}
  &&
     F_{21}\left (1,-\ep,1-\ep,-\frac{1}{\kappa_{Ij} \xi}
     \right ).
  &&
\end{align}
To obtain it, we use standard transformation rules to express these functions
through the hypergeometric functions with the argument $-\kappa_{Ij} \xi$, in
which case taking the limit $\xi \to 0$ becomes elementary. Once the singularity
is extracted, integration of the real-virtual contribution over $\xi$ becomes
straightforward.

The last contribution that needs to be discussed is the double-real one. Since
$\tilde F_{Ij}$ depends on $x$ and $\xi$, extraction of singularities in this
case is more involved and we discuss it in Appendix~\ref{sec:app-ps-ints-simp}. The final
result for the coefficient $A^{(3)}_{Ij}$ reads
\begin{equation}
  \begin{split}
    A^{(3)}_{Ij}
    & =
      \frac{1}{2}\bigg[\left ( \frac{11}{36} C_A - \frac{1}{9} n_f T_R \right ) L_{Ij}^3 + \left (
      \left(\frac{17}{18} - \frac{\pi^2}{12} \right) C_A - \frac{2}{9 } n_f T_R \right ) L_{Ij}^2
    \\
    & + \left (
      C_A \left(\frac{55}{54} + \frac{7 \pi^2}{9} -\frac{9}{2} \zeta_3 \right)
      -  n_f T_R
      \left ( \frac{2\pi^2}{9}
      -\frac{2}{27}
      \right )
      \right )
      L_{Ij}
    \\
    &+ C_A
      \left ( \frac{89}{144} \pi^2 + \frac{163}{162} +\frac{283}{36} \zeta_3 -\frac{107}{720} \pi^4
      \right )
      + n_f T_R \left ( - \frac{7 \pi^2}{36} +\frac{2}{81} -
      \frac{11}{9} \zeta_3
      \right )\bigg].
  \end{split}
\end{equation}

Putting the different terms together, we find the final result for the case of one massive and one massless emitter
\begin{align}
  \tilde{W}_{Ij}^{(2)}
  &=  \frac{1}{2}\bigg[\left ( \frac{11}{36} C_A - \frac{1}{9} n_f T_R \right ) L_{Ij}^3
    + \left (\left(\frac{17}{18} - \frac{\pi^2}{12} \right) C_A - \frac{2}{9 } n_f T_R \right ) L_{Ij}^2
    \nonumber \\
  & + \left (
    C_A \left(\frac{55}{54} + \frac{7 \pi^2}{9} -\frac{9}{2} \zeta_3 \right)
    -  n_f T_R
    \left ( \frac{2\pi^2}{9}
    -\frac{2}{27}
    \right )
    \right )
    L_{Ij}
    \nonumber \\
  &+ C_A
    \left ( \frac{89}{144} \pi^2 + \frac{163}{162} +\frac{283}{36} \zeta_3 -\frac{107}{720} \pi^4
    \right )
    + n_f T_R \left ( - \frac{7 \pi^2}{36} +\frac{2}{81} -
    \frac{11}{9} \zeta_3
    \right ) \bigg]
    \nonumber \\
  &+ \int \limits_{0}^{1} {\rm d} x \frac{{\rm d} \xi}{\xi}  \;  \int \limits_{0}^{2 \pi} \frac{{\rm d} \varphi}{2 \pi} \Bigg \{ \frac{1}{x} \left( \frac{C_A}{4} \tilde F_{Ij}^{gg,(0)}(1,\xi,x) - \frac{T_R n_f}{2} F_{Ij}^{q \bar q,(0)}(1,\xi,x) \right)\ln \left(\frac{{ T}_{N,Ij}(1,\xi,x)}{{ T}_{N,Ij}(1,\xi,0) }\right)
    \label{eq:dip-nnlo-M0-final} \\
  &+ \left(\frac{11}{6} C_A- \frac{2}{3}T_R n_f \right)R_{Ij}(1,\xi)\ln^2 \left(\frac{{ T}_{N,Ij}(1,\xi,0)Q_j}{ p_I p_j}\right)
    \nonumber
  \\
  &+ \bigg[\Big[\frac{67}{9} + \frac{11}{3} \left( L_{Ij}+ \ln u \right) +2 u \;{\rm Li}_2(-u) -u \ln^2 u - 2(1+u)\ln(1+u)
    \nonumber \\
  &- 2 (1-u) \left( \ln(1+u) \ln u + \frac{u \ln u}{1+u} + \frac{\pi^2}{6} \right) \Big] \frac{C_A}{2}  - \left(\frac{10}{9} + \frac{2}{3}( L_{Ij}+ \ln u) \right) T_R n_f \bigg]
    \nonumber \\
  & \times R_{Ij}(1,\xi)\ln \left(\frac{{ T}_{N,Ij}(1,\xi,0)Q_j}{ (p_I p_j)}\right) + (\xi \rightarrow \xi^{-1})\bigg  \}  +\Delta W_{RR,Ij}^{\rm dip}.
    \nonumber
\end{align}
We note that, to simplify the above formula, we made use of the homogeneity
property of the $N$-jettiness function which allows us to write
\begin{equation}
  {\cal T}_{N,Ij}(1,\xi,x) = \xi \; {\cal T}_{N,Ij}(\xi^{-1},1,x).
\end{equation}

%%%%%%%%%%%%%%%
\subsection{Two massive emitters}
We continue with the discussion of two massive emitters and write $\gI = I$ and
$\gJ = J$. Then,
\begin{align}
  p_I = \tilde p_I + \kappa_{IJ} \tilde p_J,
  && p_J =  \kappa_{JI} \tilde p_I + \tilde p_J,
  && R_{I J}(\alpha, \beta)  =
  \left( \frac{2 \alpha \beta v_{IJ}}{(1+v_{IJ})(\alpha \kappa_{IJ} + \beta) (\alpha + \beta \kappa_{JI} )}\right)^2.
\end{align}
We choose
\begin{align}
  && C_\gJ = C_\gI =  Q^{-1},
  && f_\gI = 1,
  && f_\gJ = 1,
  &&
\end{align}
where $Q$ does not need to be specified.

Similar to the previous cases, we require the double-real and real-virtual
contributions. We begin with the discussion of the double-real one and write
\begin{equation}
  \tilde F_{I J}
  = -\frac{C_A}{4} \tilde F_{I J}^{gg} + \frac{T_R n_f}{2}
  \tilde F_{IJ}^{q \bar q}.
\end{equation}
where
\begin{equation}
  \tilde F_{IJ}^{q \bar q}(q, q_I, q_J)= -\frac{4(1-\ep)}{(3-2\ep)} \frac{\alpha \beta \;(p_I p_J) (1+v_{IJ})}{4 q_I^2 q_J^2}
  \left ( 2 q_I q_J\;p_{IJ} -  m_I^2 q_J^2 - m_J^2 q_I^2
  \right ),
\end{equation}
and
\begin{equation}
  \tilde F_{I J}^{gg} = \alpha \beta \;(p_I p_J) (1 + v_{IJ}) f^{gg}_{IJ},
\end{equation}
with

\begin{equation}
  \begin{split}
    f^{gg}_{IJ}(q, q_I, q_J)
    & = - m_I^2 q^2 \frac{ p_{IJ} \; q_I - m_I^2 \; q_J}{q_I^2 \;q_J\; (q_I^2 - m_I^2 q^2)} - m_J^2 q^2 \frac{ p_{IJ} \; q_J - p_J^2 \; q_I}{q_J^2 \;q_I\; (q_J^2 - m_J^2 q^2)} \\
    &+ \frac{  (11 - 7 \ep) (m_I^2 \;q_J^2 + m_J^2 \;q_I^2 - 2 q_I q_J p_{IJ}  )}{2 (3 -
      2 \ep) q_I^2 \;q_J^2 }
    \\
    & +  \left(\frac{2 q_I^2 - m_I^2 q^2}{q_I^2 - m_I^2 q^2 }\right) \frac{p_{IJ} q_I^3 -m_I^4 q^2 q_J }{q_I^4 \;q_J } I_{1}^{(1)}\left[ \frac{m_I^2 q^2}{q_I^2}\right] \\
    & +  \left(\frac{2 q_J^2 - m_J^2 q^2}{q_J^2 - m_J^2 q^2 }\right) \frac{p_{IJ} q_J^3 -m_J^4 q^2 q_I }{q_J^4 \;q_I } I_{1}^{(1)}\left[ \frac{m_J^2 q^2}{q_J^2}\right]\\
    & + 2 p_{IJ} \;q^2\frac{(m_I^2 q_J^2 + m_J^2 q_I^2  - 2p_{IJ} q_I q_J)}{q_I^3 \;q_J^3} I_{1,1}^{(2)}\left[\frac{p_{IJ} q^2}{q_I q_J}, \frac{m_I^2 q^2}{q_I^2}, \frac{m_J^2 q^2}{q_J^2}\right]  \\
    & - 2 \frac{(m_I^2 q_J^2 + m_J^2 q_I^2  - p_{IJ} q_I q_J)(2 q_I  q_J - p_{IJ} q^2)}{ q_I^3 \;q_J^3 } I_{1,1}^{(2)}\left[2-\frac{p_{IJ} q^2}{q_I q_J}, \frac{m_I^2 q^2}{q_I^2}, \frac{m_J^2 q^2}{q_J^2}\right].
  \end{split}
\end{equation}
In the above equation, $I_{1,1}^{(2)}$ is the double-massive angular integral presented in Appendix~\ref{sec:app-angular}. We again use the  notation
\begin{align}
  && p_{IJ} = p_I \cdot p_J,
  && \tilde p_{IJ} = \tilde p_I \cdot \tilde p_J,
  && q_I = q \cdot p_I,
  && q_J = q \cdot p_J,
  &&
\end{align}
as well as the Sudakov decomposition of the vector $q$
\begin{equation}
  q = \alpha \tilde p_I + \beta \tilde p_J + q_\perp.
\end{equation}

Computing the scalar products, we easily find
\begin{align}
  && p_{IJ} = \frac{2}{1+v_{IJ}}\tilde p_{IJ},
  && q^2 = 2 \tilde p_{IJ}   \alpha \beta x,
  && q_I = \tilde p_{IJ} \left ( \beta + \alpha  \kappa_{IJ}  \right),
  && q_J = \tilde p_{IJ} \left ( \alpha + \beta  \kappa_{JI}  \right).
\end{align}
Similar to the previous cases, we proceed to calculate the three contributions
$A^{(i)}_{IJ}$, with $i = 1,2,3$, introduced earlier. The first quantity that we
require is the function $\tilde F_{IJ}$ expanded in $\ep$. The $\ep$-expansion
of $I_{1,1}^{(2)}$ is computed similar to the case of one massive and one
massless emitter, discussed in Appendix~\ref{sec:app-ps-ints-simp}. For $gg$ and $q \bar q$
contributions, we obtain
\begin{equation}
  F_{IJ}^{q \bar q,(0)}(\alpha,\beta,q^2) = -\frac{4}{3} R_{IJ}(\alpha,\beta),
\end{equation}
and
\begin{equation}
  \begin{split}
    F_{IJ}^{gg,(0)}(\alpha,
    & \beta,q^2) =  \frac{2}{v_{IJ}} R_{IJ}(\alpha,\beta) \ln \left(\frac{1-v_{IJ}}{1+ v_{IJ}}\right) + \frac{A_{IJ}}{2} \left(\frac{2-c_I}{(1-c_I)^{3/2}}\right)\frac{p_{IJ} q_I^3 - m_I^4 q_J q^2}{q_I^4 q_J} L_{c_I}\\
    &+ \left(2 R_{IJ}(\alpha,\beta) - \frac{A_{IJ}}{2}\frac{p_{IJ}}{q_I\; q_J}\right)\frac{(2 q_I q_J - p_{IJ} q^2)}{q_I \;q_J \;\eta_{IJ}} \log \left(\frac{p_{IJ} q^2- (2 + \eta_{IJ})\;q_I \; q_J}{p_{IJ} q^2- (2 - \eta_{IJ})\;q_I \; q_J} \right) \\
    & - \frac{11}{3} R_{IJ}(\alpha,\beta) +  A_{IJ} \;q^2 \frac{m_I^2}{1-c_I} \frac{m_I^2 \; q_J- p_{IJ}q_I}{q_I^4 \;q_J} +\{ I \leftrightarrow J\},
  \end{split}
\end{equation}
where
\begin{align}
  && A_{IJ} = \alpha \beta \; p_{IJ}(1+ v_{IJ}),
  && c_I = \frac{m_I^2 q^2}{q_I^2},
  && \eta_{IJ} = \sqrt{4- \frac{4p_{IJ}q^2}{q_I\; q_J}
     + \left(\frac{p_{IJ} \;q^2 v_{IJ}}{q_I\; q_J}\right)^2},
  &&
\end{align}
and
\begin{equation}
  L_{c_I} = \ln \left(\frac{1+ \sqrt{1-c_I}}{1- \sqrt{1-c_I}}\right).
\end{equation}

We continue with the discussion of the quantity $A_{IJ}^{(2)}$ for the massive-massive case where the double-real and real-virtual  contributions
are required.   We deal with the double-real contribution similarly  to the massive-massless case, isolating  the $x \to 0$ asymptotic
from the rest. We write
\begin{equation}
  \tilde F_{IJ}^{gg}(q,p_I,p_J)
  = \tilde F_{IJ}^{gg, x \sim 0} (q,p_I,p_J)
  + \Delta \tilde F_{IJ}^{gg} (q,p_I,p_J),
\end{equation}
where the $\Delta \tilde F_{IJ}^{gg} (q,p_I,p_J)$ vanishes at $x = 0$. To
extract the $x \sim 0$ asymptotic from the function $\tilde F_{IJ}^{gg}$ we
follow the same steps as discussed in Appendix~\ref{sec:app-ps-ints-simp} for the
massive-massless case and find
\begin{equation}
  \begin{split}
    F_{IJ}^{gg,x \sim 0} (q,p_I,p_J)
    &
      =  B_{IJ} \;\alpha\beta \;\frac{p_{IJ}(1+v_{IJ}) }{4}
      \Bigg \{ -\frac{6(1-\ep)(4-3\ep)}{\ep (3-2\ep)} \\
    &+ \frac{2^{2+2\ep}\Gamma(1+\ep) \Gamma(2-2\ep)}{\ep\;\Gamma(1-\ep)}
      \left ( \frac{ m_I^2 \;\alpha \beta \;p_{IJ} (1+ v_{IJ})x }{q_I^2} \right )^{-\ep} G_{IJ} \Bigg \},
  \end{split}
\end{equation}
where
\begin{equation}
  \begin{split}
    & G_{IJ}=
      1+
      \left(
      \frac{m_I^2q_J^2}{m_J^2q_I^2}
      \right)^\ep
    \\
    &+
      \frac{1}{v_{IJ}}
      \left(
      \frac{2p_{IJ}v_{IJ}X_F}
      {m_I^2B_{IJ}q_Iq_J^2}
      \right)^{-2\ep}
      {}_2F_1\left(
      -2\ep,-\ep;1-\ep;
      -\frac{X_F^2}
      {m_I^2B_{IJ}q_I^2q_J^2}
      \right)
    \\
    &-
      \frac{1}{v_{IJ}}
      \left(
      \frac{2m_Jv_{IJ}X_F}
      {m_I(1+v_{IJ})B_{IJ}q_Iq_J^2}
      \right)^{-2\ep}
      {}_2F_1\left(
      -2\ep,-\ep;1-\ep;
      -\frac{m_J^2X_F^2}
      {p_{IJ}^2(1+v_{IJ})^2B_{IJ}q_I^2q_J^2}
      \right).
  \end{split}
\end{equation}
and
\begin{equation}
  B_{IJ} = \frac{2 q_I q_J p_{IJ}-m_I^2 q_J^2- m_J^2 q_I^2}{ q_I^2 q_J^2  },
  \quad
  X_F = p_{IJ} q_I(1+ v_{IJ})-m_I^2 q_J.
\end{equation}
We note that in the massive-massive case, the $\xi \to 0$ limits are
non-singular and do not need to be extracted. We also note that similar to the
massive-massless case $\Delta \tilde F_{IJ}^{q \bar q} = 0$. Finally, the
function $F_{IJ}^{RR}$ required for the calculation of $A^{(2)}_{IJ}$ is
computed as follows
\begin{equation}
  F_{I J}^{RR}(\alpha,\beta)
  = \frac{\Gamma(1-\ep)^2}{\Gamma(2-2\ep)} \int \limits_{0}^{1} {\rm d} x \;
  x^{-\ep-1} (1-x)^{-\ep} \tilde F_{I J}^{x \sim 0}(\alpha,\beta,x) + \int \limits_{0}^{1} \frac{{\rm d} x}{x} \;\Delta \tilde F_{I J}(\alpha,\beta,x).
\end{equation}
The subtracted term $\Delta \tilde F_{I J}(\alpha,\beta,x)$ is straightforward
to obtain. It reads
\begin{equation}
  \begin{split}
    \Delta \tilde F_{I J}(\alpha,\beta,x)
    & =\frac{1}{4 \alpha \beta x}\Bigg\{
      (2X_{IJ}-X_I-X_J)\ln\!\left(\frac{X_I X_J}{16}\right)
    \\
    &-2(X_I+X_J-X_{IJ})\,
      H\left(\frac{X_I X_J}{(2-X_{IJ})^2}\right)
      -F(X_I;X_{IJ})-F(X_J;X_{IJ})\Bigg\},
  \end{split}
  \label{eq:delta-tld-FIJ-MM}
\end{equation}
where
\begin{align}
  &&
     H(z)=\frac{\operatorname{arctanh}\sqrt{1-z}}{\sqrt{1-z}},
  &&
     F(z;d)=\frac{z(d-z)+(2-z)(z^2-d)H(z)}{1-z},
  &&
\end{align}
and
\begin{align}
  X_I=\frac{m_I^2 \alpha\beta x \;p_{IJ}(1+v_{IJ})}{q_I^2},
  && X_J=  \frac{m_J^2  \;\alpha\beta x\; p_{IJ}(1+v_{IJ})}{q_J^2},
  && X_{IJ}=\frac{p_{IJ}^2  \;\alpha\beta x \;(1+v_{IJ})}{q_Iq_J}.
\end{align}

We extract $\tilde F^{\rm RV}_{IJ}$, from the one-loop soft current for two
massive partons, computed in Refs~\cite{Bierenbaum:2011gg, Czakon:2018iev}. The
result reads\footnote{ We consider the case when two massive quarks are in the
final state; for this reason, the time-like current is needed.}
\begin{equation}
  \begin{split}
    \tilde F^{\rm RV}_{IJ} (\alpha, \beta) =
    & 2C_A  A_{IJ}^{1+\ep} \left(\frac{2 p_{IJ}q_I q_J - m_I^2 q_J^2 - m_J^2 q_I^2}{q_I^2 \;q_J^2}\right) \left(\frac{p_{IJ}}{2 q_I q_J}\right)^\ep \Bigg\{ -\frac{1}{2\ep^2} \\
    & + \frac{1}{2\ep}\left( \frac{1}{v_{IJ}}-1 \right)\frac{1}{2}\ln(\kappa_{IJ} \kappa_{JI}) + \frac{1}{2\ep} \ln(1+ \kappa_{IJ} \kappa_{JI})\\
    & + R^{(0)[TL]}_{IJ} \left( \frac{m_I^2 q_J}{2 p_{IJ}q_I}, \frac{m_J^2 q_I}{2 p_{IJ}q_J} \right)+ \ep R^{(1)[TL]}_{IJ} \left( \frac{m_I^2 q_J}{2 p_{IJ}q_I}, \frac{m_J^2 q_I}{2 p_{IJ}q_J} \right) \bigg \},
  \end{split}
  \label{eq:tldFRV-MM}
\end{equation}
where $R^{(0)[TL]}_{IJ}(\alpha_I,\alpha_J)$ and
$R^{(1)[TL]}_{IJ}(\alpha_I,\alpha_J)$ are the functions introduced in Eq. (4) of
Ref.~\cite{Czakon:2018iev}. Using these results, we compute the function
$K_{IJ}$ and find
\begin{equation}
  \begin{aligned}
    K_{IJ}(\alpha,\beta)
    &= A_{IJ}\Bigg\{
      \frac{67}{72}+\frac{\pi^2}{4}-\frac{\pi^2}{3v_{IJ}}
      -\frac{L_0^2}{4}
      +\frac{(2v_{IJ}-1)(L_I+L_J)L_0}{8v_{IJ}}
      -\frac{(L_I+L_J)L_v}{4v_{IJ}}
    \\
    &+\frac{1}{8v_{IJ}}\left[
      \ln z_I\ln z_J-L_I L_J
      -\frac{q_I L_I^2}{\beta p_{IJ} v_{IJ}}
      -\frac{q_J L_J^2}{\alpha p_{IJ} v_{IJ}}
      \right]
    \\
    &+\frac{1}{4v_{IJ}}\left[
      \operatorname{Li}_2\!\left(\kappa_{IJ} \kappa_{JI}\right)
      -\operatorname{Li}_2(-z_I)-\operatorname{Li}_2(-z_J)
      \right]\Bigg\} + \int \limits_{0}^{1} \frac{{\rm d} x}{x} \;\Delta \tilde F_{I J}(\alpha,\beta,x) .
  \end{aligned}
\end{equation}
where 
\begin{equation}
  \begin{aligned}
    A_{IJ}&=\left(\frac{4\alpha\beta \;\tilde{p}_{IJ}^2 v_{IJ}}{q_Iq_J(1+ v_{IJ})}\right)^2,
    &z_I&=\frac{\beta }{\alpha\; \kappa_{IJ}},
    &z_J&=\frac{\alpha }{\beta\; \kappa_{JI}},
  \end{aligned}
\end{equation}
\begin{align}
  && L_I=\ln\!\left(\frac{q_I}{\kappa_{IJ}\;q_J}\right),
  && L_J=\ln\!\left(\frac{q_J}{\kappa_{JI}\;q_I}\right),
  && L_0=\ln\!\left(\frac{\alpha\beta \tilde p_{IJ}^2}{q_Iq_J}\right),
  && L_v=\ln\!\left(\frac{2v_{IJ}}{1+v_{IJ}}\right),
  &&
\end{align}
and $\Delta \tilde F_{I J}(\alpha,\beta,x)$ is given in Eq. \eqref{eq:delta-tld-FIJ-MM}.

It remains to compute $A^{(3)}_{IJ}$. Since for two massive emitters integration
over $\xi$ is not singular, we expand functions $\tilde K_{IJ}(\xi)$ and
$\tilde K_{JI}(\xi)$ in $\ep$ before integration. After the cancellation of
$1/\ep$ divergencies, the remainder is integrated numerically over $x$ and $\xi$
for the two-gluon final state and analytically for the $q \bar q$ one. We write
\begin{equation}
  A_{IJ}^{(3)} =   C_A \;A_{IJ}^{(3), gg}  + n_f T_R \;A_{IJ}^{(3), q \bar q}.
\end{equation}
We keep $A_{IJ}^{(3),gg}$ in the form of two- and one-dimensional integrals,
\begin{equation}
  \begin{split}
    A_{IJ}^{(3), gg}
    & = \int\limits_0^1 \mathrm{d} \xi \int\limits_0^1\mathrm{d} x \, \left[  f_{\delta}(1,\xi,x) + f_{\delta}(\xi,1,x)\right] \\
    &   + \int\limits_0^1 \mathrm{d} \xi \left[
      f_g(1,\xi) + f_g(\xi,1)
      +
      f_{gg}(1,\xi) + f_{gg}(\xi,1)
      \right]  \\
    & + \frac{11 \pi^2}{72} \left( 
    \frac{1}{2 v_{IJ}} \ln \frac{1+v_{IJ}}{1-v_{IJ}} 
%    \frac{\ln{(1 + v_{IJ})} - \ln{(1 - v_{IJ})}}{2 v_{IJ}}  
- 1\right).
  \end{split}
  \label{eq:AIJgg3ints}
\end{equation}
The integrands $f_{\delta}(\alpha,\beta,x)$, $f_g(\alpha,\beta)$, and
$f_{gg}(\alpha,\beta)$ are given in an ancillary file accompanying the arXiv
version of this paper.\footnote{The file \texttt{A3gg\_integrands.wl} contains
the functions \texttt{fDelta[al,be,x]}, \texttt{fG[al,be]}, and
\texttt{fGG[al,be]}, which are ready for numerical integration in
\texttt{Mathematica}, or for exporting to external numerical programs.} The
corresponding expression for $A_{IJ}^{(3),q\bar q}$ reads
\begin{equation}
  \begin{split}
    A_{IJ}^{(3), q \bar q}
    & = \frac{1}{6 v_{IJ}} \Bigg\{ \left(\ln \kappa_{IJ}+v_{IJ} \right) L_{IJ}^2  - L_{IJ}\bigg [4
      {\rm Li}_2(-\kappa_{IJ})+\ln^2 \kappa_{IJ}
    \\
    &    + \left(2v_{IJ}-\frac{10}{3}\right) \ln \kappa_{IJ} - 4 v_{IJ} \ln(1+\kappa_{IJ})-\frac{10}{3} v_{IJ} + \frac{\pi^2}{3}\bigg] + \frac{28}{9} v_{IJ} \\
    & +  \left( \frac{28}{9}  + \frac{2}{3}\pi^2 -\frac{10}{3} v_{IJ} \right) \ln \kappa_{IJ} + \left(\frac{2}{3} v_{IJ}-\frac{5}{9}\right)\pi^2  +\frac{1}{3}\ln^3\kappa_{IJ} \\
    & + \left( v_{IJ} - \frac{5}{3} \right) \ln^2\kappa_{IJ} - \frac{20}{3} \left({\rm Li}_2(-\kappa_{IJ}) - v_{IJ} \ln(1+\kappa_{IJ}) \right) \Bigg \} + ( I\leftrightarrow J).
  \end{split}
\end{equation}

Putting everything together, we derive  the final result for the case of two massive emitters
\begin{equation}
  \begin{split}
    \tilde{W}_{IJ}^{(2)}
    &=  C_A \;A_{IJ}^{(3), gg} + \frac{n_f T_R}{6 v_{IJ}}  \Bigg\{ \left(\ln \kappa_{IJ}+v_{IJ} \right) L_{IJ}^2  - L_{IJ}\bigg [4
      {\rm Li}_2(-\kappa_{IJ})+\ln^2 \kappa_{IJ}
    \\
    &    + \left(2v_{IJ}-\frac{10}{3}\right) \ln \kappa_{IJ} - 4 v_{IJ} \ln(1+\kappa_{IJ})-\frac{10}{3} v_{IJ} + \frac{\pi^2}{3}\bigg] + \frac{28}{9} v_{IJ} \\
    & +  \left( \frac{28}{9}  + \frac{2}{3}\pi^2 -\frac{10}{3} v_{IJ} \right) \ln \kappa_{IJ} + \left(\frac{2}{3} v_{IJ}-\frac{5}{9}\right)\pi^2  +\frac{1}{3}\ln^3\kappa_{IJ} \\
    & + \left( v_{IJ} - \frac{5}{3} \right) \ln^2\kappa_{IJ} - \frac{20}{3} \left({\rm Li}_2(-\kappa_{IJ}) - v_{IJ} \ln(1+\kappa_{IJ}) \right) \Bigg \} + ( I\leftrightarrow J) \\
    &+ \int \limits_{0}^{1} {\rm d} x \frac{{\rm d} \xi}{\xi}  \; \int \limits_{0}^{2 \pi} \frac{{\rm d} \varphi}{2 \pi} \Bigg \{ \frac{1}{x} \tilde F_{IJ}^{(0)}(1,\xi,x) \ln \left(\frac{{ T}_{N,IJ}(1,\xi,x) }{{ T}_{N,IJ}(1,\xi,0) }\right)  \\
    &+ \left(\frac{11}{6} C_A- \frac{2}{3}T_R n_f \right)R_{IJ}(1,\xi)\ln^2 \left(\frac{{ T}_{N,IJ}(1,\xi,0)Q}{ p_I p_J v_{IJ}^+ }\right)
      + \bigg[K_{IJ}(1, \xi)
    \\
    & + \left(\frac{11}{6}C_A - \frac{2}{3}n_f T_R\right)( L_{IJ} - \log\xi)\bigg]R_{IJ}(1,\xi)\ln \left(\frac{{ T}_{N,IJ}(1,\xi,0)Q}{ \; p_I p_J v_{IJ}^+ } \right)  \\
    & + (I \leftrightarrow J) \bigg  \}  +\Delta W_{RR,IJ}^{\rm dip} ,
  \end{split}
  \label{eq:dip-nnlo-MM-final}
\end{equation}
where
\begin{equation}
  \tilde F_{IJ}^{(0)} =  \frac{C_A}{4} \tilde F_{IJ}^{gg,(0)}(1,\xi,x) - \frac{T_R n_f}{2} F_{IJ}^{q \bar q,(0)}(1,\xi,x).
\end{equation}
and $v_{IJ}^\pm = (1 \pm v_{IJ})/2$. This completes the evaluation of the dipole
contributions to the $N$-jettiness soft function for the three cases of mass
assignments.

\section{NNLO \texorpdfstring{$N$}{N}-jettiness soft function: the tripole contribution}
\label{sec:nnlo-tripole}

\subsection{General considerations}
We continue with the discussion of the triple-color-correlated contribution to
the $N$-jettiness soft function at NNLO. The corresponding expression is given
in Eq.~\eqref{eq:S2-trip-ren} and we repeat it here for convenience
\begin{equation}
  {\tilde S}_2^{\rm tri}
  = \frac{1}{2} [Z_1,Z_1^+] +
  \frac{1}{2} [S_1,Z_1 - Z_1^+] +
  S_{RV,T^3} - Z_{2,\rm tri}
  - Z_{2, \rm tri}^+.
  \label{eq:S2tri-ren-def}
\end{equation}
We note that $S_1$ in this expression is given in Eq.~\eqref{eq:S1-sum-00-M0-MM}; it involves
an integral of the modified eikonal factor $\tilde S_{\gI \gJ}$ over a
single-gluon phase-space. However, for computing $\tilde {S_2}^{\rm tri}$ it is
more convenient to replace $\tilde S_{\gI \gJ}$ with $S_{\gI \gJ}$ in the
expression for $S_1$ using the color conservation. We will denote the
corresponding quantity as $S_{1,t}$. Then,
\begin{equation}
  {\tilde S}_2^{\rm tri}
  = \frac{1}{2} [Z_1,Z_1^+] +
  \frac{1}{2} [S_{1,t},Z_1 - Z_1^+] +
  S_{RV,T^3} - Z_{2,\rm tri}
  - Z_{2, \rm tri}^+.
  \label{eq:S2tri-ren-nodiag}
\end{equation}
We modify the above equation by expressing it through the analog of the
renormalized one-loop soft function
\begin{equation}
  \tilde S_{1,t} = S_{1,t} - Z_1 - Z_1^+,
\end{equation}
and  find
\begin{equation}
  {\tilde S}_2^{\rm tri}
  = -\frac{1}{2} [Z_1,Z_1^+] +
  \frac{1}{2} [\tilde S_{1,t},Z_1 - Z_1^+] +
  S_{RV,T^3} - Z_{2,\rm tri}
  - Z_{2, \rm tri}^+.
  \label{eq:S2tri-ren-def-used}
\end{equation}
The usefulness of this equation stems from the fact that its various terms have
quite distinct dependencies on the $1/\ep$, a feature that becomes very helpful
for tracking the cancellation of divergences when massive partons are involved.
In this respect, we note that
\begin{itemize}
  \item individual contributions to $S_{RV,T^3}$ have divergencies starting at  $1/\ep^3$;
  \item $[Z_1,Z_1^+]$ is proportional to  $1/\ep^2$ and there are no  other $\ep$-dependent contributions to it;
  \item on account of the color-conservation,  $[\tilde S_{1,t},Z_1-Z_1^+]$  has ${\cal O}(\ep^{-1})$ and finite contributions;
  \item the tripole renormalization constants $Z_{2, \rm tri}$ and $Z_{2, \rm tri}^+$,  which are non-vanishing for processes which involve two or more massive partons, contain only $1/\ep$ divergencies.
\end{itemize}
Therefore, the ${\cal O}(1/\ep^3)$ contribution should disappear from
$S_{RV,T^3}$ on its own, the ${\cal O}(1/\ep^2)$ term in $S_{RV,T^3}$ should
cancel with $[Z_1,Z_1^+]$, the ${\cal O}(1/\ep)$ from $S_{RV,T^3}$ should cancel
with $[\tilde S_{1,t}, Z_1 - Z_1^+]$ and the tripole contributions to the
renormalization constants, and the finite contribution to $\tilde S_{2,\rm tri}$
is given by the sum of the ${\cal O}(\ep^0)$ terms in
$[\tilde S_{1,t}, Z_1 - Z_1^+]$ and $S_{RV,T^3}$.

The first two terms on the right-hand side of Eq.~\eqref{eq:S2tri-ren-def-used} involve
commutators of two quantities, each being a sum of products of color charges,
i.e.
\begin{align}
  \label{eq:tripoles-nlo-part}
  &&
     Z_1 = \sum \limits_{\gI \gJ} {\bf T}_\gI  \cdot  {\bf T}_\gJ \; Z_{1,\gI \gJ},
  &&
     S_{1,t} =\sum \limits_{\gI \gJ} {\bf T}_\gI \cdot {\bf T}_\gJ \; W_{\gI \gJ}.
  &&
\end{align}
In principle, one has to be careful with the diagonal terms in the definition of
$Z_{1,\gI \gJ}$ etc. but, as we will see they do not contribute to commutators
relevant for Eq.~\eqref{eq:S2tri-ren-def-used}.

To facilitate further discussion, we consider two operators
\begin{align}
  && \hat A = \sum \limits_{\gI \gJ} A_{ \gI \gJ} {\bf T}_\gI \cdot {\bf T}_\gJ,
  && \hat B = \sum \limits_{\gI \gJ} B_{\gI \gJ} {\bf T}_\gI \cdot {\bf T}_\gJ,
  &&
\end{align}
and we assume that $A_{\gI \gJ} = A_{\gJ \gI}$ and $B_{\gI \gJ} = B_{\gJ \gI}$.
Computing the commutator, we obtain
\begin{equation}
  [ \hat A, \hat B]
  = i\sum \limits_{\gI \gJ \gK}
  F_{\gI \gJ \gK}
  \left (
    -A_{\gI \gJ} B_{\gJ \gK}
    + A_{\gI \gK} B_{\gJ \gK}
    -A_{\gI \gK} B_{\gI \gJ}
    + A_{\gK \gJ} B_{\gI \gJ}
  \right ),
  \label{eq:commAB}
\end{equation}
where $F_{\gI \gJ \gK} = f_{abc} T_\gI^a T_\gJ^b T_\gK^c$. To proceed further,
it is convenient to introduce a fully antisymmetric tensor $H_{\gI \gJ \gK}$
defined as
\begin{equation}
  H_{\gI \gJ \gK}
  = \frac{1}{2}
  f_{abc} \left (
    T_\gI^a T_\gJ^b T_\gK^c
    + T_\gK^c T_\gJ^b T_\gI^a
  \right ).
  \label{eq:tripHTTT-def}
\end{equation}
The relation between $F_{\gI \gJ \gK}$ and $H_{\gI \gJ \gK}$ reads
\begin{equation}
  F_{\gI \gJ \gK} =
  H_{\gI \gJ \gK}
  + \frac{i C_A}{2}
  \left ( \delta_{\gI \gJ} D_{\gI \gK} - \delta_{\gI \gK} D_{\gJ \gK} + \delta_{\gJ \gK} D_{\gI \gJ}
  \right ),
\end{equation}
where $D_{\gI \gK} = \{T^a_\gI,T^a_\gK \}/2$. It is easy to show that only
$H_{\gI \gJ \gK}$ contributes to Eq.~\eqref{eq:commAB}. Then, using the antisymmetry
of $H_{\gI \gJ \gK}$, we find
\begin{equation}
  [ \hat A, \hat B]
  = -4 i \sum _{\gI \gJ \gK}
  H_{\gI \gJ \gK} A_{\gI \gJ} B_{\gJ \gK}.
  \label{eq:commAB-H}
\end{equation}

Next, we use the representation in Eq.~\eqref{eq:tripoles-nlo-part} and the above result to write
\begin{equation}
  \frac{1}{2} [Z_1,Z_1^+] =- 2i \sum \limits_{\gI \gJ \gK} H_{\gI \gJ \gK} Z_{1,\gI \gJ} Z_{1,\gJ \gK}^+.
\end{equation}
To simplify this expression  further, we make the following ansatz for the renormalization constants
\begin{equation}
  Z_{1,\gI \gJ}
  = a_s
  \left ( \frac{a_{\gI \gJ}}{2 \ep^2}
    +
    \frac{b_{\gI \gJ}+i\pi  c_{
        \gI
        \gJ}
    }{2 \ep}
  \right ).
  \label{eq:Z1-complex-ansatz}
\end{equation}
It follows  that
\begin{equation}
  \frac{1}{2}
  [Z_1,Z_1^+]
  = -a_s^2 \pi \sum \limits_{\gI \gJ \gK} H_{\gI \gJ \gK} \left (
    \frac{a_{\gI \gJ} }{\ep^3}
    +\frac{b_{\gI \gJ} }{\ep^2}
  \right ) c_{\gJ \gK}.
  \label{eq:commZ1Z1-anz}
\end{equation}
We note that tensors $a_{\gI \gJ}$, $b_{\gI \gJ}$ and $c_{\gJ \gK}$ are
symmetric. To proceed further, it is convenient to introduce a new tensor
\begin{equation}
  {\bar \sigma}_{\gJ \gK} = c_{\gJ \gK} - \lambda_{\gJ {\rm fin}} - \lambda_{\gK {\rm fin}},
  \label{eq:sigma-bar-def}
\end{equation}
where $\lambda_{\gJ \rm fin} = 1$ if the parton $\gJ$ is in the final state and
$\lambda_{\gJ \rm fin} = 0$ otherwise. Then, using the color-conservation
arguments (for $\lambda_{\gJ \rm fin}$) and the symmetry arguments (for
$\lambda_{\gK \rm fin}$), we rewrite Eq.~\eqref{eq:commZ1Z1-anz} as follows
\begin{equation}
  \frac{1}{2}
  [Z_1,Z_1^+]
  = -\frac{a_s^2  \pi}{\ep^2}  \sum H_{\gI \gJ \gK}
  \left (\frac{a_{\gI \gJ}}{\ep} + b_{\gI \gJ}
  \right )
  {\bar \sigma}_{\gJ \gK}.
  \label{eq:commZ1Z1-sigma}
\end{equation}
\\

The second commutator that we require involves $\tilde S_{1,t}$ and
$Z_1 - Z_1^+$. We use Eqs~(\ref{eq:tripoles-nlo-part},\ref{eq:commAB},\ref{eq:commAB-H}) and
Eqs~(\ref{eq:Z1-complex-ansatz},\ref{eq:sigma-bar-def}) to obtain
\begin{equation}
  \frac{1}{2}
  [\tilde S_{1,t},Z_1-Z_1^+] =
  \frac{2 a_s \pi}{\ep}
  \sum H_{\gI \gJ \gK}
  \tilde S^{1,t}_{\gI \gJ} c_{\gJ \gK}
  = \frac{2 a_s \pi}{\ep}
  \sum H_{\gI \gJ \gK}
  S^{1,t}_{\gI \gJ} \; {\bar \sigma}_{\gJ \gK}.
  \label{eq:commZ1S1}
\end{equation}
The quantities $S^{1,t}_{\gI \gJ}$ evaluate to
\begin{equation}
  S^{1,t}_{\gI \gJ} = a_s\left (
    \frac{ d_{\gI \gJ}}{\ep}
    +
    S^0_{\gI \gJ} + S_{\gI \gJ}^{{\cal T}}
  \right ),
  \label{eq:S1t-ep-exp}
\end{equation}
where $S^{(0)}_{\gI \gJ}$ is the $N$-jettiness-independent and
$S_{\gI \gJ}^{{\cal T}}$ is the $N$-jettiness-dependent term. The individual
contributions in Eq.~\eqref{eq:S1t-ep-exp} depend on the hard-parton mass assignments.
The divergent parts read
\begin{align}
  && d_{ij} = 0,
  && d_{Ij} = d_{jI} =  \frac{1}{2},
  && d_{IJ} = 1,
  &&
\end{align}
and the $N$-jettiness-independent contributions evaluate to
\begin{equation}
  \begin{split}
    S^0_{i j} =& \frac{1}{2} L_{ij}^2 + \frac{1}{2} \ln^2 \left( \frac{C_j}{C_i} \right) + \frac{\pi^2}{4}
                 + \ep \left[ \frac{1}{6} L_{ij}^3 + \frac{\pi^2}{4} L_{ij} + \frac{1}{2} L_{ij}\ln^2 \left( \frac{C_j}{C_i} \right) + \frac{7}{3} \zeta_3\right] + {\cal O}(\ep^2), \\
    S^0_{I j} =&  \frac{1}{4} L_{Ij}^2 + \frac{5\pi^2}{24}+ \ep \left( \frac{1}{12} L_{Ij}^3 + \frac{5\pi^2}{24} L_{Ij} + \frac{7}{6} \zeta_3\right) +{\cal O}(\ep^2), \\
    S^0_{I J} =&  \frac{1}{v_{IJ}} \Big[ -\frac{L_{v_{IJ}}}{2} L_{IJ} + \frac{L^2_{v_{IJ}}}{4} + \frac{\pi^2}{6} - \frac{1}{2} \ln \kappa_{IJ} \ln \kappa_{JI} +
                 {\rm Li}_2(-\kappa_{IJ})+ {\rm Li}_2(-\kappa_{JI}) \Big] \\
               & + \frac{\ep}{v_{IJ}} \bigg\{ -\frac{L_{v_{IJ}}}{4} L_{IJ}^2 + \left[ \frac{L^2_{v_{IJ}}}{4} + \frac{\pi^2}{6} - \frac{1}{2} \ln \kappa_{IJ} \ln \kappa_{JI} +
                 {\rm Li}_2(-\kappa_{IJ})+ {\rm Li}_2(-\kappa_{JI}) \right] L_{IJ} \\
               & - \frac{L_{v_{IJ}}}{2} \left( \frac{L^2_{v_{IJ}}}{6} - \frac{1}{2} \ln \kappa_{IJ} \ln \kappa_{JI} + \frac{5\pi^2}{12}  \right)   \bigg\} + {\cal O}(\ep^2).
  \end{split}
  \label{eq:S0-non-TN}
\end{equation}
The $N$-jettiness-dependent
contribution  reads
\begin{equation}
  \begin{split}
    S_{\gI \gJ}^{\cal T}
    &= \frac{2 N_1(u)}{a_s} \left(\frac{\tilde p_\gI \tilde p_\gJ}{2} \right)^\ep \int \limits_0^1
    % \frac{ d\xi}{\xi} \frac{{\rm d} \phi}{2 \pi}
      \frac{ {\rm d} \xi}{\xi^{1+\ep}} \dOm
      \left [ C_\gI^{2 \ep} f_\gI^{2 \ep} g_{\gI \gJ}^{(2)}(\xi) \; {R}^{t}_{\gI \gJ}(1,\xi)
      +
      C_\gJ^{2 \ep} f_\gJ^{2 \ep} g_{\gJ \gI}^{(2)}(\xi) \; {R}^{t}_{\gI \gJ}(\xi,1)
      \right ],
  \end{split}
\end{equation}
where
\begin{equation}
  R^t_{\gI \gJ}(\alpha,\beta) = \alpha \beta \tilde p_I \tilde p_J S_{\gI \gJ}(q)
  = \frac{\alpha \beta (1+\kappa_{\gI \gJ} \kappa_{\gJ \gI} )}{
    (\beta+ \kappa_{\gI \gJ} \alpha)
    (\alpha + \kappa_{\gJ \gI} \beta)
  },
  \label{eq:Rt-def}
\end{equation}
and all other quantities are defined in Section~\ref{sec:nlo-soft-function}.
\\

We proceed with the computation of the one-loop contribution that involves the
triple-color correlator. It reads
\begin{equation}
  S_{RV,T^3}
  =
  \frac{2^{1-\ep} \pi }{\ep}
  \sum \limits_{\gI \gJ \gK}
  H_{\gI \gJ \gK} \;
  \sigma_{\gJ \gK} \;
  I_{\gI \gJ \gK},
  \label{eq:SRV-T3-def}
\end{equation}
where
\begin{equation}
  \sigma_{\gJ \gK} = \lambda_{\gJ \gK}
  -\lambda_{\gJ {\rm fin}} - \lambda_{\gK {\rm fin}},
  \label{eq:sigmaJK-def}
\end{equation}
and $\lambda_{\gI \gJ} = 1$, if $\gI, \gJ $ are both in the initial or in the
final state, and $\lambda_{\gI \gJ} = 0$ otherwise. Furthermore,
\begin{equation}
  I_{\gI \gJ \gK}  =
  [\alpha_s] \; g_s^2 \int [{\rm d} q]
  \; S_{\gI \gJ}(q)
  \left ( S_{\gJ \gK}(q) \right )^\ep
  {\cal I}_{\gJ \gK}(p_\gJ,p_\gK,q)
  \; \delta(\tau - {\cal T}_N),
\end{equation}
where ${\cal T}_N$ is the $N$-jettiness variable.

The current ${\cal I}_{\gJ \gK}$ depends on whether partons $\gJ, \gK$ are
massive or massless. We will consider the case when all massive partons are in
the final state, and massless partons can be either in the initial or in the
final state. Then, the following currents are relevant:
\begin{enumerate}

  \item if $\gJ, \gK$ are massless,
        \begin{equation}
          {\cal I}_{\gJ \gK} = \frac{\Gamma(1+\ep) \Gamma(1-\ep)^3 }{\Gamma(1-2\ep) } = 1 +{\cal O}(\ep^3),
        \end{equation}
        and Eq.~\eqref{eq:SRV-T3-def} matches the corresponding result for the one-loop soft
        current in Ref.~\cite{Catani:2000pi}. We note that the ${\cal O}(\ep^3)$ term in
        the current is irrelevant for the tripole contribution; for this reason, we will
        simply replace the massless current with one in what follows.

  \item If  one or both of the $\gJ, \gK$ partons are massive, we define
        \begin{equation}
          {\cal I}_{\gJ \gK} = \ep \Gamma(1-\ep) e^{-\ep \gamma_E}  {\cal I}_{\gJ \gK}^{CM},
        \end{equation}
        where ${\cal I}_{\gJ \gK}^{ CM}$ is the imaginary part of the soft current
        computed in Ref.~\cite{Czakon:2018iev}\footnote{Quantities $I_{\gJ \gK}^{(n)}$
        should be taken from Ref.~\cite{Czakon:2018iev} and not from the related
        reference \cite{Bierenbaum:2011gg}. This distinction is important because
        current normalizations are different in the two papers.}

        \begin{equation}
          {\cal I}_{\gJ \gK}^{ CM} =  \sum_{n=-1} \ep^n I^{(n)}_{\gJ \gK}.
        \end{equation}

        Since we focus on the case with massive partons in the final state and since we
        already accounted for the sign change in the imaginary part by including
        $\sigma_{\gJ \gK}$ in the sum of Eq.~\eqref{eq:SRV-T3-def}, we should use results for
        the relevant \emph{time-like} currents from Ref.~\cite{Czakon:2018iev}.\\
\end{enumerate}

To simplify the calculation of the real-virtual contribution, we use the
parametrization of the gluon momentum as in Section~\ref{sec:nlo-soft-function},
\begin{equation}
  q = \alpha \tilde p_\gI
  + \beta \tilde p_\gJ + \sqrt{2 \alpha \beta  \tilde p_\gI \tilde p_\gJ} \;n_\perp,\quad
  n_\perp^2 = -1.
  \label{eq:q-sudakov-nT}
\end{equation}
The quantity $I_{\gI \gJ \gK}$ evaluates to
\begin{equation}
  I_{\gI \gJ \gK}
  = [\alpha_s]^2
  \left (
    \frac{  p_\gJ p_\gK}{ 2
      ( \tilde p_{\gJ} \tilde p_{\gI})^3
    }
  \right )^\ep
  \int
  \frac{  {\rm d} \alpha {\rm d}
    \beta \; \dOm   }{(\alpha \beta )^{1+2\ep} } R^{(t)}_{\gI \gJ}(\alpha,\beta)
  \left ( V_{\gJ \gK}(\alpha,\beta)
  \right )^\ep
  {\cal I}_{\gJ \gK}(\alpha,\beta) \delta( \tau - {\cal T}_N).
\end{equation}
The function $R_{\gI \gJ}^{(t)}$ is defined in Eq.~\eqref{eq:Rt-def}. The function
$V_{\gJ \gK}$ reads
\begin{equation}
  V_{\gJ \gK}
  = \frac{\alpha \beta}{
    (\alpha+\kappa_{\gJ \gI} \beta)
    (\alpha_{\gK} \beta +
    \beta_{\gK} \alpha - 2 \sqrt{\alpha \beta \alpha_{\gK} \beta_{\gK}}
    v_{\gK, \gI \gJ}
    \cos \varphi) },
\end{equation}
where we have used
\begin{equation}
  p_\gK = \alpha_\gK \tilde p_{\gI} + \beta_\gK \tilde p_{\gJ}
  + \sqrt{2 \alpha_\gK \beta_{\gK} \tilde p_\gI \tilde p_\gJ} v_{\gK, \gI \gJ}\; n_{\gK, \perp},
  \quad n_\gK^2 = -1,
\end{equation}
with
\begin{equation}
  v_{\gK, \gI \gJ} =
  \sqrt{1 - \frac{m_{\gK}^2}{2 \tilde p_{\gI} \tilde p_{\gJ} \alpha_{\gK} \beta_{\gK}}},
  \quad n_\perp \cdot n_\gK = -\cos \varphi.
\end{equation}
Finally, the notation ${\cal I}_{\gJ \gK}(\alpha,\beta)$ implies that the
current ${\cal I}_{\gJ \gK}$ should be evaluated with the gluon momentum as in
Eq.~\eqref{eq:q-sudakov-nT}. In addition to $\alpha$ and $\beta$, the current also
depends on the azimuthal angle $\varphi$ of the gluon, but we do not show this
dependence for simplicity.

Functions $R^t_{\gI \gJ}$, $V_{\gI \gJ}$ and ${\cal I}_{\gI \gJ}$ are uniform
functions of $\alpha $ and $ \beta$ of degree zero. Therefore, with the usual
rescaling of $\alpha$ and $\beta$, we can integrate over one of these
parameters. Splitting the integration region into two sectors as in
Eq.~\eqref{eq:one-al-be-thetas}, we find
\begin{equation}
  \begin{split}
    I_{\gI \gJ  \gK}
    & = -\frac{ N_2(u)}{\ep}
      \left (
      \frac{  (p_\gJ p_\gK) (\tilde p_\gI \tilde p_\gJ)}{2
      }
      \right )^\ep
      \int \limits_{0}^{1}  \frac{{\rm d} \xi}{ \xi^{1+2\ep} }
      \dOm  \Bigg (
      \frac{ R^t_{\gI \gJ}(1,\xi)}{
      \left ( V_{\gJ \gK}(1,\xi)
      \right )^{-\ep}}
      {\cal I}_{\gJ \gK}(1,\xi) C_I^{4 \ep} f_I^{4\ep}
    \\
    & \times \left (
      \frac{T_N(1,\xi)}{ \tilde p_I \tilde p_J C_\gI f_\gI}
      \right )^{4\ep}
      +
      \frac{R^t_{\gI \gJ}(\xi,1)}{
      \left ( V_{\gJ \gK}(\xi,1)
      \right )^{-\ep} }
      {\cal I}_{\gJ \gK}(\xi,1)
      C_\gJ^{4 \ep} f_\gJ^{4\ep}
      \left (
      \frac{T_N(\xi,1)}{ \tilde p_\gI \tilde p_J C_\gJ f_\gJ}
      \right )^{4\ep}
      \Bigg ).
  \end{split}
\end{equation}
As we already explained, the function ${\cal I}_{\gJ \gK}$ is written as an
expansion in $\ep$, starting with the ${\cal O}(1)$ term. Importantly, it
possesses a smooth $\xi \to 0$ limit.
\\

It is convenient to separate the $N$-jettiness-dependent and
$N$-jettiness-independent contributions to $I_{\gI \gJ \gK}$. To do this, we
repeat what has been done in the NLO case and write
\begin{align}
  \label{eq:g4-def-RV-trip}
  &&  \left (
     \frac{T_N(1,\xi)}{ \tilde p_I \tilde p_J C_\gI f_\gI}
     \right )^{4\ep}
     = 1 + 4 \ep g_{
     \gI
     \gJ}^{(4)}(\xi),
  &&    \left (
     \frac{T_N(\xi,1)}{ \tilde p_\gI \tilde p_J C_\gJ f_\gJ}
     \right )^{4\ep}
     =
     1+4 \ep g_{
     \gJ \gI}^{(4)}(\xi).
  &&
\end{align}
Then,
\begin{equation}
  I_{\gI \gJ \gK} = I_{\gI \gJ \gK}^{(0)} + I_{\gI \gJ \gK}^{(g)},
\end{equation}
where the $g^{(4)}$-independent terms in Eq.~\eqref{eq:g4-def-RV-trip} give
$I^{(0)}_{ \gI \gJ, \gK}, $ and the $g^{(4)}$-dependent ones provide
$I^{(g)}_{ \gI \gJ, \gK}$. The first one reads
\begin{equation}
  \begin{split}
    I_{\gI \gJ, \gK}^{(0)} =
    & -\frac{ N_2(u)}{\ep}
      \left (
      \frac{  (p_\gJ p_\gK) (\tilde p_\gI \tilde p_\gJ)}{2
      }
      \right )^\ep
      \int \limits_{0}^{1} \frac{{\rm d} \xi}{ \xi^{1+2\ep} }
      \dOm  \Bigg (
      \frac{ R^t_{\gI \gJ}(1,\xi)}{
      \left ( V_{\gJ \gK}(1,\xi)
      \right )^{-\ep}}
      {\cal I}_{\gJ \gK}(1,\xi) C_I^{4 \ep} f_I^{4\ep}
    \\
    & +
      \frac{R^t_{\gI \gJ}(\xi,1)}{
      \left ( V_{\gJ \gK}(\xi,1)
      \right )^{-\ep} }
      {\cal I}_{\gJ \gK}(\xi,1)
      C_\gJ^{4 \ep} f_\gJ^{4\ep}
      \Bigg ),
  \end{split}
  \label{eq:Iijk0-def}
\end{equation}
and the second one evaluates to
\begin{equation}
  \begin{split}
    I^{(g)}_{\gI \gJ, \gK} =
    & -4  N_2(u)
      \left (
      \frac{  (p_\gJ p_\gK) (\tilde p_\gI \tilde p_\gJ)}{2
      }
      \right )^\ep
      \int \limits_{0}^{1}  \frac{{\rm d} \xi}{ \xi^{1+2\ep} } \dOm  \Bigg (
      \frac{ R^{t}_{\gI \gJ}(1,\xi)}{
      \left ( V_{\gJ \gK}(\xi)
      \right )^{-\ep}}
      {\cal I}_{\gJ \gK}(1,\xi) C_\gI^{4 \ep} f_\gI^{4\ep} g_{
      \gI \gJ}^{(4)}(
      \xi)
    \\
    & +
      \frac{R^{t}_{\gI \gJ}(\xi,1)}{
      \left ( V_{\gJ \gK}(\xi,1)
      \right )^{-\ep} }
      {\cal I}_{\gJ \gK}(\xi,1)
      C_\gJ^{4 \ep} f_\gJ^{4\ep}
      g_{
      \gJ \gI}^{(4)}(
      \xi)
      \Bigg ).
  \end{split}
\end{equation}
\\

We continue with the discussion of the \emph{$N$-jettiness-independent} part
$I^{(0)}_{\gI \gJ \gK}$ defined in Eq.~\eqref{eq:Iijk0-def}. To simplify it, we note
that, thanks to the uniformity of the functions $R^t_{\gI \gJ}$, $V_{\gJ \gK}$
and ${\cal I}_{\gJ \gK}$ that appear in that equation, the following relations
hold
\begin{align}
  && R^t_{\gI \gJ}(\xi,1) = R^t_{\gI \gJ}(1,1/\xi),
  && V_{\gI \gJ}(\xi,1) = V_{\gI \gJ}(1,1/\xi),
  && {\cal I}_{\gI \gJ}(\xi,1) = {\cal I}_{\gI \gJ}(1,1/\xi).
  &&
\end{align}
We use these equations to rewrite Eq.~\eqref{eq:Iijk0-def} through functions where
$\xi$ appears as the second argument by changing $\xi \to 1/\xi$ where
appropriate. We find
\begin{equation}
  \begin{split}
    I_{\gI \gJ  \gK}^{(0)}
    & = -\frac{ N_2(u)}{\ep}
      \left (
      \frac{  (p_\gJ p_\gK) (\tilde p_\gI \tilde p_\gJ)}{2
      }
      \right )^\ep
      \int \limits_{0}^{\infty} \frac{{\rm d} \xi}{ \xi} \; \dOm
      R^t_{\gI \gJ}(1,\xi)
      \left ( V_{\gJ \gK}(1,\xi)
      \right )^{\ep}
      {\cal I}_{\gJ \gK}(1,\xi)
    \\
    & \times \Bigg ( \theta(1-\xi)
      \xi^{-2\ep} C_\gI^{4 \ep} f_\gI^{4\ep}
      +
      \theta(\xi-1) \xi^{2\ep}
      C_\gJ^{4 \ep} {\bar f}_\gJ^{4\ep}
      \Bigg ),
  \end{split}
\end{equation}
where we have defined
\begin{equation}
  f_{\gJ}(1/\xi) = \bar f_{\gJ}(\xi).
\end{equation}

To further simplify this expression, we write
\begin{equation}
  I_{\gI \gJ \gK}^{(0)}
  = -\frac{ N_2(u)}{\ep}
  \left (
    \frac{  (p_\gJ p_\gK) (\tilde p_\gI \tilde p_\gJ)}{2
    }
  \right )^\ep \left (A_{\gI \gJ \gK}
    + B_{\gI \gJ \gK} \right ),
\end{equation}
where
\begin{equation}
  \begin{split}
    A_{\gI \gJ \gK}
    & = \int \limits_{0}^{1} \frac{{\rm d} \xi}{ \xi} \;
      \dOm  \;
      R^t_{\gI \gJ}(1,\xi)
      \left ( V_{\gJ \gK}(1,\xi)
      \right )^{\ep}
      {\cal I}_{\gJ \gK}(1,\xi)
      \Bigg (
      \xi^{-2\ep} C_I^{4 \ep} f_I^{4\ep}
      - \xi^{2\ep}
      C_\gJ^{4 \ep} {\bar f}_\gJ^{4\ep}
      \Bigg ),
    \\
    B_{\gI \gJ \gK}
    & = \int \limits_{0}^{\infty} \frac{{\rm d} \xi}{ \xi} \;
      \dOm \;
      R^t_{\gI \gJ}(1,\xi)
      \left ( V_{\gJ \gK}(1,\xi)
      \right )^{\ep} \;
      {\cal I}_{\gJ \gK}(1,\xi) \;  \xi^{2\ep}
      C_\gJ^{4 \ep} {\bar f}_\gJ^{4\ep}.
      \label{eq:Aijk-Bijk-def}
  \end{split}
\end{equation}
We find the above representation of $I^{(0)}_{\gI \gJ \gK}$ useful because in
the case of $A_{\gI \gJ \gK}$, the expression in brackets provides an immediate
$\ep$-suppression simplifying the calculation of this term. In the case of
$B_{\gI \gJ \gK}$, the infinite integration range over $\xi$ offers additional
opportunities to simplify the integral through rescaling of the integration
variable. We will see many examples of these simplifications in the follow-up
discussion.

\subsection{Organizing the
cancellation of singularities}

Calculation of the tripole color terms can be organized in such a way that
tracing the cancellation of divergent terms and deriving finite remainders is
significantly simplified. To this end, we note that it is very convenient to
write $\tilde S_{2}^{\rm tri}$ in Eq.~\eqref{eq:S2tri-ren-def-used} as a sum of
$N$-jettiness-dependent and $N$-jettiness-independent terms
\begin{equation}
  \tilde S_{2}^{\rm tri}
  =
  \tilde S_{2}^{{\rm tri},(g)}
  +
  \tilde S_{2}^{{\rm tri},(0)}.
  \label{eq:S2triGand0}                 %Former eq7.44a (1)
\end{equation}
The $N$-jettiness-dependent contribution $\tilde S_2^{{\rm tri},(g)}$ is given
by the sum of two terms
\begin{equation}
  \tilde S_2^{{\rm tri}, (g)} = \frac{2 a^2_s \pi}{\ep}
  \sum H_{\gI \gJ \gK}
  S_{\gI \gJ}^{{\cal T}} \bar \sigma_{\gJ \gK}
  + \frac{2^{1-\ep} \pi}{\ep}
  \sum H_{\gI \gJ \gK}
  \sigma_{\gJ \gK}
  I_{\gI \gJ \gK}^{(g)}.
  \label{eq:S2TriG}                     %Former eq7.44a (2)
\end{equation}
This expression is free from $1/\ep$ singularities. To see this, we note that
both $S_{\gI \gJ}^{{\cal T}}$ and $I_{\gI \gJ,\gK}^{(g)}$ are finite in the
$\ep \to 0$ limit; hence, the ${\cal O}(1/\ep)$ contribution to
$\tilde S_2^{{\rm tri}, (g)}$ immediately follows from the limiting behavior of
the two quantities.

It is straightforward to compute the required limits, except for a few terms
that we discuss below. First, we note that
\begin{equation}
  \lim_{\ep \to 0} N_2(u) = \frac{a_s^2}{4}.
  \label{eq:N2ep0}
\end{equation}
Second, the limit of the current is
\begin{equation}
  \lim_{\ep \to 0}
  {\cal I}_{
    \gJ \gK}
  =
  \left \{
    \begin{array}{cc}
      2 - 1/v_{JK}, \; & \gJ,\gK\; {\rm both ~massive}, \\
      1, & {\rm otherwise.}
    \end{array}
  \right.
  \label{eq:Irv-ep0}
\end{equation}
The quantity $v_{JK}$ is the relative velocity of particles $J,K$, see
Eq.~\eqref{eq:vIJ-def}. Using the expression for the imaginary part of the
renormalization constants, we find
\begin{equation}
  \bar \sigma_{\gJ \gK}
  = \sigma_{\gJ \gK}
  \lim_{\ep \to 0} {\cal I}_{\gJ \gK},
  \label{eq:sigma2sigmabar}
\end{equation}
where $\bar \sigma_{\gJ \gK}$ and $\sigma_{\gJ \gK}$ are defined in
Eqs~\eqref{eq:sigma-bar-def} and \eqref{eq:sigmaJK-def}, respectively. It is then easy to see that
Eqs~(\ref{eq:N2ep0},\ref{eq:sigma2sigmabar}) lead to the cancellation of the
${ \cal O}(\ep^{-1})$ contribution to Eq.~\eqref{eq:S2TriG}.
The finite remainder evaluates to
\begin{equation}
  \begin{split}
    \tilde S_2^{{\rm tri}, (g)}
    & = -2 a^2_s \pi  \sum H_{\gI \gJ \gK} \sigma_{\gJ \gK} \int \limits_{0}^{1}  \frac{{\rm d} \xi}{\xi} \; \int\limits_0^{2\pi} \frac{{\rm d} \varphi}{2 \pi} \\
    & \times \Bigg \{ R^{t}_{\gI \gJ}(1,\xi) \mathcal{L}_{N,\gI \gJ} (\xi)
      \bigg[
      {\cal I}^{(0)}_{\gJ \gK} \left (
      \mathcal{L}_{N,\gI \gJ} (\xi) + \ln \left(\frac{C_\gI^2 f_\gI^2 \mu^2 \bar u^2 p_\gJ p_\gK}{2 \xi V_{\gJ \gK}^{-1}(1,\xi)} \right)
      \right )+ {\cal I}_{\gJ \gK}^{(1)} (1,\xi) \bigg] \\
    & + R^{t}_{\gI \gJ}(\xi,1) \mathcal{L}_{N,\gJ \gI} (\xi)
      \bigg[
      {\cal I}_{\gJ \gK}^{(0)}
      \left (
      \mathcal{L}_{N,\gJ \gI} (\xi) + \ln \left(\frac{C_\gJ^2 f_\gJ^2 \mu^2 \bar u^2 p_\gJ p_\gK}{2 \xi V_{\gJ \gK}^{-1}(\xi,1)} \right)
      \right ) + {\cal I}_{\gJ \gK}^{(1)} (\xi,1) \bigg] \Bigg\},
  \end{split}
  \label{eq:S2tri-g-part}
\end{equation}
where
\begin{equation}
  \mathcal{L}_{N,\gI \gJ} (\xi)= \ln \left ( \frac{{\cal T}_{N,\gI \gJ }(1,\xi,0) }{(\tilde p_\gI \tilde p_\gJ )  C_\gI f_\gI(\xi) } \right ), \quad
  \mathcal{L}_{N,\gJ \gI} (\xi) = \ln \left ( \frac{{\cal  T}_{N,\gI \gJ }(\xi,1,0)}{(\tilde p_\gI \tilde p_\gJ) C_{\gJ} f_{\gJ}(\xi)  } \right ).
\end{equation}
We also introduced  the $\ep$-expansion of the current
\begin{equation}
  {\cal I}_{\gJ \gK}
  = {\cal I}_{\gJ \gK}^{(0)}
  + \ep \; {\cal I}_{\gJ \gK}^{(1)} +  {\cal O} (\ep^2).
\end{equation}
The first term in this expansion is $\xi$-independent; it can be found in
Eq.~\eqref{eq:Irv-ep0}.
\\

We continue with the discussion of the \emph{$N$-jettiness-independent}
contribution $\tilde S_2^{{\rm tri},(0)}$. It reads
\begin{equation}
  \begin{split}
    \tilde S_2^{{\rm tri},(0)}
    & = \frac{2^{1-\ep} \pi}{\ep} \sum H_{\gI \gJ \gK}
      I^{(0)}_{\gI\gJ \gK} \sigma_{\gJ \gK}
      + \frac{2 a_s^2 \pi}{\ep}
      \sum H_{\gI \gJ \gK}
      \left ( \frac{d_{\gI \gJ}}{\ep} +
      S_{\gI \gJ}^{(0)}
      \right )
      \bar \sigma_{\gJ \gK}
    \\
    & - \frac{a_s^2 \pi}{\ep^2}
      \sum H_{\gI \gJ \gK}
      \left ( \frac{a_{\gI \gJ}}{\ep}
      +
      b_{\gI \gJ}
      \right ) \bar
      \sigma_{\gJ \gK}
      - Z_{2,\rm tri} - Z_{2,\rm tri}^+.
  \end{split}
  \label{eq:S2tri-no-g-part}
\end{equation}
To simplify the calculation of $\tilde S_2^{\rm tri, (0)}$, we split the sum
over final-state particles into three groups, that involve {\it i}) no massive
emitters, {\it ii}) one and only one massive emitter, and {\it iii}) two massive
emitters. We then write
\begin{equation}
  \tilde S_2^{\rm tri, (0)}
  = \tilde S_{2,0m}^{\rm tri, (0)}
  +
  \tilde S_{2,1m}^{\rm tri, (0)}
  +
  \tilde S_{2,2m}^{\rm tri, (0)},
\end{equation}
where three terms on the right-hand side refer to the contributions of the three
groups. As we will show, the cancellation of $1/\ep$ singularities occurs
\emph{in each of these groups separately}. Hence, we will discuss them in turn.

Once the cancellation of $1/\ep$ singularities is established, we apply the same
splitting to the finite remainder of the tripole contribution writing
\begin{equation}
  \tilde S_2^{\rm tri}
  = \tilde S_{2,0m}^{\rm tri}+\tilde S_{2,1m}^{\rm tri}+\tilde S_{2,2m}^{\rm tri} = a_s^2 \; \Bigg\{ \sum \limits_{i,j,k} H_{ijk} \; U_{i,jk}
  + \sum \limits_{I,j,k} H_{Ijk} \; U_{I,jk}
  + \sum \limits_{I,J,k} H_{IJk} \; U_{IJ,k} \Bigg\}.
  \label{eq:S2tri-U}
\end{equation}
The quantities $U_{\gI, \gJ \gK}$ are the primary objects that one needs to
describe the full tripole contribution.

\subsection{The result for the fully massless case}
For massless partons,  the $N$-jettiness-independent triple-color contribution  reads
\begin{equation}
  \tilde S_{2,0m}^{{\rm tri},(0)}
  = \frac{2^{1-\ep} \pi}{\ep} \sum H_{ijk}
  I^{(0)}_{ijk} \sigma_{jk}
  + \frac{2 a_s^2 \pi}{\ep}
  \sum H_{ijk} \;
  S_{ij}^{(0)} \;
  \sigma_{jk}
  - \frac{a_s^2 \pi}{\ep^2}
  \sum H_{ijk}
  b_{ij}
  \sigma_{jk}.
\end{equation}
where we have used the fact that in the massless case $d_{ij} = 0$,
$\bar \sigma_{ij} = \sigma_{ij}$, and $a_{ij} = 1$. Furthermore, we have also
dropped the tripole renormalization constant $Z_{2,\rm tri}$ since it does not
contribute in the massless case.

We start with the real-virtual contribution. Using notations introduced when
discussing the general case, we write
\begin{align}
  R^t_{ij}(1,\xi) = I_{j k}(1,\xi) = 1,
  && f_i = \xi,
  && \bar f_j = \frac{1}{\xi},
  && V_{jk}(1,\xi) = \frac{\xi}{(\alpha_k \xi + \beta_k -2 \sqrt{\alpha_k \beta_k \xi} \cos \varphi)},
\end{align}
where $\alpha_k, \beta_k$ are the coefficients of the Sudakov decomposition of
the momentum $p_k$
\begin{equation}
  p_k = \alpha_k p_i + \beta_k p_j+p_{k,\perp}.
\end{equation}

We start with computing the quantity $B_{ijk}$ defined in Eq.~\eqref{eq:Aijk-Bijk-def}.
It reads
\begin{equation}
  B_{ijk} = C_j^{4 \ep}
  \int \limits_{0}^{\infty}
  \frac{{\rm d} \xi}{\xi^{1+\ep}}
  \dOm
  \frac{1}{ \left ( \alpha_k \xi + \beta_k -
      2 \sqrt{\alpha_k \beta_k \xi} \cos \varphi \right )^\ep}.
\end{equation}
We write $\xi = \beta_k x/\alpha_k$, and obtain
\begin{equation}
  B_{ijk} = C_j^{4 \ep}
  \left ( \frac{\alpha_k}{\beta^2_k} \right )^\ep
  \int \limits_{0}^{\infty}
  \frac{{\rm d} x}{x^{1+\ep}}
  \frac{\dOm}{ \left ( x + 1 -
      2 \sqrt{x} \cos \varphi \right )^\ep}.
\end{equation}

The integral over the azimuthal angle is straightforward to compute. The result
for the general integral of this type reads\footnote{Writing
$\cos \varphi = -1 + 2f$ and integrating over $f$, one obtains a hypergeometric
function with non-integer arguments. It can be simplified using Eq.~(5.3.17) in
Ref.~\cite{abramowitz1965handbook}.}
\begin{equation}
  \begin{split}
    \int
    \frac{\dOm}{ \left ( x + 1 -2 \sqrt{x}  \cos \varphi \right )^a}
    & =
      \theta(1-x)
      F_{21}\left (a , a+\ep, 1-\ep,x
      \right )
    \\
    & + \theta(x-1) x^{-a} F_{21}\left ( a, a+\ep,1-\ep,\frac{1}{x} \right ).
  \end{split}
  \label{eq:cosphi-int}
\end{equation}

Changing $x \to 1/x$ in the second term and specializing to $a = \ep$, we find
\begin{equation}
  B_{ijk}
  = C_j^{4\ep}
  \left ( \frac{\alpha_k}{\beta^2_k}
  \right )^{\ep}
  \int \limits_{0}^{1}
  \frac{{\rm d}x}{x} \;  \left ( x^{-\ep} + x^{2\ep}
  \right ) F_{21}(\ep,2\ep,1-\ep,x).
\end{equation}
It is straightforward to compute this integral.  With the required accuracy, we find
\begin{equation}
  B_{ijk}=C_j^{4\ep}
  \left ( \frac{\alpha_k}{\beta_k^2} \right )^\ep
  \left (
    -\frac{1}{2\ep}
    +4 \zeta_3 \ep^2 +{\cal O}(\ep^3)
  \right ).
  \label{eq:Bijk-000}
\end{equation}

We continue with the analysis of the function $A_{ijk}$. In this case,
\begin{equation}
  A_{ijk} =
  \int \limits_{0}^{1}
  {\rm d} \xi \;
  \frac{ \dOm}{
    \left ( \alpha_k \xi + \beta_k -2 \sqrt{\alpha_k \beta_k \xi}
      \cos \varphi
    \right )^\ep }
  \left ( C_i^{4 \ep} \xi^{3 \ep -1} - C_j^{4\ep} \xi^{-\ep -1}
  \right ).
\end{equation}
We rescale $\xi = \beta_k x/\alpha_k$, compute  the angular integral and find
\begin{equation}
  \begin{split}
    A_{ijk} =
    &
      \int \limits_{0}^{\alpha_k/\beta_k}
      \frac{{\rm d} x}{x} \;
      \left ( C_i^{4 \ep} \; \frac{\beta_k^{2\ep}} {\alpha_k^{3\ep}} \; x^{3 \ep} - C_j^{4\ep} \; \frac{\alpha_k^\ep}{\beta_k^{2\ep} }
      \; x^{-\ep}
      \right ) G(x),
  \end{split}
  \label{eq:Aijk-000}
\end{equation}
where
\begin{equation}
  G(x) = \theta(1-x) F_{21}(\ep,2\ep,1-\ep,x)
  +\theta(x-1) x^{-\ep} F_{21}(\ep,2\ep,1-\ep, x^{-1}).
  \label{eq:Gx-F21-def}
\end{equation}
This function can be written as follows
\begin{equation}
  G(x) = 1 + \theta(x-1)(x^{-\ep}-1) +\ep^2 G_2(x),
\end{equation}
with additional constraint $G_2(x) \to 0$ as $x \to 0$. Since the integrand in
Eq.~\eqref{eq:Aijk-000} is already ${\cal O}(\ep)$ for finite $x$, the function
$G_2(x)$ does not contribute to the tripole term. Hence, we find
\begin{equation}
  A_{ijk} =
  \beta_k^{-\ep}
  \left ( \frac{C_i^{4\ep}}{3 \ep}
    + \frac{C_j^{4\ep}}{\ep}
    + \ep^2 F^{(0)}_{ij,k}
  \right ),
  \label{eq:Aijk-000-ep-exp}
\end{equation}
where
\begin{equation}
  F^{(0)}_{ijk} = \theta(p_k p_j - p_k p_i )  \ln^2 \left (
    \frac{p_k p_j}{p_k p_i} \right )
  \left ( \frac{2}{3} \ln
    \frac{p_k p_j}{p_k p_i}
    - 2 \ln \frac{C_i}{C_j}
  \right ).
\end{equation}

We are now in position to write the combined contribution of $B_{ijk}$ and
$A_{ijk}$ to $\tilde S^{\rm tri,(0)}_{2,0m}$. We obtain
\begin{equation}
  \begin{split}
    \frac{2^{1-\ep} \pi}{\ep}
    I^{(0)}_{ijk}
    & = -\frac{\pi a_s^2}{2 \ep^2}
      \frac{e^{-2\gamma_E \epsilon} \Gamma(1-4\ep)
      }{\Gamma^2(1-\ep)}
      \Bigg [
      e^{2\ep (L_{ij} + L_{jk} - L_{ik})}
      \left (-\frac{1}{2\ep} +
      4 \zeta_3 \ep^2
      \right )
    \\
    & + e^{\ep(2  L_{ij} +
      L_{jk} - L_{ik})}
      \left (
      \frac{1}{3\ep}
      \left ( \frac{C_i}{C_j }\right )^{3 \ep}
      + \frac{1}{\ep}
      \left ( \frac{C_j}{C_i }\right )^{\ep}
      \right )
      +\ep^2F_{0,ijk}
      \Bigg ].
  \end{split}
\end{equation}

To proceed further, we note that thanks to the anti-symmetry of $H_{ijk}$ and
the symmetry of $\sigma_{jk}$ with respect to the permutations
$j \leftrightarrow k$, the following equation holds
\begin{equation}
  \sum \limits_{ijk}
  H_{ijk} I^{(0)}_{ijk} \sigma_{jk}
  = \sum \limits_{ijk}
  H_{ijk} \frac{1}{2} \left (
    I^{(0)}_{ijk} - I^{(0)}_{ikj} \right )
  \sigma_{jk}.
\end{equation}
This anti-symmetrization of $I_{ijk}$ is very useful, as it typically removes
the highest pole and ensures the local cancellation between different tripole
contributions, provided that the same is done in the commutator of $Z_1$ with
$Z_1^+$ and in the commutator of $Z_1-Z_1^+$ with $S_1$. Combining the different
contributions, we find
\begin{equation}
  \tilde S^{\rm tri,(0)}_{2,0m} = \frac{\pi a_s^2}{2}
  \sum H_{ijk} T^{(0)}_{ijk} \sigma_{jk},
  \label{eq:Stri0-000-final-def}
\end{equation}
where
\begin{equation}
  \begin{split}
    T^{(0)}_{ijk}
    & = -2 \left( S_{ij}^{0,(1)} -  S_{ik}^{0,(1)} \right) - \frac{\pi^2}{6} \left( L_{ij} - L_{ik}\right) +\frac{1}{2} \left (
      \tilde F_{0,ijk}-\tilde F_{0,ikj}
      \right ).
  \end{split}
\end{equation}
The functions $S_{ij}^{0,(1)}$ correspond to the $\mathcal{O}(\ep)$ contribution
to $S_{ij}^{(1)}$ in Eq.~\eqref{eq:S0-non-TN}
\begin{equation}
  S_{ij}^{0,(1)} =  \frac{1}{6} L_{ij}^3 + \frac{\pi^2}{4} L_{ij} + \frac{1}{2} L_{ij}\ln^2 \left( \frac{C_j}{C_i} \right) + \frac{7}{3} \zeta_3 ,
\end{equation}
and
\begin{equation}
  \tilde F_{0,ijk} = \theta(p_k p_i - p_k p_j )  \ln^2 \left (
    \frac{p_k p_j}{p_k p_i} \right )
  \left ( \frac{2}{3} \ln
    \frac{p_k p_j}{p_k p_i}
    - 2 \ln \frac{C_i}{C_j}
  \right ).
\end{equation}
The result for the $N$-jettiness-dependent part of the tripole contribution can be found in
Eq.~\eqref{eq:S2tri-g-part}.

\subsection{The case of one massive and two massless partons}

We continue with the case of one massive and two massless partons with  the massive parton in the final state.  We require
\begin{equation}
  \begin{split}
    \tilde S_{2,1m}^{{\rm tri},(0)}
    & = \frac{2^{1-\ep} \pi}{\ep} \sum H_{\gI \gJ \gK}
      I^{(0)}_{\gI\gJ,\gK} \sigma_{\gJ \gK}
      + \frac{2 a_s^2 \pi}{\ep}
      \sum H_{\gI \gJ \gK}
      \left ( \frac{d_{\gI \gJ}}{\ep} +
      S_{\gI \gJ}^{(0)}
      \right )
      \bar \sigma_{\gJ \gK}
    \\
    & - \frac{a_s^2 \pi}{\ep^2}
      \sum H_{\gI \gJ \gK}
      \left ( \frac{a_{\gI \gJ}}{\ep}
      +
      b_{\gI \gJ}
      \right ) \bar
      \sigma_{\gJ \gK},
  \end{split}
\end{equation}
where one of the indices should refer to a massive and the other two indices to
massless partons. We note that we have dropped the tripole renormalization
constant since it only appears when two or more emitters are massive.

We first discuss how spurious singularities associated with tensors
$d_{\gI \gJ}$ and $a_{\gI \gJ}$ disappear in this case. Consider the
contribution of the tensor $d_{\gI \gJ}$. Writing the sum explicitly, we find
\begin{equation}
  \sum \limits_{Ijk} H_{Ijk} ( d_{Ij} \sigma_{jk} - d_{jI} \sigma_{Ik} +d_{jk} \sigma_{kI} ).
  \label{eq:dij-M00-cancel}
\end{equation}
Since $d_{jk} = 0$, the last term drops out. In the first term, we antisymmetrize  $j \leftrightarrow  k$ and find
\begin{equation}
  \sum H_{Ijk} (d_{Ij} - d_{Ik}) \sigma_{jk} = 0,
\end{equation}
because $d_{Ij}= 1/2$, independent of the indices $I$ and $j$.

To show that the second term in Eq.~\eqref{eq:dij-M00-cancel} vanishes as well, we antisymmetrize it  and  find
\begin{equation}
  \sum H_{Ijk} (d_{Ij} \sigma_{Ik} - d_{Ik} \sigma_{Ij} ) = 0,
\end{equation}
where the last step follows from the fact that
\begin{equation}
  \sigma_{Ik} = -1,
  \label{eq:sigmaIk-1}
\end{equation}
independent of the index $k$, provided that the heavy quark $I$ is in the final
state. Similar arguments can be used to show that the contribution involving
tensor $a_{\gI \gJ}$ vanishes as well. It follows that we need to evaluate
\begin{equation}
  \begin{split}
    \tilde S_{2,1m}^{{\rm tri},(0)}
    & = \frac{2^{1-\ep} \pi}{\ep} \sum H_{\gI \gJ \gK}
      I^{(0)}_{\gI\gJ \gK} \sigma_{\gJ \gK}
      + \frac{2 a_s^2 \pi}{\ep}
      \sum H_{\gI \gJ \gK}
      S_{\gI \gJ}^{(0)}
      \bar \sigma_{\gJ \gK}
      - \frac{a_s^2 \pi}{\ep^2}
      \sum H_{\gI \gJ \gK}
      b_{\gI \gJ}
      \bar
      \sigma_{\gJ \gK}.
      \label{eq:Stri0-M00-final-def}
  \end{split}
\end{equation}
We will simplify the computation by anti-symmetrizing with respect to indices of
the massless particles and using Eq.~\eqref{eq:sigmaIk-1} where appropriate.

We begin with the discussion of the real-virtual contributions. Since one of the
indices refers to the massive parton and two indices to the massless ones, there
are three cases to consider.

 \subsubsection{Case \texorpdfstring{$1$}{1}: \texorpdfstring{$I$}{I} massive, \texorpdfstring{$j$}{j} and \texorpdfstring{$k$}{k} massless}
 Using the notations introduced during the general discussion, we find that in
this case
\begin{align}
  && R^t_{Ij}(1, \xi) = \frac{\xi}{\xi+\kappa_{Ij}},
  && I_{j k}(1,\xi) = 1,
  && V_{jk}(1,\xi) = \frac{\xi}{\alpha_k \xi + \beta_k - 2\sqrt{\alpha_k \beta_k \xi} \cos \varphi},
  &&
\end{align}
and $f_I = 1, \bar f_j = \xi^{-1}$. Furthermore, $\alpha_k, \beta_k$ are the
coefficients of the Sudakov decomposition of the momentum $p_k$
\begin{equation}
  p_k = \alpha_k \tilde p_I + \beta_k p_j+p_{k,\perp}.
\end{equation}
We choose to use the same normalization for the massless and massive legs
involved in the Sudakov decomposition, i.e. $C_I = C_j$.\footnote{One has to
consistently make the same choice when computing the $N$-jettiness-dependent
contribution $\tilde S_2^{(g),\rm tri }$.} We then find
\begin{equation}
  \begin{split}
    & A_{Ijk} = \left ( C_I^{4\ep}-C_j^{4\ep} \right ) \int \limits_{0}^{1}
      {\rm d} \xi \; \dOm \;
      \frac{\xi^{-\ep}}{\xi+ \kappa_{Ij} }
      \frac{1}{ \left ( \alpha_k \xi
      + \beta_k - 2\sqrt{\alpha_k \beta_k \xi} \cos \varphi
      \right )^\ep} = 0,
    \\
    & B_{Ijk} = C_j^{4 \ep}
      \int \limits_{0}^{\infty}
      {\rm d} \xi \;  \dOm \;
      \frac{\xi^{-\ep}}{\xi+ \kappa_{Ij} }
      \frac{1}{  \left (  \alpha_k \xi
      + \beta_k - 2\sqrt{\alpha_k \beta_k \xi} \cos \varphi
      \right  )^\ep}.
  \end{split}
\end{equation}

To compute $B_{Ijk}$, we rescale $\xi = \beta_k/\alpha_k x$ and obtain
\begin{equation}
  B_{Ijk} =  C_j^{4\ep}
  \left (
    \frac{\alpha_k}{\beta^2_k}
  \right )^\ep
  \int \limits_{0}^{\infty}
  {\rm d} x \; \dOm \;
  \frac{x^{-\ep}}{x + \Delta_{Ij,k}}
  \frac{1}{\left ( 1+ x - 2\sqrt{x} \cos \varphi \right )^\ep },
\end{equation}
where
\begin{equation}
  \Delta_{Ijk} = \frac{\kappa_{Ij} \alpha_k}{\beta_k} = \frac{y_{Ijk}}{1-y_{Ijk}},
  \label{eq:DeltaIjk-def}
\end{equation}
with
\begin{equation}
  y_{Ijk} = \frac{m_I^2 (p_j p_k)}{2 (p_I p_j) (p_I p_k) }.
  \label{eq:yIjk-def}
\end{equation}

We then integrate over the azimuthal angle $\dOm$ using Eq.~\eqref{eq:cosphi-int} and
find
\begin{equation}
  B_{Ijk} =  C_j^{4\ep}
  \left (
    \frac{\alpha_k}{\beta^2_k}
  \right )^\ep
  \int \limits_{0}^{\infty}
  {\rm d} x
  \frac{x^{-\ep}}{x + \Delta_{Ij,k}}
  G(x),
\end{equation}
where the function $G(x)$ is given in Eq.~\eqref{eq:Gx-F21-def}. It is convenient to
map the integration region to the interval $x \in [0,1]$. We obtain
\begin{equation}
  B_{Ijk} = C_j^{4 \ep}
  \left ( \frac{\alpha_k}{\beta_k^2} \right )^\ep F_{Ijk},
\end{equation}
where
\begin{equation}
  F_{Ijk} = \int \limits_{0}^{1} {\rm d} x
  \left ( \frac{x^{-\ep}}{x+ \Delta_{Ijk}}
    + \frac{x^{2 \ep-1}}{1+ \Delta_{Ijk} x}
  \right )F_{21}(\ep,2\ep,1-\ep,x).
  \label{eq:FIjk-def}
\end{equation}

Instead of computing the function $F_{Ijk}$, we anti-symmetrize $B_{Ijk}$ by
interchanging labels $j$ and $k$. Since, as follows from
Eqs~(\ref{eq:DeltaIjk-def},\ref{eq:yIjk-def}),
\begin{equation}
  \Delta_{Ijk} = \Delta_{Ikj},
\end{equation}
the function $F_{Ijk}$ is \emph{invariant} under the $j \leftrightarrow k$  transformation.  It follows that
\begin{equation}
  \begin{split}
    \frac{2^{1-\ep} \pi}{\ep} \left ( I^{(0)}_{Ijk}
    - I^{(0)}_{Ikj}
    \right )
    &  = -\frac{2 \pi N_2(u)}{\ep^2}
      \left (\frac{p_j p_k}{2(1-y_{Ijk})} \right )^{2 \ep}
      \left [ C_j^{4 \ep} \left (  \frac{p_I p_j}{p_I p_k} \right )^{2 \ep} - (j \leftrightarrow k)
      \right ] F_{Ijk}
    \\
    &
      =-\frac{\pi a^2_s}{2 \ep^2}
      \frac{ e^{-2 \ep \gamma_E }\Gamma(1-4\ep)}{\Gamma^2(1-\ep)}
      \left (
      \frac{y_{Ijk}}{1-y_{Ijk}}
      \right )^{2\ep}
      \left ( e^{2 \ep L_{Ij}}
      -e^{2 \ep L_{Ik}}
      \right ) F_{Ijk}.
  \end{split}
  \label{eq:Ijk0-Ikj0}
\end{equation}

The above equation clearly shows that, similarly to the massless case, the
anti-symmetrization reduces the degree of infrared divergence by one. To
finalize the computation, we need to calculate the function $F_{Ijk}$ which,
thanks to the above remark, is only needed through ${\cal O}(\ep)$. Since
\begin{equation}
  F_{21}(\ep,2\ep,1-\ep,x) = 1 + 2\ep^2 {\rm Li}_2(x)+{\cal O}(\ep^3),
\end{equation}
and since ${\rm Li}_2(x) \to 0$ when $x \to 0$, for computing $F_{Ijk}$ with the
required accuracy we can replace the hypergeometric function with one in
Eq.~\eqref{eq:FIjk-def}. The remaining integration is straightforward and the result
can be expressed in terms of hypergeometric functions. We obtain
\begin{equation}
  F_{Ijk} =
  \frac{1}{(1-\ep)\Delta_{Ijk}} F_{21}\left (1,1-\ep,2-\ep,-\Delta^{-1}_{Ijk}
  \right )
  +
  \frac{1}{2\ep} F_{21}\left (1,2\ep,1+2\ep,-\Delta_{Ijk}
  \right ).
\end{equation}
We use this result in Eq.~\eqref{eq:Ijk0-Ikj0}, expand it in powers of $\ep$ and find
\begin{equation}
  \begin{split}
    \frac{2^{1-\ep} \pi}{\ep} \left ( I^{(0)}_{Ijk}
    - I^{(0)}_{Ikj}
    \right )
    & = -\frac{\pi a^2_s}{2 \ep^2}
      \frac{ e^{-2 \ep \gamma_E }\Gamma(1-4\ep)}{\Gamma^2(1-\ep)}
      \left ( e^{2 \ep L_{Ij}}
      -e^{2 \ep L_{Ik}}
      \right )
    \\
    & \times \left \{
      \frac{1}{2\ep} + \ep
      \left ( \frac{\pi^2}{6}
      +\ln(1-y_{Ijk}) \ln y_{Ijk} - \frac{1}{2} \ln^2 y_{Ijk} +
      {\rm Li}_2(y_{Ijk})
      \right )
      \right \}.
  \end{split}
\end{equation}

\subsubsection{Case \texorpdfstring{$2$}{2}: \texorpdfstring{$i$}{i} massless, \texorpdfstring{$J$}{J} massive, \texorpdfstring{$k$}{k} massless}

In this case we find
\begin{align}
  && R^t_{iJ}(1,\xi) = \frac{1}{1+\kappa_{Ji} \xi},
  && V_{Jk}(1,\xi) = \frac{\xi}{1+\kappa_{Ji} \xi}\frac{1}{\Delta_k(\xi,\varphi)},
  && f_i = \xi,
  && \bar f_J = 1,
  &&
\end{align}
where
\begin{equation}
  \Delta_k(\xi,\varphi) = \alpha_k \xi + \beta_k - 2
  \sqrt{\alpha_k \beta_k \xi} \cos \varphi.
\end{equation}
The parameters $\alpha_k, \beta_k$ are the coefficients of the Sudakov
decomposition of the momentum $p_k$
\begin{equation}
  p_k = \alpha_k p_i + \beta_k \tilde p_J + p_{k,\perp}.
\end{equation}
The current that we need in this case is the one-mass current, which depends on
a particular combination of scalar products
\cite{Bierenbaum:2011gg,Czakon:2018iev}
\begin{equation}
  {\cal I}_{Jk}(q)  = F_{1m}\left ( \frac{m_J^2 (p_k q)}{2 (p_k p_J) (p_J q)} \right ).
  \label{eq:IJk-def}
\end{equation}
The gluon momentum in the above formula reads
\begin{equation}
  q= p_i + \xi \tilde p_J + q_\perp.
\end{equation}

Adapting the general discussion to this case, we find
\begin{equation}
  \begin{split}
    & A_{iJk} = \left (C_i^{4\ep} - C_J^{4\ep}
      \right )
      \int \limits_{0}^{1} \frac{{\rm d} \xi
      \; \dOm }{\xi^{1-3\ep}}
      \frac{1}{(1+\kappa_{Ji} \xi)^{1+\ep}}
      \frac{1}{\Delta^\ep_{k}(\xi,\varphi)} \; {\cal I}_{Jk}(q) = 0,
    \\
    & B_{iJk} = C_i^{4 \ep}
      \int \limits_{0}^{\infty} \frac{{\rm d} \xi
      \; \dOm }{\xi^{1-3\ep}}
      \frac{1}{(1+\kappa_{Ji} \xi)^{1+\ep}}
      \frac{1}{\Delta^\ep_{k}(\xi,\varphi)}
      {\cal I}_{Jk}(q).
  \end{split}
  \label{eq:AiJk-BiJk-def}
\end{equation}
We note that $A_{iJk}$ vanishes because we use $C_J= C_i$, as discussed earlier.
For the same reason, $B_{iJ,k} \sim C_i^{4 \ep}$.

The above representation of the function $B_{iJ,k}$ is not optimal for the
analytic integration because the current ${\cal I}_{Jk}(q)$ depends on the
azimuthal angle $\varphi$. This problem can be addressed by changing the basis
vectors for the Sudakov decomposition of the gluon momentum.

To explain how this works, we rewrite the integral over ${\rm d} \xi$ and $\dOm$
in Eq.~\eqref{eq:AiJk-BiJk-def} as an integral over the on-shell gluon momentum $q$ with
an additional constraint. Following the discussion of the phase-space
parametrization in Section~\ref{sec:nlo-master-formula}, it is easy to see that the following
formula holds
\begin{equation}
  \frac{{\rm d} \xi }{\xi}\;
  \dOm
  = \frac{[{\rm d} q]}{N_{iJ} (q  p_i/(p_i p_J))^{1-\ep}}
  \delta \left ( \frac{q \tilde p_J}{p_i p_J} -1
  \right ),
\end{equation}
where
\begin{equation}
  N_{iJ} = \frac{\Omega^{(d-2)}}{2 (2\pi)^{d-1}} 2^{-\ep} (p_i p_J)^{1-\ep}.
\end{equation}
Since
\begin{align}
  && \frac{1}{1+\kappa_{Ji} \xi} = \frac{p_i p_J}{q p_J},
  && \frac{1}{\Delta_k(\xi,\varphi)} = \frac{p_i p_J}{p_k q},
  &&
\end{align}
we find $B_{iJk}$
\begin{equation}
  B_{iJk} =
  \frac{C_i^{4\ep}}{N_{iJ}} \int
  \frac{[{\rm d} q]}{ (q  p_i/(p_i p_J))^{1-3 \ep}}
  \delta \left ( \frac{q \tilde p_J}{p_i p_J} -1
  \right )
  \left ( \frac{p_i p_J}{p_J q} \right )
  \left (
    \frac{(p_i q) (p_i p_J)}{(p_J q) (p_k q)}
  \right )^\ep {\cal I}_{Jk}(q).
\end{equation}

To avoid the appearance of the gluon azimuthal angle in the function
$I_{Jk}(q)$, we choose $p_k$ and $p_J$ as the basis vectors and write
\begin{equation}
  q = \alpha p_k + \beta r_J + q_\perp,
% \label{eq:q-sudakov-rJ-pk}
\end{equation}
where
\begin{equation}
  p_J = r_J + \kappa_{Jk} p_k,\quad r_J^2 = 0.
\end{equation}
We then find
\begin{equation}
  B_{iJk}
  = \frac{C_i^{4\ep} N_{kJ}}{N_{iJ}}
  \int \frac{{\rm d} \alpha {\rm d} \beta
    \dOm (\alpha \beta)^{-\ep} }{ (q  p_i/(p_i p_J))^{1-3 \ep}}
  \delta \left ( \frac{q \tilde p_J}{p_i p_J} -1
  \right )
  \left ( \frac{p_i p_J}{p_J q} \right )
  \left (
    \frac{(p_i q) (p_i p_J)}{(p_J q) (p_k q)}
  \right )^\ep {\cal I}_{Jk}(q).
\end{equation}
Next, we redefine $\beta = \alpha \eta$, where the new integration variable
$\eta$ goes from zero to infinity. With this variable transformation, the gluon
momentum becomes
\begin{equation}
  q = \alpha \tilde q,
\end{equation}
where
\begin{equation}
  \tilde q =  p_k + \eta r_J + {\tilde q}_\perp,
\end{equation}
and $|{\tilde q}_\perp| = \sqrt{2 p_k p_J \; \eta }$.
We obtain
\begin{equation}
  B_{iJk}
  = \frac{C_i^{4\ep} N_{kJ}}{N_{iJ}}
  \int \frac{{\rm d} \alpha}{\alpha}
  \frac{{\rm d} \eta \; \eta^{-\ep} \;
    \dOm   }{ (\tilde q  p_i/(p_i p_J))^{1-3 \ep}}
  \delta \left ( \alpha \frac{ \tilde q \tilde p_J}{p_i p_J} -1
  \right )
  \left ( \frac{p_i p_J}{p_J \tilde q} \right )
  \left (
    \frac{(p_i \tilde q) (p_i p_J)}{(p_J \tilde q) (p_k \tilde q)}
  \right )^\ep {\cal I}_{Jk}(\tilde q),
\end{equation}
where integrations over $\alpha$ and $\eta$ extend from zero to infinity. Since
\begin{equation}
  \int \limits_{0}^{\infty} \frac{{\rm d} \alpha}{\alpha}
  \delta \left ( \alpha \frac{\tilde q \tilde p_J}{p_i \tilde p_J} -1
  \right ) =1,
\end{equation}
we finally find
\begin{equation}
  B_{iJk}
  = \frac{C_i^{4\ep} N_{kJ}}{N_{iJ}}
  \int  \limits_{0}^{\infty}
  \frac{{\rm d} \eta \; \eta^{-\ep}
    [{\rm d} \Omega_\perp ]   }{ (\tilde q  p_i/(p_i p_J))^{1-3 \ep}}
  \left ( \frac{p_i p_J}{p_J \tilde q} \right )
  \left (
    \frac{(p_i \tilde q) (p_i p_J)}{(p_J \tilde q) (p_k \tilde q)}
  \right )^\ep {\cal I}_{Jk}(\tilde q).
\end{equation}

It is straightforward to express all scalar products in the above formula
through the integration variables. Writing
\begin{equation}
  p_i = \alpha_i p_k + \beta_i r_J + p_{i,\perp},
\end{equation}
we find
\begin{equation}
  \frac{\tilde qp_i}{p_i p_J}
  = \frac{p_k p_J}{p_i p_J}
  \left ( \alpha_i \eta + \beta_i - 2\sqrt{\alpha_i \beta_i \eta} \cos \varphi
  \right ),
\end{equation}
and
\begin{align}
  && \frac{p_i p_J}{p_J \tilde q }= \frac{p_i p_J}{p_k p_J(1+\kappa_{Jk} \eta )},
  && \frac{( p_i \tilde q) (p_i p_J) }{( p_J \tilde q) (p_k \tilde q) }
     = \frac{p_i \tilde q}{p_i p_J}
     \left ( \frac{p_i p_J}{p_k p_J} \right )^2
     \frac{1}{(1+\kappa_{Jk} \eta) \eta}. \;
  &&
\end{align}
The one-mass current depends on the following variable (c.f. Eq.~\eqref{eq:IJk-def})
\begin{equation}
  \frac{m_J^2 (p_k q)}{2 (p_k p_J) p_J q }
  = \frac{\kappa_{Jk} \eta}{1+\kappa_{Jk} \eta}.
\end{equation}
Putting everything together, we find
\begin{equation}
  B_{iJ,k}
  = C_i^{4\ep} \left (
    \frac{p_i p_J}{p_k p_J}
  \right )^{1-\ep}
  \int  \limits_{0}^{\infty}
  \frac{{\rm d} \eta \; \eta^{-2\ep}
    \dOm \; ( 1+\kappa_{Jk} \eta)^{-1-\ep}  }{ \left ( \alpha_i \eta + \beta_i - 2\sqrt{\alpha_i \beta_i \eta} \cos \varphi \right )^{1-4\ep}}
  F_{1m}
  \left (\frac{\kappa_{Jk} \eta}{1+\kappa_{Jk} \eta} \right ).
\end{equation}
We use Eq.~\eqref{eq:cosphi-int} to integrate over the angle $\varphi$, change the
remaining integration variable $\eta = \beta_i/\alpha_i x$ and map the
$x$-integration region to the interval $[0,1]$. We find
\begin{equation}
  \begin{split}
    B_{iJ,k}
    & = C_i^{4\ep} \left (
      \frac{p_i p_J}{p_k p_J}
      \right )^{1-\ep} \frac{\beta_i^{2 \ep}}{
      \alpha_i^{1-2\ep}
      }
      \int  \limits_{0}^{1}
      {\rm d} x\; G(x,\Delta_{Jki}) \; F_{21}(1-4\ep,1-3\ep,1-\ep,x),
      \label{eq:BiJk-ang-int}
  \end{split}
\end{equation}
where $\Delta_{Jki} = \kappa_{Jk}\beta_i/\alpha_i$ and
\begin{equation}
  \begin{split}
    G(x,\Delta)
    & =
      \frac{x^{-2 \ep}}{
      ( 1+\Delta  \; x)^{1+\ep}}
      F_{1m}
      \left (\frac{\Delta  x }{1+ \Delta x} \right )
      +
      \frac{x^{-\ep}}{
      ( x+\Delta )^{1+\ep}}
      F_{1m}
      \left (\frac{\Delta }{x+\Delta } \right ).
  \end{split}
\end{equation}

The hypergeometric function in the integrand of Eq.~\eqref{eq:BiJk-ang-int} contains a
non-integrable singularity at $x =1$, reflecting the collinear $q || p_i$
singularity of the original $B_{iJk}$ integral. Extracting it, we find
\begin{equation}
  B_{iJ,k}
  = C_i^{4\ep} \left (
    \frac{p_i p_J}{p_k p_J}
  \right )^{1-\ep} \frac{\beta_i^{2 \ep}}{
    \alpha_i^{1-2\ep}
  }
  \int  \limits_{0}^{1}
  {\rm d} x\; (1-x)^{-1+6\ep} \; G(x,\Delta_{Jk,i}) \;
  F_{21}(3\ep,2\ep,1-\ep,x).
  \label{eq:BiJk-ang-int-trans}
\end{equation}
Since
\begin{equation}
  \Delta_{Jki}=\frac{y_{Jki}}{1-y_{Jki}},
\end{equation}
with $y_{Jki}$ defined in Eq.~\eqref{eq:yIjk-def}, the integral in
Eq.~\eqref{eq:BiJk-ang-int-trans} is invariant under the $k \leftrightarrow i$ replacement.
Because of this, it becomes straightforward to form the required antisymmetric
combination that contributes to the triple-color part of the soft function. We
find
\begin{equation}
  \frac{2^{1-\ep} \pi}{\ep} \left ( I^{(0)}_{iJk}
    - I^{(0)}_{kJi}
  \right )
  = -\frac{\pi a_s^2}{2 \ep^2}
  \frac{ e^{-2 \ep \gamma_E }\Gamma(1-4\ep)}{\Gamma^2(1-\ep)}
  \left ( e^{2 \ep L_{Ji}}
    -e^{2 \ep L_{Jk}}
  \right )
  \frac{\Gamma(1-\ep) \Gamma(1-6 \ep)}{\Gamma(1-3\ep) \Gamma(1-4\ep)} \;
  F_{Jki}.
\end{equation}
where
\begin{equation}
  F_{Jki} = \frac{y_{Jki}^{2 \ep}}{
    (1-y_{Jki})^{1-2\ep}} \int  \limits_{0}^{1}
  {\rm d} x\; (1-x)^{-1+6\ep} G(x,\Delta_{Jki}).
  \label{eq:FJki-def}
\end{equation}
We note that in order to arrive at Eq.~\eqref{eq:FJki-def}, we have replaced the
hypergeometric function, appearing in Eq.~\eqref{eq:BiJk-ang-int-trans}, with its value at
$x =1$. This is allowed, since retaining the full dependence introduces terms
that have too strong an $\ep$-suppression to contribute to the final result for
the triple-color correlated term. It remains to compute the function $F_{Jki}$
to first order in $\ep$. Isolating the singularity at $x=1$, we find
\begin{equation}
  F_{Jki}
  =
  \frac{y_{Jik}^{2\ep} F_{1m}( y_{Jik}  )}{3 \ep (1-y_{Jki})^{-3\ep} }
  +
  \frac{y_{Jik}^{2\ep}}{(1-y_{Jik} )^{1-2\ep}}
  \int \limits_{0}^{1}
  {\rm d} x \; (1-x)^{-1+6 \ep}
  \left ( G(x,\Delta_{Jki})
    - G(1,\Delta_{Jki})
  \right ).
\end{equation}
It is straightforward to compute this function to the required order in the
$\ep$-expansion using the known result for the one-mass current
\cite{Czakon:2018iev}. We obtain
\begin{equation}
  \begin{split}
    F_{Jki}
    = &
        \frac{1}{3 \ep }-\frac{(1-2 y_{Jki})  }{3 (1-y_{Jki})} \ln (y_{Jki})
        + \ep \Bigg (
        \frac{\pi ^2}{9} -\frac{(1-2 y_{Jki}) \ln^2(y_{Jki})}{3 (1-y_{Jki})}
    \\
      & -
        \frac{4 (2-y_{Jki}) \text{Li}_2(1-y_{Jki})}{3 (1-y_{Jki})}
        \Bigg ).
  \end{split}
\end{equation}

\subsubsection{Case \texorpdfstring{$3$}{3}: \texorpdfstring{$i$}{i} and \texorpdfstring{$j$}{j} massless, \texorpdfstring{$K$}{K} massive }

In this case,
\begin{align}
  && R^t_{ij}=1,
  && f_i = \xi,
  && \bar f_j = \frac{1}{\xi},
  && V_{jK} = \frac{\xi}{\Delta_K(\xi, \varphi)},
  &&
\end{align}
where
\begin{equation}
  \Delta_K(\xi,\varphi) = \alpha_K \xi + \beta_K
  -2 \sqrt{\alpha_K \beta_K \xi} \; v_{K,ij} \cos \varphi.
\end{equation}
As already explained in the previous subsection, ${\cal I}_{jK}$ is a function
of a particular combination of scalar products
\begin{equation}
  {\cal I}_{jK}(q) = F_{1m}\left ( \frac{m_K^2 (p_j q)}{2 (p_K p_j)  (p_K q)} \right ),
\end{equation}
where one has to use
\begin{align}
  \label{eq:q-sudakov-pi-pK}
  && q = p_i + \xi p_j + q_\perp,
  && p_K = \alpha_K p_i + \beta_K p_j + p_{K, \perp}.
  &&
\end{align}
We find
\begin{equation}
  \begin{split}
    & A_{ijK} = \int \limits_{0}^{1}
      \frac{{\rm d} \xi \; \dOm }{\xi^{1-\ep}}
      \frac{1}{ \left ( \Delta_K(\xi,\varphi) \right )^\ep} \; {\cal I}_{jK}(q) \left (C_i^{4\ep} \xi^{2 \ep} - C_j^{4\ep} \xi^{-2\ep}
      \right ),
    \\
    & B_{ijK} = C_j^{4 \ep}
      \int \limits_{0}^{\infty} \frac{{\rm d} \xi \;
      \dOm }{\xi^{1+\ep}}
      \frac{1}{\left ( \Delta_K(\xi,\varphi) \right )^\ep} \; {\cal I}_{jK}(q).
  \end{split}
  \label{eq:AijK-BijK-def}
\end{equation}

At variance with the previous cases, $A_{ijK}$ does not vanish and we need to
discuss it. The singularities of the $A_{ijK}$ integrand are at $\xi = 0$ and,
according to Eq.~\eqref{eq:q-sudakov-pi-pK}, $q = p_i$ at $\xi = 0$. Isolating the
$\xi = 0$ singularity, we write
\begin{equation}
  \begin{split}
    & A_{ijK}= \beta_K^{-\ep} F_{1m}(y_{Kij})
      \int \limits_{0}^{1} \frac{{\rm d} \xi}{\xi}
      \left ( C_i^{4\ep} \xi^{3 \ep} - C_j^{4\ep} \xi^{-\ep}
      \right )
    \\
    & +
      \int \limits_{0}^{1}
      \frac{{\rm d} \xi}{\xi^{1-\ep}} \;
      \dOm  \left [
      \frac{1}{ \left ( \Delta_K(\xi, \varphi) \right )^\ep} \; {\cal I}_{jK}(q)
      -
      \beta_K^{-\ep} {\cal I}_{jK}(p_i)
      \right ]
      \left (C_i^{4\ep} \xi^{2 \ep} - C_j^{4\ep} \xi^{-2\ep}
      \right ) .
  \end{split}
\end{equation}
The first term can be integrated over $\xi$ and the second one can be expanded
in $\ep$ since all singularities are regularized. Hence, we obtain the following
representation for $A_{ijK}$
\begin{equation}
  A_{ijK}  = \beta_K^{-\ep}
  F_{1m}(y_{Kji}) \left ( \frac{C_i^{4\ep}}{3\ep}
    + \frac{C_j^{4\ep}}{\ep} \right ) + \ep^2 F^{(A)}_{ijK},
\end{equation}
with
\\
\begin{equation}
  F^{(A)}_{ijK}   =
  4 \int \limits_{0}^{1}
  \frac{{\rm d} \xi}{\xi}
  \int \limits_{0}^{2 \pi} \frac{{\rm d} \varphi}{2 \pi}
  \ln \left (
    \frac{C_i }{C_j} \xi
  \right )
  \left \{
    \ln \left [ \frac{\beta_K}{\Delta_K(\xi,\varphi)}
    \right ]
    + {\cal I}_{jK}^{(1)}(q) - {\cal I}_{jK}^{(1)}(p_i)
  \right \},
  \label{eq:FA_ijK-def}
\end{equation}
where we have used
\begin{equation}
  {\cal I}_{jK}(q) = 1 + \ep {\cal I}_{jK}^{(1)}(q) +{\cal O}(\ep^2).
\end{equation}

To compute $B_{ijK}$, we again change basis vectors for the Sudakov
decomposition of the gluon momentum $q$. Following the discussion of the
$B_{iJk}$ case, we write the integral over ${ \rm d} \xi$ and
${\rm d} \Omega_\perp$ as an integral over the gluon momentum with the
additional constraint
\begin{equation}
  \frac{{\rm d} \xi }{\xi}
  [{\rm d} \Omega_\perp ]
  = \frac{[{\rm d} q]}{N_{ij} (q  p_i/(p_i p_j))^{1-\ep}}
  \delta \left ( \frac{qp_j}{p_i p_j} -1
  \right ),
  % \label{eq:dxi-to-dq}
\end{equation}
and express all quantities that appear in the formula for $B_{ijK}$ in
Eq.~\eqref{eq:AijK-BijK-def} as scalar products of the gluon momentum $q$ with other
momenta. We find
\begin{align}
  && \xi = \frac{p_i q}{p_i p_j},
  && \Delta_K(\xi,\varphi) = \frac{ p_K q}{p_i p_j}.
  &&
\end{align}
Choosing $p_K$ and $p_j$ as the basis vectors for the Sudakov decomposition, we
write
\begin{equation}
  q = \alpha \tilde p_K + \beta p_j + q_\perp,
\end{equation}
where $\tilde p_K = p_K - \kappa_{Kj} p_j$.
Putting everything together,  we obtain
\begin{equation}
  B_{ijK} =
  C_j^{4\ep}
  \left ( \frac{p_i p_j}{p_K p_j} \right )^{2\ep}
  \int \limits_{0}^{\infty}
  \frac{{\rm d} \alpha {\rm d} \beta (\alpha \beta )^{-\ep}
    [{\rm d} \Omega ]}{ (qp_i/(p_K p_j))} \;
  \delta \left ( \alpha \frac{ p_K p_j}{p_i p_j} -1 \right )
  \left (
    \frac{p_K p_j}{p_K q}
  \right )^\ep
  {\cal I}_{jK}(q).
\end{equation}
Changing  the integration variable
\begin{equation}
  \beta = \alpha \eta,
\end{equation}
and using  the scaling properties of $q$,
\begin{align}
  && q = \alpha \tilde q,
  && \tilde q = \tilde p_K + \eta p_j + \tilde q_\perp,
  &&
\end{align}
we integrate over $\alpha$ and find
\begin{equation}
  B_{ijK} =
  C_j^{4\ep}
  \left ( \frac{p_i p_j}{p_K p_j} \right )^{1-\ep}
  \int \limits_{0}^{\infty}
  \frac{{\rm d}  \eta \; \eta^{-\ep}
    [{\rm d} \Omega ]}{ (\tilde  qp_i/(p_K p_j))}
  \left (
    \frac{p_K p_j}{p_K \tilde  q}
  \right )^\ep
  {\cal I}_{jK}(\tilde q).
\end{equation}
Writing
\begin{equation}
  p_i = \alpha_i \tilde p_K + \beta_i p_j + p_{i,\perp},
\end{equation}
we compute the scalar products of various momenta, rescale
$\eta = \beta_i/\alpha_i x$, and obtain
\begin{equation}
  B_{ijK} = C_j^{4\ep}
  \left (
    \frac{( p_i p_j) (p_K p_j)}{(p_i \tilde p_K)^2}
  \right )^{\ep}
  \int \limits_{0}^{\infty}
  \frac{{\rm d} x \;
    \; \dOm }{
    ( 1+ x
    -2 \sqrt{x} \cos \varphi )
  }
  \frac{x^{-\ep}}{(x + \Delta_{Kji} )^\ep} F_{1m} \left (
    \frac{\Delta_{Kji}}{
      \Delta_{Kji} + x}
  \right ),
\end{equation}
with
\begin{equation}
  \Delta_{Kji} = \frac{\kappa_{K j} \alpha_i}{\beta_i}
  =\frac{y_{Kji}}{1-y_{Kji}}.
\end{equation}

Integrating over $\varphi$  we find
\begin{equation}
  B_{ijK}
  =
  C_j^{4\ep}
  \left (
    \frac{( p_i p_j) (p_K p_j)}{(p_i \tilde p_K)^2}
  \right )^{\ep}
  F_{ijK},
\end{equation}
where
\begin{equation}
  \begin{split}
    F_{ijK} & =  \int \limits_{0}^{\infty}
              {\rm d} x \;
              \frac{x^{-\ep}}{(x+\Delta_{Kj,i})^\ep} \;
              F_{1m}\left (\frac{\Delta_{Kj,i}}{\Delta_{Kj,i}+x} \right )
              \Big ( \theta(1-x)
              F_{21}(1,1+\ep,1-\ep,x)
    \\
            &  +\theta(x-1) x^{-1}
              F_{21}(1,1+\ep,1-\ep,x^{-1} )
              \Big ).
  \end{split}
\end{equation}
The function $F_{ijK}$ is invariant under $i \leftrightarrow j$ permutation;
therefore, we can easily form an antisymmetric combination for this mass
assignment. The result reads
\begin{equation}
  \begin{split}
    & \frac{2^{1-\ep} \pi}{\ep}  \left ( I^{(0)}_{ijK}
      - I^{(0)}_{jiK}
      \right )
      = -\frac{\pi a_s^2}{2 \ep^2}
      \frac{ e^{-2 \ep \gamma_E }\Gamma(1-4\ep)}{\Gamma^2(1-\ep)}
      \Bigg [
      \left ( e^{2 \ep L_{Kj}}
      -e^{2 \ep L_{Ki}}
      \right ) \frac{y_{Kij}^{2\ep}}{(1-y_{Kij})^{2\ep}}
      F_{ijK}
    \\
    & + e^{2\ep L_{ij} }\Big [e^{\frac{\ep}{2} (L_{Kj}-L_{Ki})}
      \left (
      \frac{1}{3\ep}
      \left ( \frac{C_i}{C_j} \right )^{3\ep}
      + \frac{1}{\ep} \left (
      \frac{C_j}{C_i}
      \right )^\ep
      \right ) F_{1m}(y_{Kij})
      + \ep^2 F^{(A)}_{ijK}
      - (i \leftrightarrow j)
      \Big ]
      \Bigg ].
  \end{split}
\end{equation}

It remains to compute the function $F_{ijK}$ through ${\cal O}(\ep)$. This
function has a complex structure of singularities. To expose it, we start by
mapping the integration region on the interval $ x \in [0,1] $ and then rewrite
the hypergeometric function to extract the singularity at $x=1$. We find
\begin{equation}
  \begin{split}
    F_{ijK} & =
              \int \limits_{0}^{1}
              \frac{ {\rm d} x}{(1-x)^{1+2\ep}} \;
              \Bigg [
              \frac{x^{-\ep} }{(x+\Delta_{Kji})^\ep} \;
              F_{1m} \left (
              \frac{\Delta_{Kji}}{x +\Delta_{Kji}}
              \right )
    \\
            & + \frac{x^{2\ep-1}}{(1+\Delta_{Kji} x)^\ep} \;
              F_{1m}\left (
              \frac{\Delta_{Kji} x }{1 + \Delta_{Kji} x}
              \right )
              \Bigg ]
              F_{21}(-\ep,-2 \ep,1-\ep,x).
  \end{split}
\end{equation}
All singularities in this equation are exposed, so that we can compute the
integral as an expansion in powers of $\ep$. To do this, we start by separating
the integrand into two terms -- one that has the singularity at $x = 0$ and
another one that has the singularity at $x = 1$. We write
\begin{equation}
  F_{ijK} = F_{ijK}^{(a)}+F_{ijK}^{(b)},
\end{equation}
where
\begin{equation}
  \begin{split}
    F_{ijK}^{(a)} & =
                    \int \limits_{0}^{1}
                    \frac{ {\rm d} x}{(1-x)^{2\ep}} \;
                    \frac{x^{2\ep-1}}{(1+\Delta_{Kji} x)^\ep} \;
                    F_{1m}\left (
                    \frac{\Delta_{Kji} x }{1 + \Delta_{Kji} x}
                    \right )
                    F_{21}(-\ep,-2 \ep,1-\ep,x),
    \\
    F^{(b)}_{ijK}
                  & =
                    \int \limits_{0}^{1}
                    \frac{ {\rm d} x}{(1-x)^{1+2\ep}} \;
                    \Bigg [
                    \frac{x^{-\ep} }{(x+\Delta_{Kji})^\ep} \;
                    F_{1m} \left (
                    \frac{\Delta_{Kji}}{x +\Delta_{Kji}}
                    \right )
    \\
                  & + \frac{x^{2\ep}}{(1+\Delta_{Kji} x)^\ep} \;
                    F_{1m} \left (
                    \frac{\Delta_{Kji} x }{1 + \Delta_{Kji} x}
                    \right )
                    \Bigg ]
                    F_{21}(-\ep,-2 \ep,1-\ep,x).
  \end{split}
\end{equation}

To compute $F^{(a)}_{ijK}$, we note that $F_{21}(-\ep,-2\ep,1-\ep,x)$ deviates
from $1$ only at order ${\cal O}(\ep^2)$ and that its ${\cal O}(\ep^2)$
contribution vanishes at $x = 0$. Therefore, with the required ${\cal O}(\ep)$
accuracy, the hypergeometric function in the integrand of $F^{(a)}_{ijK}$ can be
replaced with one. Furthermore, since $F_{1m}(0)=1$ we can write
\begin{equation}
  F^{(a)}_{ijK}
  =
  \Gamma(2\ep) \Gamma(1-2\ep) F_{21}(\ep,2\ep,1,-\Delta_{Kji})
  + \int \limits_{0}^{1}
  \frac{{\rm d} x}{x}
  \left [ F_{1m}\left (
      \frac{\Delta_{Kji} \; x }{1 + \Delta_{Kji} \; x}
    \right ) - 1
  \right ] +{\cal O}(\ep^2).
\end{equation}
We note that we only need to retain contributions of order $\ep$ in the second
term on the right-hand side of the above equation.

We continue with the calculation of $F^{(b)}_{ijK}$; in this case, the
singularity is at $x=1$ and we can replace the hypergeometric function in the
integrand with its value at $x=1$ with the required accuracy. We find
\begin{equation}
  F_{Kij}^{(b)}
  = \frac{\Gamma(-\ep) \Gamma(1+2\ep)}{\Gamma(1+\ep)}
  (1-y_{Kij})^\ep \;
  F_{1m} \left ( y_{Kij}
  \right )
  +\int \limits_{0}^{1}
  \frac{ {\rm d} x}{(1-x)^{1+2\ep}} \; G^{(b)}_{Kij}(x),
\end{equation}
where
\begin{equation}
  \begin{split}
    G^{(b)}_{Kij} & =
                    \frac{x^{-\ep} }{(x+\Delta_{Kji})^\ep}
                    F_{1m} \left (
                    \frac{\Delta_{Kji}}{x +\Delta_{Kji}}
                    \right )
                    + \frac{x^{2\ep}}{(1+\Delta_{Kji} x)^\ep} \;
                    F_{1m}\left (
                    \frac{\Delta_{Kji} x }{1 + \Delta_{Kji} x}
                    \right )
    \\
                  & - \frac{2}{(1+\Delta_{Kji})^\ep}
                    F_{1m}\left (
                    \frac{\Delta_{Kji}}{1 +\Delta_{Kji}}
                    \right ).
  \end{split}
\end{equation}

It is straightforward to compute the required integrals and to expand the result
through ${\cal O}(\ep)$. Putting everything together, we arrive at the following
simple result for the function $F_{Kij}$
\begin{equation}
  \frac{y_{Kij}^{2\ep}}{(1-y_{Kij})^{2\ep} } F_{Kij}  =
  -\frac{1}{2 \ep}
  -\frac{\ln y_{Kij}}{1-y_{Kij}}
  -\ep \frac{(1+2 y_{Kij}) \ln^2 y_{Kij}}{1-y_{Kij}}.
\end{equation}

\subsubsection{The result for the case with one massive and two massless partons }

We are now in a position to present the result for the case of two massless and
one massive partons. Combining the results for the three cases discussed in the
previous sections and adding the renormalization constants following
Eq.~\eqref{eq:Stri0-M00-final-def}, we obtain a remarkably simple expression for the
$N$-jettiness-independent term
\begin{equation}
  \begin{split}
    \tilde S_{2,1m}^{\rm tri, (0)}
    & = \alpha_s^2 \pi
      \sum \limits_{I,j,k}
      H_{Ijk} \; \Bigg \{ \sigma_{jk}
      \Bigg [
      \frac{1}{12}
      \left ( L_{Ik}-L_{Ij}
      \right )^3 + \frac{1}{4}
      \left ( L_{Ik}-L_{Ij}
      \right ) L_{Ij} L_{Ik}
    \\
    & + \frac{\left ( L_{Ik}-L_{Ij}
      \right )}{4} \left [ \pi^2  - \ln^2 (y_{Ijk}) - 2 {\rm Li}_2(1-y_{Ijk}) \right ] \Bigg]  + \frac{1}{24} (L_{Ik} - L_{Ij})^3 \\
    &+
      \frac{2}{3}
      \ln \left (\frac{C_j}{C_k} \right  )^3 +
      \frac{1}{4} L_{Ij} L_{Ik}
      (L_{Ik} - L_{Ij}) -\frac{1}{4} \left ( F^{(A)}_{kj,I} -F^{(A)}_{jk,I} \right )
    \\
    & +
      \frac{(L_{Ik} - L_{Ij} )}{2} \left ( \frac{\pi^2}{2}
      + \ln^2 \frac{C_j}{C_k}
      -
      \frac{
      y_{Ijk} \ln^2(y_{Ijk}) + 2
      {\rm Li}_2(1 - y_{Ijk})}{1-y_{Ijk}}
      \right )
      \Bigg \}.
  \end{split}
  \label{eq:Stri0-M00-final}
\end{equation}
We remind the reader that the function $F^{(A)}_{kjI}$ is defined in
Eq.~\eqref{eq:FA_ijK-def} and the quantities $L_{Ij}$ can be found in
Appendix~\ref{sec:app-ren}.

\subsection{The case of two massive and one massless partons}
\label{sec:trip-MM0}

We continue with the case of two massive and one massless partons; both massive
partons are in the final state. Similar to the preceding mass assignment, it is
easy to see that all spurious singularities related to $a_{\gI \gJ}$ and
$d_{\gI \gJ}$ tensors cancel in this case as well. We therefore require
\begin{equation}
  \begin{split}
    \tilde S_{2,2m}^{{\rm tri},(0)}
    & = \frac{2^{1-\ep} \pi}{\ep} \sum H_{\gI \gJ \gK}
      I^{(0)}_{\gI\gJ\gK} \sigma_{\gJ \gK}
      + \frac{2 a_s^2 \pi}{\ep}
      \sum H_{\gI \gJ \gK}
      S_{\gI \gJ}^{(0)}
      \bar \sigma_{\gJ \gK}
      - \frac{a_s^2 \pi}{\ep^2}
      \sum H_{\gI \gJ \gK}
      b_{\gI \gJ}
      \bar
      \sigma_{\gJ \gK}
      +Z_{\rm tri},
  \end{split}
\end{equation}
where the sum must involve two massive and one massless parton. We note that in
this case a dedicated renormalization constant is needed to make the result
finite; it can be found in Appendix~\ref{sec:app-ren}.

We emphasize that the computation can be significantly simplified by
anti-symmetrizing with respect to the indices of massive particles and using
Eq.~\eqref{eq:sigmaIk-1} where appropriate. There are three distinct mass
assignments in this case that we consider below.

\subsubsection{Case \texorpdfstring{$I$}{I} massive, \texorpdfstring{$J$}{J} massive, \texorpdfstring{$k$}{k} massless }

Functions $R^t_{IJ}$ and $ V_{Jk}$, that we introduced earlier, read
\begin{align}
  && R^t_{IJ} = \frac{\xi (1 + \kappa_{IJ} \kappa_{JI})}{(\xi + \kappa_{IJ})(1+\kappa_{JI} \xi)},
  && V_{Jk} = \frac{\xi}{(1+\kappa_{JI} \xi) \Delta_k(\xi,\varphi)}.
  &&
\end{align}
The function $\Delta_k$ is
\begin{equation}
  \Delta_k = \alpha_k \xi + \beta_k -
  2 \sqrt{\alpha_k \beta_k \xi} \cos \varphi.
\end{equation}
These expressions correspond to the following parametrizations of the gluon
momentum
\begin{equation}
  q = \tilde p_I + \xi \tilde p_J + q_\perp,
  \label{eq:q-sudakov-pI-pJ}
\end{equation}
and the momentum of the massless  parton $k$
\begin{equation}
  p_k = \alpha_k \tilde p_I + \beta_k \tilde p_J + p_{k,\perp}.
\end{equation}
We also choose $C_I = C_J = C$ in  this case.  Furthermore, we choose
\begin{align}
  && f_{I} = \xi^{1/2}, &&  \bar f_{J} = \xi^{-1/2}. &&
\end{align}

With this choice, we find that $A_{IJk} = 0$, and
\begin{equation}
  \begin{split}
    & B_{IJk} = C^{4\ep}
      \int \limits_{0}^{\infty}
      \frac{{\rm d} \xi}{\xi} \; \dOm \;
      R^t_{IJ}
      V_{Jk}^\ep
      \; {\cal I}_{Jk}(q).
  \end{split}
\end{equation}

We note that for the purpose of the anti-symmetrization, it is convenient to
consider the following quantity
\begin{equation}
  T_{IJk} = \left (
    \frac{\tilde p_I \tilde p_J}{p_I p_J} \;
    \frac{p_J p_k}{p_I p_J}
  \right )^\ep B_{I J k}   = C^{4\ep}
  \left ( \frac{v^+_{IJ} \;  p_J p_k}{p_I p_J}
  \right )^\ep \int \limits_{0}^{\infty} \frac{{\rm d} \xi}{ \xi} \;
  \dOm \;
  R^t_{I J}
  V^\ep_{J k} \;
  {\cal I}_{J k}(q),
\end{equation}
where $v_{IJ}^+=(1+v_{IJ})/2$. We then rewrite it as
\begin{equation}
  T_{IJk}= C^{4\ep} \left ( \frac{ v_{IJ}^+ \; p_J p_k}{p_I p_J}
  \right )^\ep  \int \limits_{0}^{\infty} \frac{{\rm d} \xi}{ \xi} \;
  \dOm \;
  R^t_{I J} \left [
    V^\ep_{J k} +  \;
    W_{IJk}^{-\ep} \tilde {\cal I}_{J k}(q)
    + (V^\ep_{Jk} - W_{IJk}^{-\ep} ) \tilde {\cal I}_{Jk}(q)
  \right ],
  \label{eq:T_IJk}
\end{equation}
where
\begin{align}
  && \tilde {\cal I}_{Jk}(q) = {\cal I}_{Jk}(q)-1 \sim {\cal O}(\ep),
  && W_{IJk} = \frac{q \tilde p_I}{\tilde p_I \tilde p_J} \;
     \left (\frac{p_J p_k}{q \; r_J} \right )^2,
  &&
\end{align}
with
\begin{equation}
  r_J = p_J - \kappa_{Jk} p_k.
  \label{eq:r2p}
\end{equation}

The different terms in the integrand in Eq.~\eqref{eq:T_IJk} have different
degrees of suppression with respect to the parameter $\ep$. The last term is
${\cal O}(\ep^2)$, so that it only contributes to the finite part of the soft
function. The first term is ${\cal O}(\ep^0)$ and the second one is
${\cal O}(\ep)$. We begin with the analysis of the first term in
Eq.~\eqref{eq:T_IJk}. It reads
\begin{equation}
  T^{(1)}_{IJk}= C^{4\ep} \left ( \frac{v^+_{IJ} \;  p_J p_k}{p_I p_J}
  \right )^\ep  \int \limits_{0}^{\infty} \frac{{\rm d} \xi}{ \xi} \;
  \dOm \;
  R^t_{I J} \;
  V^\ep_{J k}.
\end{equation}
We anti-symmetrize $T^{(1)}_{IJk}$ in indices $I$ and $J$, and find
\begin{equation}
  T^{(1)}_{[IJ]k}
  =\left (  C^4 \; v^+_{IJ}
  \right )^\ep
  \int \limits_{0}^{\infty} \frac{{\rm d} \xi}{\xi}  \;
  \dOm \;
  \left \{
    R^t_{IJ}(\xi)  \left ( V_{J k}(\xi) \;\frac{p_J p_k}{p_I p_J}
    \right )^\ep - R^t_{JI}(\xi)
    \left (  V_{I k}(\xi) \;\frac{p_I p_k}{p_I p_J}
    \right )^\ep
  \right \} ,
  \label{eq:T_IJk-T_JIk}
\end{equation}
where
\begin{equation}
  T^{(1)}_{[IJ]k} = T_{IJk}^{(1)}-T_{JIk}^{(1)}.
\end{equation}
We then perform the $\xi \to 1/\xi$ variable transformation in the second term
of the integrand in Eq.~\eqref{eq:T_IJk-T_JIk}. Since
\begin{align}
  && R^t_{JI}(\xi^{-1}) = R^t_{IJ}(\xi),
  && V_{Ik}(\xi^{-1}) = \frac{1+\kappa_{JI} \xi}{\xi + \kappa_{IJ}} \; V_{Jk}(\xi),
  &&
\end{align}
we obtain
\begin{equation}
  T^{(1)}_{[IJ]k}
  = ( C^{4} \;  v^+_{IJ} )^\ep\;
  \int \limits_{0}^{\infty} \frac{{\rm d} \xi}{\xi}  \;
  \dOm   \;
  R^t_{IJ}  V_{J k}^\ep
  \left \{
    \left ( \frac{p_J p_k}{p_I p_J}
    \right )^\ep -
    \left (  \frac{1+\kappa_{JI} \xi}{\xi + \kappa_{IJ}}\;\frac{p_I p_k}{p_I p_J}
    \right )^\ep
  \right \}.
\end{equation}
Finally, rescaling the integration variable as $ \xi = x \beta_k/\alpha_k$, we
find
\begin{equation}
  \begin{split}
    & T^{(1)}_{[IJ]k}
      = (C^4 v_{IJ}^+)^\ep
      \int \limits_{0}^{\infty} {\rm d}x  \; \dOm   \;
      \frac{1+\kappa_{IJ} \kappa_{JI} }{(x+\Delta_{IJk})(1+\Delta_{JIk} x)}  \left ( \frac{x}{(1+\Delta_{JIk} x) \Delta(x,\varphi)}
      \right )^\ep
    \\
    & \times  \left ( \frac{v_{IJ}}{1-y_{JIk}}
      \right )^\ep \left \{
      1 -
      \left (  \frac{(1+\Delta_{JIk} x)(1-y_{JIk})}{(x + \Delta_{IJk})(1-y_{IJk}) }
      \right )^\ep
      \right \},
  \end{split}
  \label{eq:T_IJk-T_JIk-trans}
\end{equation}
where
\begin{equation}
  \Delta(x,\varphi) =
  1+x - 2\sqrt{x} \cos \varphi.
\end{equation}
Furthermore, we have used
\begin{align}
  && \Delta_{IJk} = \kappa_{IJ} \frac{\alpha_k}{\beta_k},
  && \Delta_{JIk} = \kappa_{JI} \frac{\beta_k}{\alpha_k},
  &&
\end{align}
and
\begin{equation}
  y_{IJk} =\frac{m_I^2}{2 \tilde p_I \tilde p_J  } \frac{p_J p_k}{p_I p_k} =  \frac{m_I^2}{p_I p_J (1+v_{IJ}) } \frac{p_J p_k}{p_I p_k}.
\end{equation}
It is easy to see that $\Delta_{IJk}$, $\Delta_{JIk}$ can be written as
\begin{align}
  && \Delta_{IJk} = \frac{y_{IJk}(1-y_{JIk})}{1-y_{IJk}},
  && \Delta_{JIk} = \frac{y_{JIk}(1-y_{IJk})}{1-y_{JIk}}.
  &&
\end{align}

We continue with integrating in Eq.~\eqref{eq:T_IJk-T_JIk-trans} over the angle $\varphi$
using Eq.~\eqref{eq:cosphi-int}, drop the hypergeometric function that appears as the
result of the $\varphi$-integration since it does not contribute to the required
orders in the $\ep$-expansion, and find
\begin{equation}
  \begin{split}
    & T^{(1)}_{[IJ]k}
      = ( C^{4} v^+_{IJ})^\ep \;
      \int \limits_{0}^{\infty} {\rm d} x
      \frac{1+\kappa_{IJ} \kappa_{JI} }{(x+\Delta_{IJk})(1+\Delta_{JIk} x)}  \left ( \frac{x}{1+\Delta_{JIk}x}
      \right )^\ep
    \\
    & \times  \left ( \frac{v_{IJ}}{1-y_{JIk}}
      \right )^\ep \left \{
      1 -
      \left (  \frac{(1+\Delta_{JIk} x)(1-y_{JIk})}{(x + \Delta_{IJ k})(1-y_{IJk}) }
      \right )^\ep
      \right \}
      \left ( 1 + \theta(x-1)(x^{-\ep} - 1) \right ).
  \end{split}
  \label{eq:T_IJk-T_JIk-trans-trans}
\end{equation}

It is straightforward to expand $T^{(1)}_{[IJ]k}$ in Eq.~\eqref{eq:T_IJk-T_JIk-trans-trans} in
powers of $\ep$. We obtain
\begin{equation}
  \begin{split}
    & T^{(1)}_{[IJ]k}
      =
      - \frac{\ep}{2v_{IJ}}
      \ln \frac{1+v_{IJ}}{1-v_{IJ}}
      \ln \frac{ y_{JIk}}{y_{IJk}}
      \;
      -\ep^2
      \int \limits_{0}^{\infty} {\rm d} x  \;
      \frac{1+\kappa_{IJ} \kappa_{JI} }{(x+\Delta_{IJk})(1+\Delta_{JIk} x)}
    \\
    & \times  \ln \left ( \frac{(1+\Delta_{IJ,k}) (1+ \Delta_{JIk} x)}{(1+\Delta_{JIk}) ( x + \Delta_{IJk})}
      \right )
      \ln
      \left (
      ( C^2 v^+_{IJ} )^2 \;  x^{\theta(1-x)}
      \sqrt{\frac{(1+\Delta_{JI,k})(1+\Delta_{IJk})}{(1+\Delta_{JIk}x)(x+\Delta_{IJk})}}\;
      \right ).
  \end{split}
\end{equation}
The remaining one-dimensional integral  over $x$  in the above expression can be  computed numerically
in a straightforward way.

We continue with the discussion of the second term in Eq.~\eqref{eq:T_IJk}. It
reads
\begin{equation}
  T^{(2)}_{IJk} =
  C^{4\ep}
  \left ( v^+_{IJ}\; \frac{p_J p_k}{p_I p_J} \right )^\ep \int \limits_{0}^{\infty} \frac{{\rm d} \xi}{ \xi} \;
  \dOm \;
  R^t_{I J}
  \; W_{IJk}^{-\ep}
  \left (
    {\cal I}_{J k}(q)  - 1 \right ).
\end{equation}

To simplify its calculation, we need to change the basis vectors employed in the
Sudakov decomposition of the gluon momentum and we do this by first writing the
integral $T^{(2)}_{IJk}$ in a parametrization-invariant form. Since the gluon
momentum $q$ is given by the following equation,
\begin{equation}
  q = \tilde p_I + \xi \tilde p_J + q_\perp,
\end{equation}
we can write $R^t_{IJ}$ as
\begin{equation}
  R^t_{IJ}= \frac{(q \tilde p_I)(q \tilde p_J) (1+\kappa_{IJ}\kappa_{JI})}{(q p_I) (q p_J) }.
\end{equation}
Next, we switch to a different parametrization of the momentum $q$,
\begin{equation}
  q = \alpha p_k + \beta r_J + q_\perp,
  \label{eq:q-sudakov-pk-rJ}
\end{equation}
with $r_J$ defined in Eq.~\eqref{eq:r2p}. We rescale $\beta = \alpha \eta$ and
obtain
\begin{equation}
  \begin{split}
    T^{(2)}_{IJk}
    = \left( C^2 v^+_{IJ} \right)^{2\ep}
    \frac{p_I p_J}{p_J p_k}
    \int  \limits_{0}^{\infty}
    {\rm d} \eta \; \eta^{-\ep} \;
    \dOm \;
    \frac{(p_k p_J)^2}{(p_I q)(p_J q) } ( {\cal I}_{Jk}(q)  - 1),
  \end{split}
\end{equation}
where $q$ is defined in Eq.~\eqref{eq:q-sudakov-pk-rJ} and $\alpha=1$, $\beta = \eta$ should
be used there.

Computing scalar products of the gluon momentum with $p_k$, $r_J$ and $p_I$, and
rescaling $\eta = \beta_I/\alpha_I x$, we find
\begin{equation}
  \begin{split}
    T^{(2)}_{IJ,k}
    = ( C^2 v^+_{IJ} )^{2\ep} \;
    \frac{p_I p_J}{p_J p_k}
    \frac{\beta_I^{- \ep}}{\alpha_I^{1-\ep}}
    & \int  \limits_{0}^{\infty}
      \frac{{\rm d} x \;
      \dOm  \; x^{-\ep}  }{ ( 1+x - 2\sqrt{x} v_{I,kJ} \cos \varphi ) }
      \frac{1}{(1+\Delta_{JkI} x )}
    \\
    & \times
      \left [ F_{1m}\left (
      \frac{\Delta_{JkI} x}{1+\Delta_{JkI} x}  \right )
      - 1
      \right ],
  \end{split}
  \label{eq:T2_IJk}
\end{equation}
where  $\Delta_{JkI} = \kappa_{Jk} \beta_I/\alpha_I$ and
\begin{equation}
  v_{I,kJ} = \sqrt{ 1- \frac{m_I^2}{2 p_k p_J \alpha_I \beta_I}}
  =\sqrt{1 - \frac{y_{IkJ}}{1-y_{JkI}}}.
  \label{eq:vIkJ-def}
\end{equation}
Furthermore, the function $F_{1m}$ in Eq.~\eqref{eq:T2_IJk} is the one-mass
current.

At this point, the integration over the azimuthal angle can be performed.  We find
\begin{equation}
  \begin{split}
    & I_\varphi = \int \frac{\dOm }{
      1 + x  -2 \sqrt{x} v \cos \varphi
      } =
      \frac{1}{1+x + 2 \sqrt{x} v}
      F_{21}\left (1,\frac{1}{2}-\ep,1-2\ep, z \right )
    \\
    & =\frac{4^{-2\ep}}{\sqrt{(1+x)^2 - 4x v^2}}
      (1+\sqrt{1-z})^{4 \ep} (1-z)^{-\ep}
      F_{21}\left (-\ep,-2\ep,1-\ep,
      \left ( \frac{1-\sqrt{1-z}}{1+\sqrt{1-z}}
      \right )^2
      \right ),
  \end{split}
  \label{eq:Iphi-def}
\end{equation}
where
\begin{equation}
  z = \frac{4 \sqrt{x} v}{1+x+2 \sqrt{x} v}.
\end{equation}

Since the hypergeometric function in Eq.~\eqref{eq:Iphi-def} deviates from unity
only at ${\cal O}(\ep^2)$, and since $T^{(2)}_{IJk}$ is already ${\cal O}(\ep)$,
it can be omitted. Finally, it is convenient to rescale $x \to x/\Delta_{JkI}$.
We find
\begin{equation}
  \begin{split}
    T^{(2)}_{IJk}
    = \left ( C^4 \; ( v^+_{IJ} )^2
    \kappa_{Jk} \right )^{\ep}
    &
      \frac{1}{1-y_{JkI}}
      \int  \limits_{0}^{\infty}
      \frac{ {\rm d} x \; x^{-\ep} }{\sqrt{(\Delta_{JkI}+x)^2 - 4x \Delta_{JkI} v_{I,Jk}^2}}
    \\
    & \times
      \left (\frac{ 1+\sqrt{1-z_1}}{2} \right )^{4 \ep}
      (1-z_1)^{-\ep} \;
      \frac{1}{(1+ x )} \;
      \left [ F_{1m}\left (
      \frac{x}{1+ x}  \right )
      - 1
      \right ],
  \end{split}
\end{equation}
where
\begin{equation}
  z_1 = \frac{4 \sqrt{x \Delta_{JkI}} v_{I,jK}}{\Delta_{JkI} + x + 2\sqrt{x \Delta_{JkI}} v_{I,jK}}.
\end{equation}
The integration in this expression is not singular. Therefore, we can expand the
integrand in powers of $\ep$. It is also convenient to change variables as
$ x \to x/(1-x)$, mapping the integration region onto the $[0,1]$ interval. We
write the result in the following way
\begin{equation}
  T^{(2)}_{IJk} = \ep T^{(2,1)}_{IJk} + \ep^2 T^{(2,2)}_{IJk},
\end{equation}
where
\begin{equation}
  T^{(2,1)}_{IJk}=\frac{1}{1-y_{JkI}} \int \limits_{0}^{1}
  \frac{ {\rm d} x  }{\sqrt{R_{IJk}(x)} }
  \; F_{1m}^{(1)}
  \left(x \right ),
\end{equation}
and
\begin{equation}
  \begin{split}
    T^{(2,2)}_{IJ,k} & =\frac{1}{1-y_{JkI}} \int \limits_{0}^{1}
                       \frac{ {\rm d} x  }{\sqrt{R_{IJk}(x)}}  \Bigg [ F_{1m}^{(2)}
                       \left (   x  \right )
                       + \ln  \left (
                       \frac{\kappa_{Jk} (1-x) (1+\sqrt{\theta_{IJk}(x)})^4}{C^{-4}  ( v^+_{IJ} )^{-2} \; 16 \; x \; \theta_{IJk}(x)}
                       \right )
                       F_{1m}^{(1)} \left (
                       x \right )
                       \Bigg  ].
  \end{split}
\end{equation}
The functions $F^{(1)}_{1m}$ and $F^{(2)}_{2m}$ are the ${\cal O}(\ep)$ and
${\cal O}(\ep^2)$ expansion coefficients of the one-mass current $F_{1m}$ and
\begin{equation}
  \begin{split}
    & R_{IJk} (x)= (\Delta_{JkI}(1-x)+x)^2 - 4x(1-x) \Delta_{JkI} v_{I,kJ}^2,
    \\
    & \theta_{IJk} (x) =
      \frac{\Delta_{JkI}(1-x) + x - 2\sqrt{(1-x) x \Delta_{JkI}} v_{I,Jk}}{\Delta_{JkI}(1-x) + x + 2\sqrt{(1-x) x \Delta_{JkI}} v_{I,kJ}}.
      \label{eq:R_IJk}
  \end{split}
\end{equation}
We note that $T^{(2,1)}_{IJk}$ contributes to the divergent part and has to be
computed analytically. This computation is relatively standard but quite
tedious; because of this we do not present the result of this computation here.

It remains to compute the last contribution to Eq.~\eqref{eq:T_IJk}. Since it is
already ${\cal O}(\ep^2)$, we can expand it in powers of $\ep$ right away. We
find
\begin{equation}
  T^{(3)}_{IJk}= \ep^2 \int \limits_{0}^{\infty} \frac{{\rm d} \xi}{ \xi} \; \int\limits_0^{2\pi} \frac{{\rm d} \varphi}{2 \pi} \;
  R^t_{I J} \;
  \ln ( V_{Jk} W_{IJk} )
  {\cal I}^{(1)}_{Jk}(q).
\end{equation}
We note that in this expression the parametrization of the gluon momentum $q$
shown in Eq.~\eqref{eq:q-sudakov-pI-pJ} should be used. In principle, one can integrate
over the azimuthal angle analytically but this produces unnecessarily
complicated expressions. Because of this, we decided to leave $T^{(3)}_{IJk}$ as
a two-dimensional integral which we evaluate numerically.

\subsubsection{Case \texorpdfstring{$I$}{I} massive, \texorpdfstring{$j$}{j} massless, \texorpdfstring{$K$}{K} massive}

In this case, the relevant quantities read
\begin{align}
  && R^t_{Ij} = \frac{\xi}{\xi+\kappa_{Ij}},
  && V_{Jk} = \frac{\xi}{\Delta_K(\xi,\varphi)},
  && f_I = 1,
  && \bar f_{j} = \xi^{-1},
  && C_I = C_j,
  &&
\end{align}
where
\begin{equation}
  \Delta_K = \alpha_K \xi + \beta_K
  - 2 \sqrt{\alpha_K \beta_K \xi} \; v_{K,Ij} \cos \varphi,
\end{equation}
with
\begin{equation}
  v_{K,Ij} = \sqrt{1 - \frac{m_K^2}{2 p_I p_j \alpha_K \beta_K}}.
\end{equation}
We note that  the following momenta  parametrizations are employed
\begin{align}
  && q = \tilde p_I + \xi p_j + q_\perp,
  && p_K = \alpha_K \tilde p_I + \beta_K p_j + p_{K,\perp}.
  &&
\end{align}
Using the above  expressions,  we find
\begin{equation}
  \begin{split}
    & A_{IjK}
      = \left ( C_I^{4 \ep} - C_j^{4 \ep}
      \right )
      \int \limits_{0}^{1} \frac{{\rm d} \xi \; \dOm }{\xi }
      \frac{\xi}{\xi + \kappa_{Ij}}
      \left ( \frac{\xi}{\Delta_K} \right )^\ep {\cal I}_{jK}(q)
      \;
      \xi^{-2\ep} = 0,
    \\
    & B_{IjK}
      =
      C_j^{4 \ep}
      \int \limits_{0}^{\infty}
      \frac{{\rm d} \xi \; \dOm} {\xi }
      \frac{\xi}{\xi + \kappa_{Ij}}
      \left ( \frac{\xi}{\Delta_K} \right )^\ep {\cal I}_{jK}(q)
      \; \xi^{-2\ep},
  \end{split}
\end{equation}
where the first equation holds because we use $C_I = C_j$.

We continue with the computation of $B_{IjK}$. To keep track of all
momenta-dependent factors relevant for the anti-symmetrization, we define
\begin{equation}
  T_{IjK} = \left (
    \frac{p_I p_j \;  p_j p_K}{(p_I p_K)^2}
  \right )^\ep B_{IjK}.
\end{equation}
We write $T_{IjK}$ in the invariant form using
\begin{align}
  && \xi = \frac{q \tilde p_I}{p_I p_j},
  && R^t_{Ij} = \frac{q \tilde p_I}{q p_I},
  && V_{jK} = \frac{q \tilde p_I}{q p_K},
  &&
\end{align}
and obtain
\begin{equation}
  \begin{split}
    T_{IjK}
    & = C_j^{4\ep} \left (
      \frac{(p_I p_j) (p_j p_K)}{(p_I p_K)^2}
      \right )^\ep
      \int \frac{[{\rm d} q] \; \delta \left [(q p_j)/(p_I p_j) -1 \right ]}{N_{Ij} ((q \tilde p_I)/(p_I p_j))^{1-\ep}}
      R_{Ij} V_{jK}^\ep {\cal I}_{jK} (q)
      \;
      \left (
      \frac{q \tilde p_I}{p_I p_j}
      \right )^{-2\ep}
    \\
    & = C_j^{4\ep}\left (
      \frac{(p_I p_j) (p_j p_K)}{(p_I p_K)^2}
      \right )^\ep
      \int \frac{[{\rm d} q]}{N_{Ij} }
      \; \delta \left (\frac{q p_j}{p_I p_j} -1 \right )
      \frac{p_I p_j}{q p_I}
      \left ( \frac{p_I p_j}{q p_K} \right )^\ep {\cal I}_{jK}(q).
  \end{split}
\end{equation}

$T_{IjK}$ contains a collinear $ q || p_j$ divergence which we need to isolate
and extract. To do this, we use the following property of the current
\begin{equation}
  I_{jK}(p_j) = 1.
\end{equation}
Then,  we split $T_{IjK}$ into two parts
\begin{equation}
  T_{IjK} = T_{IjK}^{(a)} + T_{IjK}^{(b)},
\end{equation}
where
\begin{equation}
  \begin{split}
    & T_{IjK}^{(a)}
      = C_j^{4\ep} \left (
      \frac{(p_I p_j) (p_j p_K)}{(p_I p_K)^2}
      \right )^\ep
      \int \frac{[{\rm d} q]}{N_{Ij} }
      \;
      \delta \left (\frac{q p_j}{p_I p_j} -1 \right )
      \frac{p_I p_j}{q p_I}
      \left ( \frac{p_I p_j}{q p_K} \right )^\ep,
    \\
    & T_{IjK}^{(b)}
      = C_j^{4\ep} \left (
      \frac{(p_I p_j) (p_j p_K)}{(p_I p_K)^2}
      \right )^\ep
      \int \frac{[{\rm d} q]}{N_{Ij} }
      \; \delta \left (\frac{q p_j}{p_I p_j} -1 \right )
      \frac{p_I p_j}{q p_I}
      \left ( \frac{p_I p_j}{q p_K} \right )^\ep ( {\cal I}_{jK}(q) -1 ).
  \end{split}
  \label{eq:T_IjK-ab}
\end{equation}
The collinear divergence resides in $T^{(a)}_{IjK}$ and $T^{(b)}_{IjK}$ is
finite.

We begin with the analysis of $T^{(a)}_{IjK}$. To compute it, we choose
$\tilde p_I$ and $\tilde p_K$ as the basis vectors and write
\begin{equation}
  q = \alpha \tilde p_K + \beta \tilde p_I + q_\perp.
  \label{eq:q-sudakov-pk-pI}
\end{equation}
Changing the integration variables $\beta = \alpha \eta$ and integrating over
$\alpha$ we find
\begin{equation}
  T_{IjK}^{(a)} = C_j^{4\ep}
  \left (
    \frac{(p_I p_j) (p_j p_K)}{(p_I p_K)^2}
  \right )^\ep \left ( \frac{p_I p_j}{\tilde p_K \tilde p_I}
  \right )^{1-\ep}
  \int \limits_{0}^{\infty} \;
  \frac{ {\rm d} \eta}{\eta^\ep}
  \; \dOm \;
  \left ( \frac{\tilde p_K \tilde p_I}{q p_j} \right )^{1-3\ep}
  \;
  \frac{\tilde p_K \tilde p_I}{q p_I}
  \; \left ( \frac{\tilde p_K \tilde p_I}{q p_K}
  \right )^\ep,
\end{equation}
where one should  use $q$ as in Eq.~\eqref{eq:q-sudakov-pk-pI} with $\alpha = 1$ and $\beta = \eta$.  Then, writing
\begin{equation}
  p_j = \alpha_j \tilde p_I + \beta_j \tilde p_K + p_{j,\perp},
\end{equation}
and rescaling $\eta = \alpha_j/\beta_j x$, we obtain
\begin{equation}
  \begin{split}
    T_{IjK}^{(a)} & = C_j^{4\ep}
                    \left (
                    \frac{(p_I p_j) (p_j p_K)}{(p_I p_K)^2}
                    \right )^\ep
                    \left ( \frac{p_I p_j}{\tilde p_K \tilde p_I}
                    \right )^{1-\ep}
                    \frac{\alpha_j^{\ep}}{\beta_j^{1-2\ep}}
                    \int \limits_{0}^{\infty}
                    {\rm d} x \;
                    \frac{x^{-\ep} }{1+\Delta_{IKj} x}
                    \; \left ( \frac{1}{x+\Delta_{KIj}}
                    \right )^\ep
    \\
                  & \times \int \frac{ \dOm }
                    {( x+1 - 2 \sqrt{x} \cos \varphi )^{1-3\ep} }
                    \;,
  \end{split}
\end{equation}
where
\begin{align}
  && \Delta_{IKj} = \frac{ \kappa_{IK} \alpha_j}{\beta_j},
  && \Delta_{KIj} = \frac{\kappa_{KI} \beta_j}{\alpha_j}.
  &&
\end{align}
Finally, integrating over $\dOm $ with the help of Eq.~\eqref{eq:cosphi-int}, we derive
the following representation for the function $T_{IjK}^{(a)}$
\begin{equation}
  \begin{split}
    & T_{IjK}^{(a)}  = C_j^{4\ep}
      \left ( \frac{p_I p_j}{\tilde p_K \tilde p_I}
      \right )^{1-\ep} \left (
      \frac{(p_I p_j) (p_j p_K)}{(p_I p_K)^2}
      \right )^\ep
      \frac{\alpha_j^{\ep}}{\beta_j^{1-2\ep}}
      \int \limits_{0}^{\infty}
      \frac{{\rm d} x \; x^{-\ep}}
      {1+\Delta_{IKj} x}
      \; \left ( \frac{1}{x+\Delta_{KIj}}
      \right )^\ep
    \\
    & \times \Bigg ( \theta(1-x) F_{21}(1-3\ep,1-2\ep,1-\ep,x)
      + \frac{\theta(x-1)}{x^{1-3\ep}}  F_{21}(1-3\ep,1-2\ep,1-\ep,x^{-1})
      \Bigg  ).
  \end{split}
\end{equation}
We map the integration region onto the $[0,1]$ interval, rewrite the
hypergeometric function to expose the $x = 1$ singularity and find
\begin{equation}
  \begin{split}
    T_{IjK}^{(a)} & = C_j^{4\ep}
                    \left ( \frac{p_I p_j}{\tilde p_K \tilde p_I}
                    \right )^{1-\ep} \left (
                    \frac{(p_I p_j) (p_j p_K)}{(p_I p_K)^2}
                    \right )^\ep
                    \frac{\alpha_j^{\ep}}{\beta_j^{1-2\ep}}
                    \int \limits_{0}^{1}
                    {\rm d} x \; x^{-\ep} \; (1-x)^{-1+4 \ep}
    \\
                  & \times  F_{21}(2\ep,\ep,1-\ep,x) \left (
                    \frac{1}{(1+\Delta_{IKj} x ) (x+\Delta_{KIj})^\ep}
                    +
                    \frac{1}{(x+\Delta_{IKj}  ) (1+\Delta_{KIj}x )^\ep}
                    \right ).
  \end{split}
  % \label{eq:T_IjK_a-def}
\end{equation}
To compute this integral, we split it into two terms: one, where
$F_{21}(2\ep,\ep,1-\ep,x)$ is replaced with its value at $x = 1$ and the other
one, where we consider the difference of two hypergeometric functions
$F_{21}(..,x) -F_{21}(..,1)$. The latter is finite at $x = 1$ and starts at
${\cal O}(\ep^2)$, so it is easy to analyze it. We write
\begin{equation}
  T_{IjK}^{(a)} = T_{IjK}^{(a,1)} + T_{IjK}^{(a,2)},
\end{equation}
where
\begin{equation}
  \begin{split}
    &T_{IjK}^{(a,1)}  = C_j^{4\ep}
      \left ( \frac{p_I p_j}{\tilde p_K \tilde p_I}
      \right )^{1-\ep} \left (
      \frac{(p_I p_j) (p_j p_K)}{(p_I p_K)^2}
      \right )^\ep
      \frac{\alpha_j^{\ep}}{\beta_j^{1-2\ep}}
      \frac{\Gamma(1-\ep)\Gamma(1-4\ep)}{\Gamma(1-3\ep) \Gamma(1-2\ep) }
    \\
    &
      \times \int \limits_{0}^{1}
      {\rm d} x \; x^{-\ep} \; (1-x)^{-1+4 \ep}
      \left (
      \frac{1}{(1+\Delta_{IKj} x ) (x+\Delta_{KIj})^\ep}
      +
      \frac{1}{(x+\Delta_{IKj}  ) (1+\Delta_{KIj}x )^\ep}
      \right ),
  \end{split}
\end{equation}
and
\begin{equation}
  \begin{split}
    T_{IjK}^{(a,2)}
    & =
      2 \ep^2
      \; \int \limits_{0}^{1}
      {\rm d}x \;
      \frac{(1+x)(1+\Delta_{IKj})^2}{(1-x) (1+\Delta_{IKj} x)(x + \Delta_{IKj})} \left ( {\rm Li}_2(x)
      -\frac{\pi^2}{6} \right ).
  \end{split}
\end{equation}
The function $T^{(a,1)}_{IjK}$ can be written as follows
\begin{equation}
  \begin{split}
    T^{(a,1)}_{IjK}
    & = \left (
      \frac{(p_j p_I)(p_j p_K)}{( p_I p_K )^2 } \right )^{2\ep}  f_{IK,j}^{2 \ep}
      \frac{\Gamma(1-\ep)\Gamma(1-4\ep)}{\Gamma(1-3\ep) \Gamma(1-2\ep) } (1+\Delta_{KIj})^{
      \ep}
    \\
    &  \times
      \Bigg \{
      \frac{\Gamma(1-\ep)  \Gamma(1+4\ep)}{4 \ep \Gamma(1+3\ep)}
      \Bigg (
      F_{21}(\ep,1-\ep,1+3\ep,-\Delta_{KIj} )
    \\
    & + \Delta_{KIj}^{-\ep} F_{21}(\ep,1-\ep,1+3\ep,-\Delta_{KIj}^{-1}
      )
      \Bigg )
    \\
    & +
      \int \limits_{0}^{1}
      {\rm d} x \; \left (
      \frac{\Delta_{IK,j}x^{-\ep} (1-x)^{4 \ep} }{(1+\Delta_{IKj} x ) (x+\Delta_{KIj})^\ep}
      +
      \frac{x^{-\ep} (1-x)^{4 \ep} }{(x+\Delta_{IKj}  ) (1+\Delta_{KIj}x )^\ep}
      \right )
      \Bigg \},
  \end{split}
\end{equation}
where
\begin{equation}
  f_{IK,j} = \frac{C_j^2 (1-y_{IKj})(1-y_{KIj})(1+v_{IK})}{2 v_{IK}^2}.
\end{equation}

The above formula admits a straightforward expansion in powers of $\ep$ since
the collinear divergence has been extracted. However, care is needed to
construct such an expansion without suffering from an uncontrollable growth of
the expression's size at intermediate steps. To this end, we separate the
contribution with hypergeometric functions from the integral over $x$ and denote
the former as $T_{IjK}^{(a,11)}$ and the latter as $T_{IjK}^{(a,12)}$,
\begin{equation}
  T_{IjK}^{(a,1)}=T_{IjK}^{(a,11)} +  T_{IjK}^{(a,12)}.
\end{equation}

It is convenient to write the antisymmetrized ($I \leftrightarrow K$)  version of $T_{IjK}^{a,11}$.  We find
\begin{equation}
  \begin{split}
    T_{[IjK]}^{a,11}
    & =
      \ep
      \Bigg [
      -  X_1(z_{KI,j}) +  X_1(z_{IK,j})
      \Bigg  ] + \ep^2 \Bigg [
      X_{2}(z_{KIj}) -  X_{2}(z_{IKj})
    \\
    & -2\ln \left (\frac{f_{IK,j} (p_jp_K) (p_jp_I)}{(p_Ip_K)^2} \right )
      \left ( X_1(z_{KI,j}) - X_1(z_{IK,j}) \right )
      \Bigg ],
  \end{split}
\end{equation}
where
\begin{equation}
  z_{KIj} = \frac{\Delta_{KIj}}{1+ \Delta_{KI,j}},
\end{equation}
and
\begin{equation}
  \begin{split}
    X_1(x) & =  \ln(1-x) \ln x,
    \\
    X_2(x) &  =
             - \ln^2(1 - x) \ln x
             +
             \ln (1- x) \left (
             \frac{\pi^2}{3}   +  \ln^2 x \right )
    \\
           &  - 2 \ln \left ( \frac{1-x}{x} \right )
             {\rm Li}_2(x)
             - 5 {\rm Li}_3(1 - x) -
             5{\rm Li}_3(x).
  \end{split}
\end{equation}

The next term reads
\begin{equation}
  \begin{split}
    T_{IjK}^{(a,12)} & =
                       \left (
                       \frac{(p_j p_K) (p_jp_I) }{(p_I p_K)^2}
                       \right )^{2\ep} f^{2 \ep}_{IK,j}
                       \frac{\Gamma(1-\ep)\Gamma(1-4\ep)}{\Gamma(1-3\ep) \Gamma(1-2\ep) } (1+\Delta_{KIj})^{
                       \ep}
    \\
                     &
                       \times \int \limits_{0}^{1}
                       {\rm d} x \;
                       \left (
                       \frac{ \Delta_{IKj} x^{-\ep} \; (1-x)^{4 \ep} }{(1+\Delta_{IKj} x ) (x+\Delta_{KIj})^\ep}
                       +
                       \frac{x^{-\ep} \; (1-x)^{4 \ep}}{(x+\Delta_{IKj}  ) (1+\Delta_{KIj}x )^\ep}
                       \right ).
  \end{split}
\end{equation}
We write its expansion in powers of $\ep$ in the following way
\begin{equation}
  T_{IjK}^{(a,12)}
  = \int \limits_{0}^{1}
  {\rm d} x
  \Bigg [
  \frac{\Delta_{IKj}}{1+
    \Delta_{IKj}x}
  \left ( 1 + \ep L_{IKj}
    + \frac{\ep^2}{2} \left( L_{IKj}^2 + \frac{2 \pi^2}{3}
    \right ) \right )
  + (\Delta_{\gI  \gK j} \to \Delta_{\gI \gK j}^{-1})
  \Bigg ],
  \label{eq:T_IjK_a12-def}
\end{equation}
where
\begin{equation}
  L_{IKj} = \ln \left [ \left ( \frac{(p_j p_K) (p_jp_I) }{(p_I p_K)^2}\right )^2  \frac{f_{IK,j}^2 (1-x)^4}{x} \frac{1+\Delta_{KIj}}{x+ \Delta_{KIj}} \right ].
  \label{eq:L_IKj-def}
\end{equation}
The transformation $\Delta_{\gI \gK j} \to 1/\Delta_{\gI \gK j}$ shown in
Eq.~\eqref{eq:T_IjK_a12-def} refers to explicit dependences on these parameters in the
integrand, including in the function $L_{IKj}$ in Eq.~\eqref{eq:L_IKj-def}, and it
should be applied to both $\Delta_{IKj}$ and $\Delta_{KIj}$.
\\

We then continue with the calculation of $T^{(b)}_{IjK}$, defined in
Eq.~\eqref{eq:T_IjK-ab}. In this case, we choose the basis vectors to be $\tilde p_K$
and $p_j$. Since the component of $q$ along $\tilde p_K$ is fixed by the
$\delta$ function, it is convenient to write
\begin{equation}
  q = \alpha \tilde p_K + \beta p_j + q_\perp,
\end{equation}
then rescale $\beta = \eta \alpha$, with $\eta \in [0, \infty]$, and integrate
over $\alpha$. We find
\begin{equation}
  T_{IjK}^{(b)}
  = C_j^{4\ep} \left (
    \frac{p_j p_K}{p_I p_K}
  \right )^{2\ep}
  \frac{p_I p_j}{p_K p_j}
  \int \limits_{0}^{\infty}
  {\rm d} \eta \; \eta^{-\ep} \;
  \dOm \;
  \;
  \frac{p_K p_j}{q p_I}
  \left ( \frac{p_K p_j}{q p_K} \right )^\ep
  \left ( {\cal I}_{jK}(q) - 1 \right ),
  \label{eq:T_IjK_b-def}
\end{equation}
with
\begin{equation}
  q = \tilde p_K + \eta p_j + q_\perp.
\end{equation}
Computing the scalar products in Eq.~\eqref{eq:T_IjK_b-def}, we obtain
\begin{equation}
  \begin{split}
    T_{IjK}^{(b)}&  = \;
                   C_j^{4 \ep} \left (
                   \frac{p_j p_K}{p_I p_K}
                   \right )^{2\ep}
                   \frac{p_I p_j}{p_K p_j}
                   \int \limits_{0}^{\infty}
                   \frac{{\rm d} \eta \; \eta^{-\ep} \;
                   \dOm }{\alpha_I \eta + \beta_I
                   -2 \sqrt{\alpha_I \beta_I \eta} v_{I,Kj} \cos \varphi}
    \\
                 & \times
                   \frac{1}{
                   \left ( \kappa_{Kj} + \eta
                   \right )^\ep }  \left (
                   F_{1m}\left ( \frac{\kappa_{Kj}}{\eta+\kappa_{Kj}}
                   \right )-1
                   \right ),
  \end{split}
\end{equation}
where  we have used
\begin{equation}
  p_I = \alpha_I \tilde p_K + \beta_I p_j + p_{I,\perp}.
\end{equation}
We also note that $v_{I,Kj}$ can be read off from Eq.~\eqref{eq:vIkJ-def}, where
obvious adjustments will have to be made.

We rescale $\eta = \beta_I/\alpha_I x$, integrate over $\dOm$, simplify the
resulting expression as discussed in the previous mass-assignment case and find
\begin{equation}
  \begin{split}
    T_{IjK}^{(b)} & = \; C_j^{4 \ep}
                    \left ( \frac{(p_K p_j) (p_I p_j)}{(p_I  p_K)^2 (1-y_{KjI}) }
                    \right )^{2\ep}
                    \int \limits_{0}^{\infty}
                    \frac{{\rm d} x \; x^{-\ep}  }{\sqrt{(x+1)^2 - 4 x v_{I,Kj}^2}}
                    \left ( \frac{1+\sqrt{1-z} }{2} \right )^{4 \ep} (1-z)^{-\ep}
    \\
                  & \times \left [
                    \frac{1}{\Delta_{KjI} + x}
                    \right ]^\ep
                    \left ( F_{1m}\left ( \frac{\Delta_{KjI}}{x+\Delta_{KjI}}
                    \right ) -1
                    \right ),
  \end{split}
\end{equation}
where
\begin{align}
  && \Delta_{KjI} = \kappa_{Kj}\frac{\alpha_I}{\beta_I},
  && z = \frac{4 \sqrt{x} v_{I,Kj}}{1+x + 2 \sqrt{x} v_{I,Kj}}.
  &&
\end{align}
It is convenient  to change $x \to 1/x$, followed by the rescaling $x \to x/\Delta_{KjI}$. We obtain
\begin{equation}
  \begin{split}
    T_{IjK}^{(b)} & = \;
                    \left ( \frac{C_j^2(p_K p_j) (p_I p_j)}{( p_I p_K)^2 y_{KjI}}
                    \right )^{2\ep}
                    \frac{y_{KjI}}{1-y_{KjI}}
                    \int \limits_{0}^{\infty}
                    \frac{{\rm d} x \; x^{2\ep-1}   }{\sqrt{(x+\Delta_{KjI})^2 - 4 x \Delta_{KjI} v_{I,Kj}^2}}
    \\
                  & \times \left ( \frac{1+\sqrt{1-z_1} }{2} \right )^{4 \ep} \frac{(1-z_1)^{-\ep}}{(1+x)^\ep}
                    \left ( F_{1m}\left ( \frac{x}{1+x}
                    \right ) -1
                    \right ),
  \end{split}
\end{equation}
with
\begin{equation}
  z_1 = \frac{4 \sqrt{x \Delta_{KjI}} v_{I,Kj}}{\Delta_{KjI}+x + 2 \sqrt{x \Delta_{KjI}} v_{I,Kj}},
\end{equation}
and we have used
\begin{equation}
  \Delta_{KjI} = \frac{y_{KjI}}{1-y_{KjI}},
\end{equation}
to rewrite the prefactor of the integral.
Finally, we map the integration region to the $[0,1]$ interval using  the $x \to x/(1-x)$ transformation  and find
\begin{equation}
  \begin{split}
    T_{Ij,K}^{(b)} & = \;
                     \left ( \frac{C_j^2(p_K p_j) (p_I p_j)}{( p_I p_K)^2 y_{KjI}}
                     \right )^{2\ep}
                     \frac{y_{KjI}}{1-y_{KjI}}
                     \int \limits_{0}^{1}
                     \frac{{\rm d} x \; x^{2\ep-1}(1-x)^{-\ep}  }{\sqrt{R_{IKj}(x)}}
    \\
                   & \times \left ( \frac{1+\sqrt{\theta_{IKj}(x)} }{2} \right )^{4 \ep}
                     ( \theta_{IKj}(x))^{-\ep}
                     \left ( F_{1m}\left ( x
                     \right ) -1
                     \right ),
  \end{split}
\end{equation}
where we have used functions introduced in Eq.~\eqref{eq:R_IJk} to present the
result.

The above equation allows a straightforward  expansion of $T_{IjK}^{(b)}$ in $\ep$. We obtain
\begin{equation}
  T_{IjK}^{(b)} = \ep T_{IjK}^{(b,1)} + \ep^2 T_{IjK}^{(b,2)},
\end{equation}
where
\begin{equation}
  T_{IjK}^{(b,1)}
  =
  \frac{y_{KjI}}{1-y_{KjI}}
  \int \limits_{0}^{1}
  \frac{{\rm d} x }{\sqrt{R_{IKj}(x)}}
  \; \frac{1}{x} \; F^{(1)}_{1m}\left ( x
  \right ),
\end{equation}
and
\begin{equation}
  \begin{split}
    T_{IjK}^{(b,2)}
    & =
      \frac{y_{KjI}}{1-y_{KjI}}
      \int \limits_{0}^{1}
      \frac{{\rm d} x }   {\sqrt{R_{IKj}(x)} }
      \; \frac{1}{x} \;
      \Bigg [
      F^{(2)}_{1m}\left ( x
      \right )
    \\
    & + 2 F^{(1)}_{1m} \left ( x
      \right )
      \ln \left ( \frac{C_j^2 (p_K p_j)}{(p_I p_K) \kappa_{Kj}}
      \frac{( 1+\sqrt{\theta_{IKj}(x)})^2 x}{4\sqrt{(1-x)\theta_{IKj}(x)}}
      \right )
      \Bigg ].
  \end{split}
\end{equation}

\subsubsection{Case \texorpdfstring{$i$}{i} massless, \texorpdfstring{$J$}{J} massive, \texorpdfstring{$K$}{K} massive}

In this case, we find
\begin{align}
  && R^t_{iJ} = \frac{1}{1+\kappa_{Ji} \xi},
  && V_{JK} = \frac{\xi}{(1+\kappa_{Ji} \xi) \Delta_K(\xi,\varphi)},
  && f_i = \xi,
  && \bar f_J =1,
  &&
\end{align}
with
\begin{equation}
  \Delta_K =
  \alpha_K \xi + \beta_K
  - 2 \sqrt{\alpha_K \beta_K \xi} v_{K,iJ} \cos \varphi.
\end{equation}
Momenta parametrizations used to obtain the above formulas read
\begin{align}
  && q = p_i + \xi \tilde p_J + q_\perp,
  && p_K = \alpha_K p_i + \beta_K \tilde p_J + p_{K,\perp}.
  &&
\end{align}
The expressions for $A_{iJK}$ and $B_{iJK}$ are
\begin{equation}
  \begin{split}
    & A_{iJK} = ( C_i^{4 \ep} - C_J^{4 \ep}
      ) \int \limits_{0}^{1}
      \frac{{\rm d} \xi \; \dOm }{\xi}
      \frac{\xi^{3 \ep}}{(1+\kappa_{Ji} \xi)^{1+\ep} }
      \left ( \frac{1}{\Delta_K} \right )^\ep {\cal I}_{JK}(q) = 0,
    \\
    & B_{iJK} = C_i^{4 \ep} \;
      \int \limits_{0}^{\infty}
      \frac{{\rm d} \xi \; \dOm }{\xi}
      \frac{\xi^{3 \ep}}{(1+\kappa_{Ji} \xi)^{1+\ep} }
      \left ( \frac{1}{\Delta_K} \right )^\ep {\cal I}_{JK}(q),
  \end{split}
\end{equation}
where the vanishing of $A_{iJK}$ is the consequence of the choice $C_J = C_i$.

Following the discussion of the two previous mass-assignments, we introduce the
quantity $T_{iJK}$
\begin{equation}
  T_{iJK} = \left (\frac{p_i p_J}{p_J p_K}  \right )^\ep B_{iJK}.
\end{equation}
To compute it, we change the basis vectors of the Sudakov decomposition of the
gluon momentum $q$ to the massless versions of $p_K$ and $p_J$. Repeating the
same analysis as before, we find
\begin{equation}
  T_{iJK}
  = C_i^{4 \ep} \left (
    \frac{p_i p_J}{\tilde p_K \tilde r_J}
  \right )^{1-\ep} \left (\frac{p_i p_J}{p_J p_K}  \right )^\ep
  \int \limits_{0}^{\infty}
  \frac{{\rm d} \eta \; \eta^{-\ep}
    \; \dOm }{ (q p_i/(\tilde p_K \tilde r_J))^{1-4\ep}} \frac{\tilde p_K \tilde r_J}{q p_J}
  \left (  \frac{(\tilde p_K  \tilde r_J)^2}{(q p_J) (q p_K)} \right )^\ep {\cal I}_{JK}(q).
\end{equation}
The gluon momentum  $q$  in the above expression reads
\begin{equation}
  q = \tilde p_K + \eta \tilde r_J + q_\perp.
\end{equation}
To integrate over ${\rm d} \Omega_\perp$, we write
\begin{equation}
  p_i = \alpha_i \tilde p_K + \beta_i \tilde r_J + p_{i,\perp},
\end{equation}
and rescale $\eta = \beta_i x/\alpha_i$.  We obtain
\begin{equation}
  \begin{split}
    T_{iJK}
    & = C_i^{4 \ep} \left (
      \frac{p_i p_J}{\tilde p_K \tilde r_J}
      \right )^{1-\ep} \left (\frac{p_i p_J}{p_J p_K}  \right )^\ep
      \frac{\beta_i^{2 \ep}}{\alpha_i^{1-2\ep}}
      \int \limits_{0}^{\infty}
      \frac{{\rm d} x \; x^{-\ep}
      \; \dOm }{ (1+x - 2 \sqrt{x} \cos \varphi)^{1-4\ep}}
    \\
    & \times \frac{1}{(1+\Delta_{JKi} x )^{1+\ep}} \;
      \frac{1}{\left ( x+\Delta_{KJi} \right )^\ep} \; {\cal I}_{JK}(q),
  \end{split}
\end{equation}
where
\begin{align}
  && \Delta_{JKi} = \frac{\kappa_{JK} \beta_i}{\alpha_i},
  && \Delta_{KJi} = \frac{\kappa_{KJ} \alpha_i}{\beta_i}.
  &&
\end{align}
The current ${\cal I}_{JK}$ is independent of $\varphi$ since it is a function
of the scalar products $q p_J$ and $q p_K$. Because of this, integration over
$\varphi$ becomes straightforward. We use Eq.~\eqref{eq:cosphi-int} and find
\begin{equation}
  \begin{split}
    & T_{iJK}
      = C_i^{4 \ep} \left (
      \frac{p_i p_J}{\tilde p_K \tilde r_J}
      \right )^{1-\ep} \left (\frac{p_i p_J}{p_J p_K}  \right )^\ep
      \frac{\beta_i^{2 \ep}}{\alpha_i^{1-2\ep}}
      \int \limits_{0}^{\infty}
      \frac{ {\rm d} x  \; x^{-\ep} }{(1+\Delta_{JKi} x )^{1+\ep}}
      \frac{1}{( x+\Delta_{KJi})^\ep}  {\cal I}_{JK}(q)
    \\
    &
      \times [\theta(1-x) F_{21}(1-4\ep,1-3\ep,1-\ep,x)
      + \theta(x-1) x^{-1+4\ep} F_{21}(1-4\ep,1-3\ep,1-\ep,x^{-1})
      ].
  \end{split}
\end{equation}
We then map the integration region to the interval $x \in [0,1]$, and find
\begin{equation}
  \begin{split}
    T_{iJK} & = C_i^{4 \ep} \left (\frac{p_i p_J}{p_J p_K}  \right )^\ep
              \left (
              \frac{p_i p_J}{\tilde p_K \tilde r_J}
              \right )^{1-\ep}
              \frac{\beta_i^{2 \ep}}{\alpha_i^{1-2\ep}}
              \int \limits_{0}^{1}
              {\rm d} x \; (1-x)^{-1+6\ep} x^{-\ep} F_{21}(3\ep,2\ep,1-\ep,x)
    \\
            & \times\Bigg (
              \frac{ F_{2m}(\alpha_{JK}(x),\beta_{JK}(x) ) }{(1+\Delta_{JKi} x)^{1+\ep} (x + \Delta_{KJi})^\ep}
              + \frac{F_{2m}\left(\alpha_{JK}(x^{-1}),\beta_{JK}(x^{-1}) \right ) }{(x+\Delta_{JKi} )^{1+\ep} (1 + \Delta_{KJi} x)^\ep}
              \Bigg ),
  \end{split}
  \label{eq:T_iJK}
\end{equation}
where $F_{2m}$ is the generic expression for the two-mass current. We note that
$F_{2m}$ is a symmetric function of its arguments, and that the two functions
$\alpha_{JK}$ and $\beta_{JK}$ read
\begin{align}
  % \label{eq:al_JK}
  && \alpha_{JK} = \frac{ \Delta_{JKi} (\Delta_{KJi} + x) }{(1+\Delta_{JKi} \Delta_{KJi} )(1+\Delta_{JKi} x)},
  && \beta_{JK}=\frac{\Delta_{KJi} (1+\Delta_{JKi} x)}{(1+\Delta_{JKi} \Delta_{KJi} )(\Delta_{KJi} + x)}.
  &&
\end{align}
To construct the expansion of the function $T_{iJK}$ in powers of $\ep$, it is
useful to note that the following equation holds
\begin{equation}
  F_{2m}\left(\alpha_{JK}(x^{-1}),\beta_{JK}(x^{-1}) \right )
  = F_{2m}\left(\alpha_{KJ}(x),\beta_{KJ}(x) \right ).
  \label{eq:F2m-trans}
\end{equation}

We proceed with replacing
$F_{21}(...,x) = (F_{21}(...,x) - F_{21}(...,1))+F_{21}(...,1)$ in
Eq.~\eqref{eq:T_iJK} and consider the two terms separately,
\begin{equation}
  T_{iJK} = T_{iJK}^{(a)} + T_{iJK}^{(b)}.
\end{equation}
With the required accuracy
\begin{equation}
  T_{iJK}^{(a)}
  = \ep^2 \left (2- \frac{1}{v_{JK}}
  \right )
  \int \limits_{0}^{1}
  {\rm d} x \; \frac{ \left ( 6 {\rm Li}_2(x) - \pi^2
    \right )}{1-x}
  \frac{(1+x)(1+\Delta_{JKi})^2}{
    (1+\Delta_{JKi} \; x) (x+\Delta_{JKi})
  },
\end{equation}
where we have used
\begin{equation}
  F_{2m}\left(\alpha_{JK}(x),\beta_{JK}(x) \right ) = 2 -\frac{1}{v_{JK} } + {\cal O}(\ep).
\end{equation}

The quantity $T^{(b)}_{iJK}$ is more complicated. It is obtained upon
replacement of $F_{21}(3\ep,2\ep,1-\ep,x)$ with its value at $x=1$ in
Eq.~\eqref{eq:T_iJK}. Simplifying prefactors, we find
\begin{equation}
  \begin{split}
    T^{(b)}_{iJK} & =
                    f_{JK}^\ep
                    \frac{\Gamma(1-6\ep)\Gamma(1-\ep)}{\Gamma(1-3\ep)\Gamma(1-4\ep)}
                    (1+\Delta_{JK,i})  \int \limits_{0}^{1}
                    {\rm d} x \; (1-x)^{-1+6\ep} x^{-\ep}
    \\
                  & \times \Bigg (
                    \frac{ F_{2m}(\alpha_{JK}(x),\beta_{JK}(x) )  }{(1+\Delta_{JKi} x)^{1+\ep} (x + \Delta_{KJi})^\ep}
                    + \frac{F_{2m}(\alpha_{JK}(x^{-1}),\beta_{JK}(x^{-1}) )}{(x+\Delta_{JKi} )^{1+\ep} (1 + \Delta_{KJi} x)^\ep}
                    \Bigg ),
  \end{split}
  \label{eq:T_iJK_b-def}
\end{equation}
where
\begin{equation}
  f_{JK} =  \left ( \frac{(p_i p_J)( p_i p_K) }{(p_J p_K)^2} \right )^2 \frac{C_i^4 (1-y_{KJi})^2(1-y_{JKi})^2 (1+v_{JK})}{ 2 v_{JK}^4}.
\end{equation}
We note that $f_{JK}$ is symmetric under the $J \leftrightarrow K$ replacement.

To proceed further, it is convenient to perform partial fractioning of the
integrand in Eq.~\eqref{eq:T_iJK_b-def}. We write
\begin{equation}
  T^{(b)}_{iJK} =  T^{(b,1)}_{iJK} +T^{(b,2)}_{iJK},
\end{equation}
where
\begin{equation}
  \begin{split}
    T^{(b,1)}_{iJK}
    & =f_{JK}^\ep
      \frac{\Gamma(1-6\ep)\Gamma(1-\ep)}{\Gamma(1-3\ep)\Gamma(1-4\ep)}
      \int \limits_{0}^{1}
      {\rm d} x \; (1-x)^{-1+6\ep} x^{-\ep}
    \\
    & \times \Bigg (
      \frac{ F_{2m}(\alpha_{JK}(x),\beta_{JK}(x) ) }{(1+\Delta_{JKi} x)^{\ep} (x + \Delta_{KJi})^\ep}
      + \frac{F_{2m}(\alpha_{JK}(x^{-1}),\beta_{JK}(x^{-1}) )}{(x+\Delta_{JKi} )^{\ep} (1 + \Delta_{KJi} x)^\ep}
      \Bigg ),
  \end{split}
\end{equation}
and
\begin{equation}
  \begin{split}
    T^{(b,2)}_{iJK}
    & =f_{JK}^\ep
      \frac{\Gamma(1-6\ep)\Gamma(1-\ep)}{\Gamma(1-3\ep)\Gamma(1-4\ep)}
      \int \limits_{0}^{1}
      {\rm d} x \; (1-x)^{6\ep} x^{-\ep}
    \\
    & \times \Bigg (
      \frac{ \Delta_{JK,i} \; F_{2m}(\alpha_{JK}(x),\beta_{JK}(x) )  }{(1+\Delta_{JKi} x)^{1+\ep} (x + \Delta_{KJi})^\ep}
      + \frac{F_{2m}(\alpha_{JK}(x^{-1}),\beta_{JK}(x^{-1}) ) }{(x+\Delta_{JKi} )^{1+\ep} (1 + \Delta_{KJi} x)^\ep}
      \Bigg ).
  \end{split}
\end{equation}

Using Eq.~\eqref{eq:F2m-trans}, it is straightforward to check that, upon
anti-symmetrization with respect to $J$ and $K$, the function $T_{iJK}^{(b,1)}$
vanishes
\begin{equation}
  T_{i[JK]}^{(b,1)} =0.
\end{equation}
The function $T_{iJK}^{(b,2)}$ can be expanded in $\ep$ directly. Upon
anti-symmetrization, we find
\begin{equation}
  \begin{split}
    T^{b,2}_{i[JK]}
    & = F^{(0)}_{JK} \ln \frac{y_{KJi}}{y_{JKi}}
      -\ep \int \limits_{0}^{1}
      {\rm d} x \Bigg [  \frac{(1-\Delta_{JKi} \Delta_{KJ,i})}{(1+\Delta_{JKi} x)(x+\Delta_{KJi})} \Bigg (
      F^{(1)}_{JK}(x) +F^{(0)}_{JK}
      {\cal L}_{JKi}(x)
    \\
    & +
      \ep \left ( F^{(2)}_{JK}(x)+\pi^2 F^{(0)}_{JK}
      + \frac{F^{(0)}_{JK}}{2}  {\cal L}^2_{JKi}(x)
      + F^{(1)}_{JK}(x) {\cal L}_{JKi}(x)
      \right ) \Bigg )
      - (J \leftrightarrow K) \Bigg  ],
  \end{split}
  \label{eq:Tb2_iJK-Tb2_iKJ}
\end{equation}
with
\begin{equation}
  {\cal L}_{JKi}(x) = \ln \left [ \frac{(1-x)^6 f_{JK}}{(1+\Delta_{JKi} x  )x(x+\Delta_{KJi})} \right ],
\end{equation}
and
\begin{equation}
  F_{2m}(\alpha_{JK}(x),\beta_{JK}(x)) = F^{(0)}_{JK}(x) + \ep F^{(1)}_{JK}(x) + \ep^2 F^{(2)}_{JK}(x).
\end{equation}
Finally, we note that 
\begin{equation}
  F^{(0)}_{JK}= 2 - \frac{1}{v_{JK}}.
\end{equation}

\subsubsection{The result for the case with two massive and one massless partons }

For the case with two massive and one massless emitters, the required
$N$-jettiness-independent real-virtual contribution can be written in the
following way
\begin{equation}
  S_{RV} = \frac{\pi a_s^2}{4 \ep^2}
  \frac{e^{-2 \gamma_E \ep} \Gamma(1-4\ep)}{\Gamma(1-\ep)^2}
  \; \sum H_{IJk}
  \left ( \frac{ (\mu \bar u)^2 (p_I p_J)}{2} \right )^{2\ep}  {\cal D}_{IJk},
\end{equation}
where
\begin{equation}
  {\cal D}_{IJk} = ( T_{IJk} - T_{JIk} ) -( T_{IkJ}-T_{JkI}) + (T_{kIJ} - T_{kJI} ).
\end{equation}
The individual contributions to $D_{IJk}$ were discussed in the preceding
subsections; here we collect them into a single formula. We write
\begin{equation}
  {\cal D}_{IJk}
  = D_{IJk}^{(0)} + \ep {\cal D}^{(1)}_{IJk}
  + \ep^2 D^{(2)}_{IJk} + {\cal O}(\ep^3).
  \label{eq:DIJk-exp-ep}                 %Former eq5.271 (1)
\end{equation}
We present the individual expansion coefficients separately. However, before we
do this, we list all the functions and parameters that we require to present the
result. They are
\begin{equation}
  \begin{split}
    & \Delta_{\gI \gJ \gK} = \frac{y_{\gI \gJ \gK }(1-y_{\gJ \gI \gK })}{1-y_{\gI \gJ \gK}},
      \;\;\;\; y_{\gI \gJ \gK} = \frac{m_\gI^2}{p_\gI p_\gJ(1+v_{\gI \gJ})} \frac{p_\gJ p_\gK}{p_\gI p_\gK},\;\;\;
      z_{IJk} =\frac{\Delta_{IJk}}{1+\Delta_{IJk}},
    \\
    & \gamma_{IJk}(x) = x + \Delta_{IJk},\qquad \beta_{IJk}(x) = 1+ \Delta_{IJk} x,
      \qquad \alpha_{IJk} = 1+ \Delta_{IJk},
    \\
    & L_{IJk} =\ln \left [ \left ( \frac{(p_I p_k)(p_J p_k)}{(p_I p_J)^2}\right )^2  \frac{C_k^4 (1-y_{IJk})^2 (1-y_{JIk})^2(1+v_{IJ})^2 (1-x)^4}{4  x v_{IJ}^4} \frac{\alpha_{JIk}}{\gamma_{JIk}(x)} \right ],
    \\
    & {\bar L}_{IJk}=L_{IJ,k}|_{\gamma_{JIk} \to \beta_{JIk} },
    \\
    & {\cal L}_{IJ,k} = \ln \left [ \left ( \frac{(p_I p_k)(p_J p_k)}{(p_I p_J)^2} \right )^2
      \frac{(1-x)^6 C_k^4(1-y_{IJk})^2(1-y_{JIk})^2 (1+v_{IJ})}{\beta_{IJk}(x) \; \gamma_{JIk}(x) \; x \; 2 v_{IJ}^4}
      \right ],
    \\
    & R_{IJk} = (\Delta_{JkI}(1-x)+x)^2 - 4x(1-x) \Delta_{JkI} v_{I,kJ}^2,
    \\
    & \theta_{IJk} =
      \frac{\Delta_{JkI}(1-x) + x - 2\sqrt{(1-x) x \Delta_{JkI}} v_{I,kJ}}{\Delta_{JkI}(1-x) + x + 2\sqrt{(1-x) x \Delta_{JkI}} v_{I,kJ}},
    \\
    & X_1(x)  =  \ln(1-x) \ln x,
    \\
    & X_2(x)   =
      - \ln^2(1 - x) \ln x
      +
      \ln (1- x) \left (
      \frac{\pi^2}{3}   +  \ln^2 x \right )
    \\
    &  \quad \quad \quad - 2 \ln \left ( \frac{1-x}{x} \right )
      {\rm Li}_2(x)
      - 5 {\rm Li}_3(1 - x) -
      5{\rm Li}_3(x).
  \end{split}
\end{equation}
We will also use the following short-hand notation
\begin{equation}
  \left \langle F_{\gI \gJ}(x) \right \rangle_{\gI \leftrightarrow \gJ}
  = \int \limits_{0}^{1} {\rm d} x
  \; ( F_{\gI \gJ}(x) - F_{\gJ \gI}(x) ).
\end{equation}

The contributions to $D^{(0)}_{IJk}$ originate from Eqs~\eqref{eq:T_IjK_a12-def} and
\eqref{eq:Tb2_iJK-Tb2_iKJ}, and the first equation needs to be anti-symmetrized. We find
\begin{equation}
  D^{(0)}_{IJk} = -\ln \frac{y_{JIk}}{y_{IJk}} + \left ( 2-
    \frac{1}{v_{IJ}} \right ) \ln \frac{y_{JIk}}{y_{IJk}}
  = \left ( 1-\frac{1}{v_{IJ}}
  \right ) \ln \frac{y_{JIk}}{y_{IJk}}.
\end{equation}
The ${\cal O}(\ep)$ and ${\cal O}(\ep^2)$ contributions to Eq.~\eqref{eq:DIJk-exp-ep}  evaluate to
\begin{equation}
  \begin{split}
    & D^{(1)}_{IJk} = -\frac{1}{2 v_{IJ}} \ln \frac{1+v_{IJ}}{1-v_{IJ}}  \;
      \ln \frac{ y_{JIk}}{ y_{IJk} }
      -\left \langle  \ln(1+\Delta_{JIk})\ln\frac{\Delta_{JIk}}{1+\Delta_{JIk}}\right \rangle_{I \leftrightarrow J}
    \\
    & - \left \langle
      \frac{\Delta_{IJk}}{\beta_{IJk}(x) }\;  L_{IJk} +
      \frac{1}{\gamma_{IJk}(x)} \; {\tilde L}_{IJk} \right \rangle_{I \leftrightarrow J} + \left \langle  \frac{1}{1-y_{JkI}}  \frac{F_{1m}^{(1)}(x)}{\sqrt{R_{IJk}(x)}} \left (
      1- \frac{y_{JkI}}{x} \right )
      \right \rangle_{I \leftrightarrow J}
    \\
    & - \frac{2 v_{IJ}}{1+v_{IJ}}
      \left \langle
      \frac{(F^{(1)}_{IJ}(x)+ F^{(0)}_{IJ}
      {\cal L}_{IJk}(x)}{\beta_{IJk}(x) \; \gamma_{JIk}(x) }
      \right \rangle_{I \leftrightarrow J} ,
  \end{split}
  \label{eq:D1_IJk}                 %Former eq5.271 (2)
\end{equation}
and
\begin{equation}
  \begin{split}
    D^{(2)}_{IJk}& =
                   -\frac{2}{1+v_{IJ}}
                   \Bigg \langle
                   \frac{1}{\gamma_{IJk}(x) \beta_{JIk}(x)}  \; \ln \left ( \frac{\alpha_{IJk} \; \beta_{JI,k}(x)}{\alpha_{JIk}  \gamma_{IJk}(x)}
                   \right )
    \\
                 & \times
                   \ln \left (
                   (C^2 v_{IJ}^+)^2 \; x\sqrt{\frac{\alpha_{JIk} \alpha_{IJk} }{\beta_{JIk}(x)\;\gamma_{IJk}(x) }} \;
                   \right )
                   \Bigg \rangle_{I \leftrightarrow J}
                   + \Bigg  \langle
                   \frac{1}{1-y_{JkI}}
                   \frac{1 }{\sqrt{R_{IJk}(x)}}
    \\
                 & \times \Bigg [ F_{1m}^{(2)}
                   \left (   x  \right )
                   + \ln  \left (
                   \frac{(C^2 v^+_{IJ} )^2 \kappa_{Jk} (1-x) (1+\sqrt{\theta_{IJk}(x)})^4}{16 \; x \;\theta_{IJk}(x)}
                   \right )
                   F_{1m}^{(1)} \left (
                   x \right )
                   \Bigg  ]
                   \Bigg  \rangle_{I \leftrightarrow J}
    \\
                 &- \left \langle
                   X_{2}(z_{JIk})-2\ln
                   \left ( \frac{C_k^2(1-y_{IJk})(1-y_{JIk})(1+v_{IJ})}{2 v_{IJ}^2} \frac{(p_I p_k)(p_J p_k)}{(p_I p_J)^2}
                   \right ) X_{1}(z_{JIk})
                   \right \rangle_{I \leftrightarrow J}
    \\
                 & - \frac{1}{2} \left \langle
                   \frac{\Delta_{IJk}}{\beta_{IJk}(x) }\;  \left (  L^2_{IJk} + \frac{2\pi^2}{3} \right )
                   +\frac{1}{\gamma_{IJk}(x)} \; \left (  {\tilde L}^2_{IJk} + \frac{2\pi^2}{3} \right )\right \rangle_{I \leftrightarrow J}
    \\
                 & - \left \langle \frac{y_{JkI}}{(1-y_{JkI}) x \sqrt{R_{IJk}(x)}}
                   \Bigg [
                   F^{(2)}_{1m}\left ( x
                   \right )
                   + 2 F^{(1)}_{1m} \left ( x
                   \right )
                   \ln \left ( \frac{C_k^2 s_{Jk}} {(p_I p_J) \kappa_{Jk}}
                   \frac{( 1+\sqrt{\theta_{IJk}(x)})^2 x}{4\sqrt{(1-x)\theta_{IJk}(x)}}
                   \right )
                   \Bigg ]
                   \right \rangle_{I \leftrightarrow J}
    \\
                 & + \left ( \frac{5}{3} -\frac{1}{v_{IJ}}  \right ) \left \langle
                   \frac{(1+x) \; \alpha_{IJk}^2}{(1-x) \; \beta_{IJk}(x) \; \gamma_{IJk}(x)} \left ( 6 {\rm Li}_2(x) - \pi^2
                   \right )
                   \right \rangle_{I \leftrightarrow J}
    \\
                 & + \pi^2 F^{(0)}_{IJ} \ln \frac{y_{JIk}}{y_{IJk}}
                   - \frac{v_{IJ}}{1+v_{IJ}}
                   \left \langle
                   \frac{(2 F^{(2)}_{IJ}(x)+ 2 F^{(1)}_{IJ}
                   {\cal L}_{IJk}(x)
                   + F^{(0)}_{IJ}
                   {\cal L}^2_{IJk}(x)
                   }{\beta_{IJk}(x) \; \gamma_{JIk}(x) }
                   \right \rangle_{I \leftrightarrow J}
    \\
                 & + G_{IJk}(q_{IJ}) - G_{JIk}(q_{JI}),
  \end{split}
\end{equation}
where
\begin{align}
  && q_{IJ} = \bar x \tilde p_I + x \tilde p_J + q_\perp, && \bar x = 1-x, &&
\end{align}
\begin{equation}
  G_{IJk}(q) = \int \limits_{0}^{1} {\rm d} x \;
  \int\limits_0^{2\pi} \frac{{\rm d} \varphi}{2 \pi}
  \frac{1+\kappa_{IJ} \kappa_{JI} }{(\bar x +x \kappa_{JI})(x + \bar x \kappa_{IJ} )}
  \ln \left [ \frac{x^2  {\bar x}^2  (\tilde p_I \tilde p_J)}{( \bar x +\kappa_{JI} x ) (p_k q)} \left ( \frac{p_J p_k}{r_J q}
    \right )^2
  \right ] I^{(1)}_{Jk}(q),
\end{equation}
and
\begin{equation}
  r_J = p_J - \kappa_{Jk} p_k.
\end{equation}
The azimuthal angle $ \varphi$ parametrizes the direction of the vector
$q_\perp$ relative to the direction of the vector $p_{k,\perp}$ in a plane
transverse to directions of $p_{I}$ and $p_J$.

We note that the analytic computation of the function $D^{(1)}_{IJk}$ is
straightforward but somewhat tedious. In fact, individual contributions shown in
Eq.~\eqref{eq:D1_IJk} contain polylogarithmic functions with complicated
arguments. However, when the different contributions are put together, a
remarkably simple result is found
\begin{equation}
  D^{(1)}_{IJk} =
  2 \ln \left ( \frac{y_{JIk}}{y_{IJk}}
  \right )
  \left (
    \ln(1+v_{IJ})
    - \frac{(1-v_{IJ})}{v_{IJ}} \ln
    \frac{2 C_k^2 (p_I p_k) (p_J p_k) }{m_I^2 m_J^2}
    - \frac{1}{v_{IJ}} \ln (2 v_{IJ})
  \right ).
\end{equation}

Finally, we can assemble the final result for the case of the two massive and
one massless parton. After putting everything together, we obtain the following
expression for the $N$-jettiness-independent term
\begin{equation}
  \begin{split}
    \tilde S_{1,2m}^{\rm tri, (0)}
    & =  \frac{a_s^2 \pi}{2}
      \sum \limits_{I,J,k}
      H_{IJk} \; \Bigg \{
      \frac{1}{2}
      D^{(2)}_{IJk} +  \ln \left( \frac{\mu^2 \bar u^2 (p_I p_J )}{2} \right) D^{(1)}_{IJk}
      + \Big [ \ln^2 \left( \frac{\mu^2 \bar u^2 (p_I p_J)}{2} \right)
    \\
    & + \frac{7 \pi^2}{12}  \Big ] \; D^{(0)}_{IJk} - \left( 1-\frac{1}{v_{IJ}}\right) \left[ \frac{1}{6} \left( L^{ 3}_{Ik} - L^{3}_{Jk}\right) +\frac{5 \pi^2}{12} \left( L_{Ik} - L_{Jk}\right)\right]
      \Bigg \}.
  \end{split}
  \label{eq:Stri0-MM0-final}
\end{equation}

\section{Results}
\label{sec:results}

We have used the semi-analytic results for the dipole and tripole contributions
presented in the previous sections to construct a {\sf FORTRAN} code which
computes the $N$-jettiness soft function for given momenta and masses of hard
emitters. The number of massive hard emitters is restricted to two. We emphasize
that, since all divergencies have been canceled analytically, the code does not
require an infrared regulator.

The code contains separate modules for evaluating the dipole and tripole
contributions. The dipole contribution requires the numerical evaluation of a
finite difference between soft functions with inclusive and regular
$N$-jettiness variables. This difference, given by a five-dimensional integral,
is the most computationally-demanding part of the calculation. In addition,
there are several contributions to the inclusive $N$-jettiness soft function
that require up to three numerical integrations; together, these contribution
also require a non-negligible amount of computing effort. The tripole
contribution is evaluated analogously, but the highest number of integrations in
this case is two. Hence, in comparison to the dipole contributions, the tripole
one requires rather modest computational effort.

In this section, we illustrate the efficacy of the derived formulas by
presenting the numerical results for the $N$-jettiness soft function for three
processes with massive final-state particles. We also compare our results with
those of Ref.~\cite{Bell:2026pou}, where the zero-jettiness soft function for
heavy quark pair production in hadron collisions was computed.

We note that for all numerical computations reported below we choose the
normalization coefficients in the $N$-jettiness function
Eq.~\eqref{eq:TN-std-def} to be $Q_i = E_i$, where $E_i$ is the energy of the
massless hard parton $i$. With this normalization, the $N$-jettiness soft
function depends on the relative angles between directions of hard massless
partons as well as directions of heavy quarks and their velocities.

To present the  results for the renormalized soft function, we write it in the following way
\be
 \tilde S = 1 + \tilde S_1 + \left(\frac{1}{2} \tilde{S}_{1}^2 + \tilde S_2^{\rm dip} + \tilde S_2^{\rm tri} \right) + \ldots,
\ee
where
\begin{equation}
  \begin{split}
  & \tilde S_1
      = a_s \; \Bigg\{ \sum \limits_{i \ne j} {\bf T}_i \cdot {\bf T}_j \; \tilde{W}_{ij}^{(1)}
      + 2 \sum \limits_{I,j} {\bf T}_I \cdot {\bf T}_j \; \tilde{W}_{Ij}^{(1)}
      + \sum \limits_{I \ne J} {\bf T}_I \cdot {\bf T}_J \; \tilde{W}_{IJ}^{(1)} \Bigg\},
    \\
    & \tilde S_2^{\rm dip}
      = a_s^2 \; \Bigg\{ \sum \limits_{i \ne j} {\bf T}_i \cdot {\bf T}_j \; \tilde{W}_{ij}^{(2)}
      + 2 \sum \limits_{I,j} {\bf T}_I \cdot {\bf T}_j \; \tilde{W}_{Ij}^{(2)}
      + \sum \limits_{I \ne J} {\bf T}_I \cdot {\bf T}_J \; \tilde{W}_{IJ}^{(2)} \Bigg\},
    \\
    & \tilde S_2^{\rm tri}
      = a_s^2 \; \Bigg\{ \sum \limits_{i,j,k} H_{ijk} \; U_{i,jk}
      + \sum \limits_{I,j,k} H_{Ijk} \; U_{I,jk}
      + \sum \limits_{I,J,k} H_{IJk} \; U_{IJ,k} \Bigg\}.
  \end{split}
  \label{eq:StldDipTrip}
\end{equation}
The tripole coefficients in Eq. \eqref{eq:StldDipTrip} satisfy the following antisymmetry relations
\begin{equation}
  U_{i,jk} = -U_{i,kj}, \quad U_{I,jk} = -U_{I,kj}, \quad U_{IJ,k} = -U_{JI,k}.
\end{equation}
We note that for sums with indices of the same type, i.e. $i,j$ or $I,J$,
symmetric combinations of indices should be included.

The coefficients $\tilde W_{ij}^{(1)}$, $\tilde W_{Ij}^{(1)}$ and $\tilde W_{IJ}^{(1)}$ entering $\tilde S_1$ are given in Eqs~\eqref{eq:dip-nlo-00-final}, \eqref{eq:dip-nlo-M0-final} and \eqref{eq:dip-nlo-MM-final}, respectively. The dipole coefficient $\tilde W_{ij}^{(2)}$, $\tilde W_{Ij}^{(2)}$ and $\tilde W_{IJ}^{(2)}$ entering $\tilde S_2^{\rm dip}$ can
be found in Eqs~\eqref{eq:dip-nnlo-00-final}, \eqref{eq:dip-nnlo-M0-final} and \eqref{eq:dip-nnlo-MM-final},
respectively. Likewise, the tripole coefficients $U_{\gI \gJ \gK}$, defined in
Eq.~\eqref{eq:S2tri-U}, can be obtained by combining the $N$-jettiness dependent
contribution of Eq.~\eqref{eq:S2tri-g-part} with the corresponding results presented in
Eqs \eqref{eq:Stri0-000-final-def}, \eqref{eq:Stri0-M00-final} and \eqref{eq:Stri0-MM0-final}.

In what follows, we present the dipole ($\tilde W_{\gI \gJ}^{(2)}$) and tripole
($U_{\gI \gJ \gK}$) coefficients for three processes: $pp \to Q \bar Q$,
$pp \to Q \bar Q + j$ and $pp \to Q \bar Q + 2j$ for selected kinematic points.
In all examples presented below, we consider two incoming massless partons with
the four-momenta
\begin{equation}
  \label{eq:p12def}
  p^\mu_{1,2} = (1,0,0,\pm 1),
\end{equation}
where we put their energies to one in accord with the above discussion of the
normalization factors in the $N$-jettiness observable. Furthermore, for all
numerical results below, we set the product of the Laplace transform parameter
$\bar u$ and the renormalization scale $\mu$ to one.

\subsection{Heavy quark  pair production \texorpdfstring{$pp \to Q \bar Q$}{pp -> QQ} }

In this section, we compute the zero-jettiness soft function for heavy quark
pair production in hadron collisions. Such a calculation was very recently
reported in Ref.~\cite{Bell:2026pou}, providing a useful cross check for our
results.

\begin{figure}[t]
  \centering
  \includegraphics[width=0.9\linewidth]{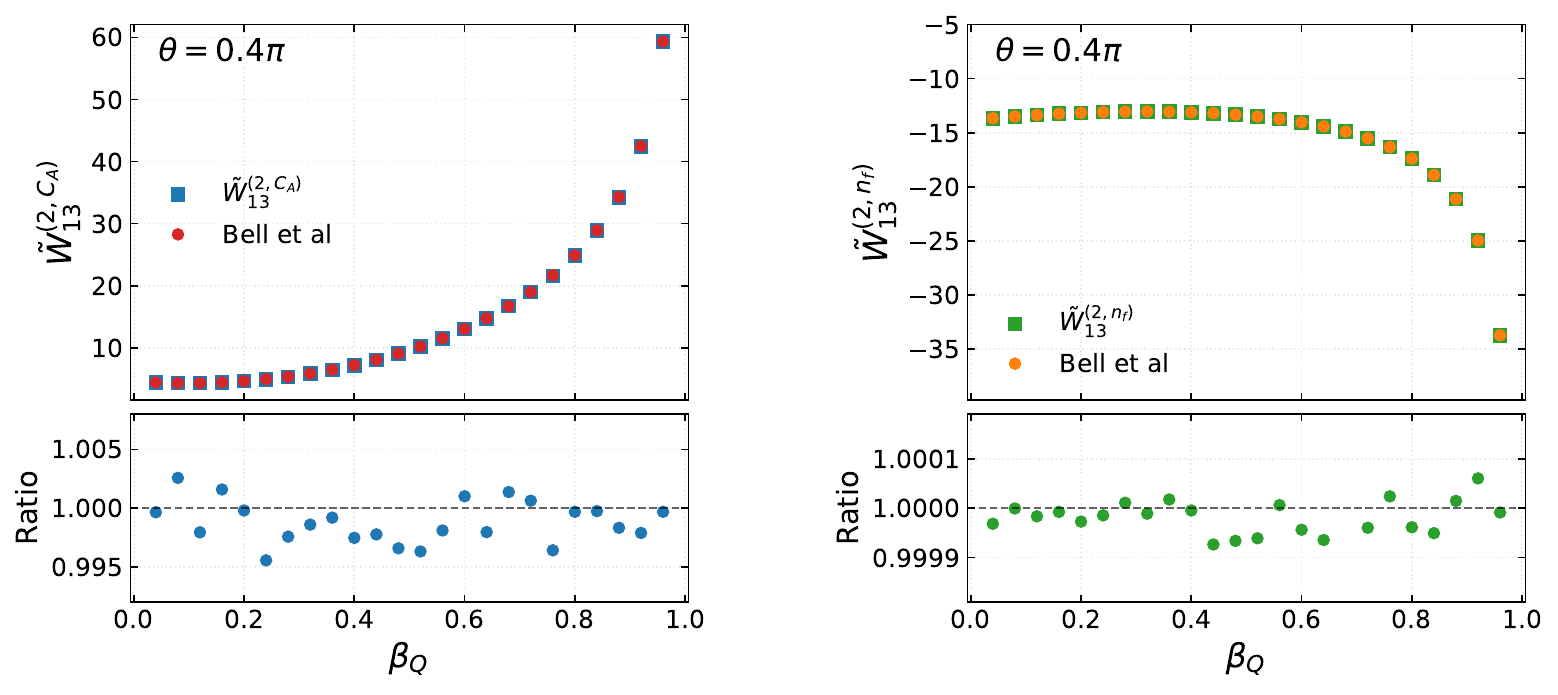}
  \caption{The dipole coefficient $\tilde W_{13}^{(2)}(\theta,\beta_Q)$ for the
    $Q \bar Q$ production process Eq.~\eqref{eq:ffToQQ} at fixed  value of the production angle $\theta = 0.4 \pi$. Comparison
    with the results of  Ref.~\cite{Bell:2026pou} is shown in lower panes.}
  \label{fig:0jet-dip13}
\end{figure}

\begin{figure}
  \centering
  \includegraphics[width=0.9\linewidth]{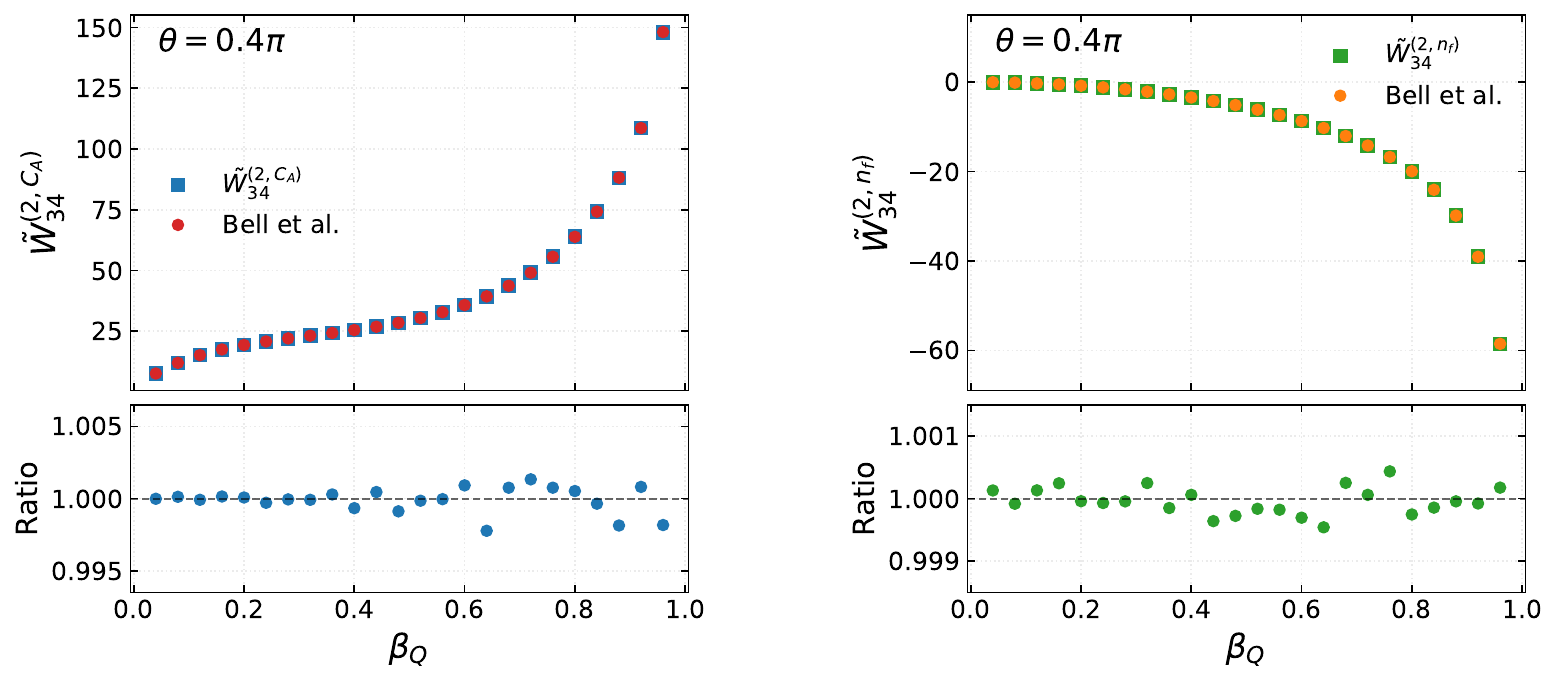}
  \caption{
    The dipole coefficient $\tilde W_{34}^{(2)}(\theta,\beta_Q)$ for the  
    $Q \bar Q$ production process Eq.~\eqref{eq:ffToQQ} at fixed  value of the production angle $\theta = 0.4 \pi$. Comparison
    with the results of  Ref.~\cite{Bell:2026pou} is shown in lower panes.
  }
  \label{fig:0jet-dip34}
\end{figure}

Following Ref.~\cite{Bell:2026pou}, we consider partonic process
\begin{equation}
  f(p_1) + f(p_2) \to Q(p_3) + \bar Q(p_4),
  \label{eq:ffToQQ}
\end{equation}
with momenta $p_{1,2}$ defined in Eq.~\eqref{eq:p12def}. We work in the
center-of-mass frame of the partonic process and parametrize the heavy-quark
momenta as
\begin{equation}
  \label{eq:QQn3n4-b2b}
  p_{3,4} = \frac{m_Q}{\sqrt{1-\beta_Q^2}}(1, \pm \beta_Q \sin \theta, 0,  \pm \beta_Q\cos \theta).
\end{equation}
It follows from Eq.~\eqref{eq:QQn3n4-b2b} that $m_Q$ and $\beta_Q$ are the mass
and the velocity of the heavy quark, respectively, and $\theta$ is it's
production angle. We note that the zero-jettiness soft function cannot depend on
$m_Q$ and we have verified that our results do indeed have this property.

\begin{table}[ht]
  \centering
  \renewcommand{\arraystretch}{1.5}
  \begin{tabular}{|c|c||l|l||l|l|}
    \hline
    & \multicolumn{1}{c||}{\rule{0pt}{3ex}$\beta_Q$}
    & \multicolumn{1}{c|}{$\tilde W_{\gI \gJ}^{(2, C_A)}$}
    & \multicolumn{1}{c||}{Ref. \cite{Bell:2026pou}}
    & \multicolumn{1}{c|}{$\tilde W_{\gI \gJ}^{(2,n_f)}$}
    & \multicolumn{1}{c|}{Ref. \cite{Bell:2026pou}} \\[0.8ex]
    \hline\hline
    \multirow{3}{*}{$\tilde W_{13}^{(2)}$}
    & $0.04$ & -12.829(7) & -12.85(2) & -2.7856(1) & -2.786(3) \\
    \cline{2-6}
    & $0.44$ & 15.567(6) & 15.58(2) & -15.4291(8)  & -15.428(2)  \\
    \cline{2-6}
    & $0.96$ & 17.914(5) & 17.92(1) & -24.4444(9) & -24.444(2) \\
    \hline \hline
    \multirow{3}{*}{$\tilde W_{34}^{(2)}$}
    & $0.04$ & -2.673(6) & -2.674(4) & -8.6446(6) & -8.6455(7)  \\
    \cline{2-6}
    & $0.44$ & 37.548(9) & 37.52(1) & -8.623(4)  & -8.621(1)  \\
    \cline{2-6}
    & $0.96$ & 0.115(6)  & 0.120(5) & -8.7332(7) & -8.7339(8) \\
    \hline
  \end{tabular}
  \caption{Dipole coefficients for heavy-quark pair production at $\theta = 0.6 \pi$, for different values of $\beta_Q$.}
  \label{tab:0jet-dip-th06-betaQ}
\end{table}
In Figs.~\ref{fig:0jet-dip13} and \ref{fig:0jet-dip34} we show the comparison of
our results for two dipole coefficients $\tilde W_{13}^{(2)}$ and $\tilde W_{34}^{(2)}$ with those of
Ref.~\cite{Bell:2026pou}. We show separately the results for two soft gluons
(proportional to $C_A$) and for two soft massless quarks (proportional to $n_f$)
for the production angle $\theta= 0.4 \pi$, as a function of the velocity
$\beta_Q$. We observe excellent agreement with the results of
Ref.~\cite{Bell:2026pou}.

In addition, we show the numerical results for $\tilde W_{13}^{(2)}$ and $\tilde W_{34}^{(2)}$ and their
comparison with Ref.~\cite{Bell:2026pou} for different values of $\beta_Q$ and
the angle $\theta=0.6\;\pi$ in Table~\ref{tab:0jet-dip-th06-betaQ}. We further note that a
grid of 576 points in the $\beta_Q-\theta$ plane for $\tilde S_2^{\rm dip}$ is
provided in Ref.~\cite{Bell:2026pou} and we have checked that we agree with
those results at the level which is similar to what is seen in
Table~\ref{tab:0jet-dip-th06-betaQ}. It is interesting to note that with out code it takes
about thirty minutes to compute the full set of 576 points with precision
commensurate with the one reported in Ref.~\cite{Bell:2026pou}, illustrating
high numerical efficiency of the code.\footnote{We emphasize that the code is
very general and is not optimized for calculations in the back-to-back
kinematics.}

We continue with the discussion of the triple-color-correlated contributions for
the heavy quark production process Eq.~\eqref{eq:ffToQQ}. As explained in the
previous sections, it is possible to construct finite tripole coefficients
$U_{ \gI \gJ \gK}$ by antisymmetrizing various contributions in a particular
way. In addition to that, the process in Eq.~\eqref{eq:ffToQQ} is quite special
because, thanks to the color conservation, there is just one independent tripole
structure. Furthermore, this process involves just two massless partons which
means that massless tripoles ($U_{i,jk}$) do not exist, and, because of the
back-to-back kinematics $U_{4,12}=-U_{3,12}$ and $U_{34,1} = -U_{34,2}$. With
this results, we obtain
\begin{equation}
  \tilde S_2^{\rm tri}
  = 4 \; a_s^2 \; H_{123} \left( U_{3,12} + U_{34,1} \right) = 16 \left( \frac{\alpha_s}{4 \pi} \right)^2 H_{123} \left( U_{3,12} + U_{34,1} \right) = \left( \frac{\alpha_s}{4 \pi} \right)^2 H_{123} \; C_{123}.
\end{equation}

In Table~\ref{tab:0jet-trip-bell}, we present numerical results for the coefficients
$U_{3,12}$ and $U_{34,1}$ and compare their sum with the tripole contribution
extracted from Ref.~\cite{Bell:2026pou} for different values of $\beta_Q$ and
two production angles $\theta=0.4\pi$ and $\theta=0.04\pi$. In both cases, we
find very good agreement with the largest deviations at the level of $0.1$
percent. 
We note that a typical
tripole point reaches ${\cal O }(0.1 \%)$ accuracy in about half a second,
making it possible to calculate all $576$ points in the $(\beta_Q-\theta)$ grid
of Ref.~\cite{Bell:2026pou} in about 5 minutes for this level of precision.

\begin{figure}[t]
  \centering
  \includegraphics[width=0.5\linewidth]{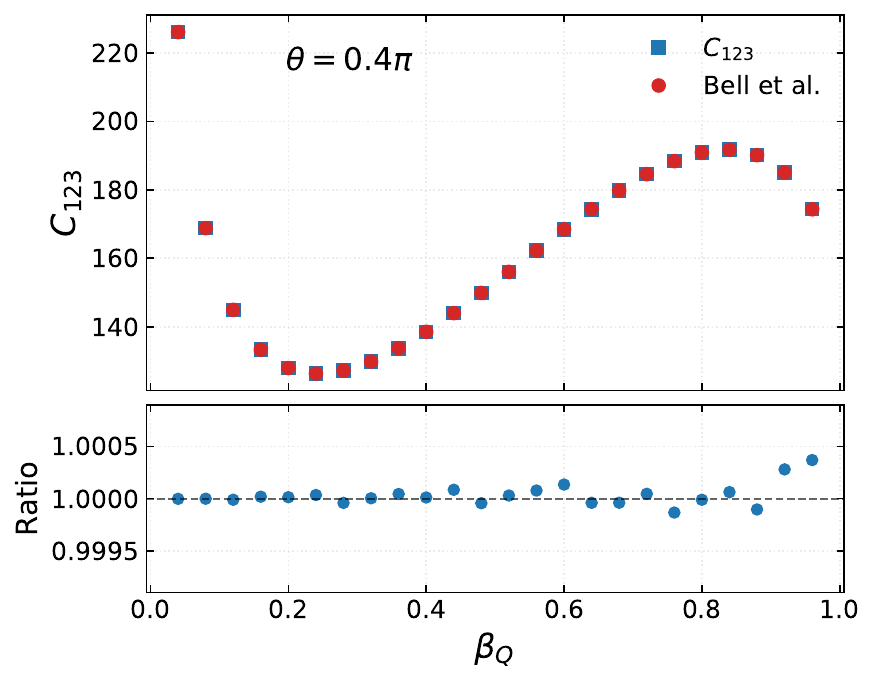}
  \caption{Tripole sum coefficient $C_{123}$ of the 0-jettiness soft function in $Q \bar Q$ production at fixed $\theta = 0.4 \pi$, and the comparison with Ref.~\cite{Bell:2026pou}.}
  \label{fig:0jet-trip}
\end{figure}

\begin{table}[ht]
  \centering
  \renewcommand{\arraystretch}{1.5}
  \begin{tabular}{|c|c|l|l|l|l|}
    \hline
    $\theta$ & $\beta_Q$ & \multicolumn{1}{c|}{$U_{3,12}$} & \multicolumn{1}{c|}{$U_{34,1}$} & \multicolumn{1}{c|}{$C_{123}$} & Ref.~\cite{Bell:2026pou}  \\
    \hline\hline
    \multirow{3}{*}{$0.4\pi$} &  0.04 & 0.766237(2)  & 13.3640(2)  & 226.084(3) & 226.084(7)  \\
    \cline{2-6}
             & 0.44 & 8.03213(6)  & 0.9667(5)  & 143.981(8) & 143.970(6)  \\
    \cline{2-6}
             & 0.96 & 13.2903(5)  & -2.390(3)  & 174.40(5) & 174.32(2)  \\
    \hline\hline
    \multirow{3}{*}{$0.04\pi$} & 0.04 & 2.462743(4)  & 42.9431(2)  &  726.493(3) & 726.493(9)  \\
    \cline{2-6}
             & 0.44 & 29.84383(6)  & 3.4502(1)  & 532.704(2) &  532.70(1)  \\
    \cline{2-6}
             & 0.96 & 164.0438(7)  & {0.409(1)} & 2631.24(2) & 2631.22(4)   \\
    \hline
  \end{tabular}
  \caption{Tripole coefficients and sum for heavy-quark pair production at $\theta = 0.4 \pi$ and $\theta = 0.04 \pi$, for selected values of $\beta_Q$.}
  \label{tab:0jet-trip-bell}
\end{table}

\subsection{Heavy quark pair production in association with jets}
\label{sec:heavy-quark-pair-jets}
The main result of this paper is Eq.~\eqref{eq:StldDipTrip}, which provides the
NNLO contribution to the $N$-jettiness soft function for processes involving up
to two massive particles and an arbitrary number of massless hard partons. In
this section, we illustrate the use of this equation by providing numerical
results for dipole and tripole contributions to the $N$-jettiness soft function
for the $Q\bar Q$ production accompanied by one or two additional jets.

For these cases, we generalize the kinematic configuration in
Eq.~\eqref{eq:QQn3n4-b2b} by allowing two massive particles to be at an
arbitrary angle to each other. Their momenta read
\begin{equation}
  \label{eq:p34-non-b2b}
  p_{I} = \frac{m}{\sqrt{1-\beta_I^2} }
  \bigl(1, \beta_I \sin \theta_{I} \cos \phi_I,\beta_I \sin \theta_I \sin \phi_I,\beta_I \cos \theta_I\bigr), \qquad I=3,4.
\end{equation}
The four-momenta of  additional massless jets are parametrized as follows 
\begin{equation}
  \label{eq:p56jetsDef}
  n_{i} = \bigl(1, \sin \theta_{i} \cos \phi_i,\sin \theta_i \sin \phi_i, \cos \theta_i\bigr), \qquad i=5,6, \dots
\end{equation}

Aside from showcasing the scope of our formulas, the results of this section are
also intended as benchmarks for validating their implementation, and for future
independent calculations of the $N$-jettiness soft function. For the dipole
contribution, we present both the full result and its separation into the
inclusive part and the finite correction
\be
\tilde W_{\gI \gJ}^{(2)}= \tilde W_{\gI \gJ}^{(2),{\rm inc}} +\Delta W_{\gI \gJ}^{\rm dip}.
\ee
\\
\noindent
{\bf For the case of $Q \bar Q$ production with an additional jet}, we
choose velocities of heavy quarks to be
\begin{equation}
  \beta_3 = 0.2,\qquad \beta_4 = 0.35.
\end{equation}
Directions of heavy quarks and the massless jet are parametrized using the following angles
\begin{align}
  &&
     \theta_3 = \frac{6\pi}{10},  && \phi_3 = \frac{3 \pi}{5},
  &&
     \theta_4 = \frac{9 \pi}{10}, && \phi_4 = \frac{6 \pi}{5},
  &&
     \theta_5 = \frac{3 \pi}{10}, && \phi_5 = 0.
  &&
\end{align}
Numerical results for dipole and tripole contributions for this kinematic point
are shown in Tables~\ref{tab:1jet-dip},~\ref{tab:1jet-trip-M00},~\ref{tab:1jet-trip-MM0}.
We restric the tables to contributions involving massive legs, as the the
calculation of the corresponding fully massless results is well established in
the literature.
\\

\begin{table}[!h]
  \centering
  \renewcommand{\arraystretch}{1.2}
  \begin{tabularx}{\textwidth}{|c|X|X|X||X|X|X|}
    \hline
    Dipoles &\multicolumn{3}{c||}{Gluons($C_A$ part)} &\multicolumn{3}{c|}{Quarks ($T_R n_f$ part)} \\ [0.5ex]
    \hline
    (M0) & \multicolumn{1}{c|}{$\tilde W_{I j}^{(2),{\rm inc}}$} & \multicolumn{1}{c|}{$\Delta W_{Ij}^{\rm dip}$} & \multicolumn{1}{c||}{$\tilde W_{Ij}^{(2)}$} & \multicolumn{1}{c|}{$\tilde W_{I j}^{(2),{\rm inc}}$} &  \multicolumn{1}{c|}{$\Delta W_{Ij}^{\rm dip}$} & \multicolumn{1}{c|}{$\tilde W_{Ij}^{(2)}$} \\
    \hline \hline
    \textbf{3}1 & 25.489(7)  & -5.559(8) & 19.930(9)   & -6.675(2) & -3.0046(6) & -9.680(2)  \\
    \hline
    \textbf{3}2 & 9.318(1) & -4.663(4)  & 4.654(4)  &  -5.4995(3) & -1.2937(2) &  -6.7932(4) \\
    \hline
    \textbf{3}5 & 29.283(7) & -5.673(9)  & 23.61(1)  &  -6.415(1) & -3.5242(6) & -9.939(1) \\
    \hline
    \textbf{4}1 & 30.2633(9) & -5.943(4)  &  24.321(4)  &   -8.356(2)  &  -3.6331(7) & -11.989(2)  \\
    \hline
    \textbf{4}2 & 3.2368(7) & -3.278(1)  &  -0.041(1)  & -3.5867(2) & -0.8166(1) &  -4.4033(2) \\
    \hline
    \textbf{4}5 & 31.712(8) & -5.669(9) & 26.04(1)   &  -7.555(2)  & -3.9638(7) & -11.519(2) \\
    \hline \hline
    (MM) & \multicolumn{1}{c|}{$\tilde{W}_{I J}^{(2),{\rm inc}}$} & \multicolumn{1}{c|}{$\Delta W_{IJ}^{\rm dip}$} & \multicolumn{1}{c||}{$\tilde W_{IJ}^{(2)}$} & \multicolumn{1}{c|}{$\tilde W_{I J}^{(2),{\rm inc}}$} &  \multicolumn{1}{c|}{$\Delta W_{IJ}^{\rm dip}$} & \multicolumn{1}{c|}{ $\tilde W_{IJ}^{(2)}$} \\
    \hline \hline
    \textbf{34} & 9.221(2) & -0.4331(1) &  8.788(2) & -0.2466(1) &  -0.1519(1) & -0.3986(1) \\
    \hline
  \end{tabularx}
  \caption{Dipole coefficients for the one-jettiness case.}
  \label{tab:1jet-dip}
\end{table}

\begin{table}[!h]
  \centering
  \renewcommand{\arraystretch}{1.2}
  \begin{tabular}{|c|l|l|l|}
    \hline
    $U_{Ijk}$ &\multicolumn{3}{c|}{$jk$} \\ [0.5ex]
    \hline
    $I$ & \multicolumn{1}{c|}{12} & \multicolumn{1}{c|}{15} & \multicolumn{1}{c|}{25}  \\
    \hline\hline
    \textbf{3} & -17.264(1)  & 0.2373(1)  & 1.562(2)
    \\
    \cline{1-4}
    \textbf{4} &  -38.748(5) & 0.1899(1)  & 1.2678(2)
    \\
    \hline
  \end{tabular}
  \caption{Massive-massless-massless  tripole coefficients for the one-jettiness case.}
  \label{tab:1jet-trip-M00}
\end{table}

\begin{table}[ht]
  \centering
  \renewcommand{\arraystretch}{1.2}
  \begin{tabular}{|c|l|l|l|}
    \hline
    $U_{IJk}$ &\multicolumn{3}{c|}{$k$} \\ [0.5ex]
    \hline
    $IJ$ & \multicolumn{1}{c|}{1} & \multicolumn{1}{c|}{2} & \multicolumn{1}{c|}{5}  \\
    \hline\hline
    \textbf{34} & 12.931(1)  & -13.920(1)  & 8.580(1)
    \\
    \hline
  \end{tabular}
  \caption{Massive-massive-massless tripole coefficients
    for the one-jettiness case.}
  \label{tab:1jet-trip-MM0}
\end{table}

\noindent
{\bf For two additional jets} we take the heavy-quark velocities to be
\begin{equation}
  \beta_3 = 0.25, \qquad \beta_4 = 0.45.
\end{equation}
Directions of heavy quarks and and massless jets are parametrized using the
following angles
\begin{align}
  &&
     \theta_3 = \frac{9 \pi}{10}, && \phi_3 = \frac{6 \pi}{5},
  &&
     \theta_4 = \frac{12 \pi}{10}, && \phi_4 = \frac{9 \pi}{5},
  &&
  \\
  &&
     \theta_5 = \frac{3 \pi}{10}, && \phi_5 = 0,
  &&
     \theta_6 = \frac{6\pi}{10},  && \phi_6 = \frac{3 \pi}{5}.
  &&
\end{align}
Numerical results for the dipole and tripole contributions for the two-jet case
are provided in Tables~\ref{tab:2jet-dip}, \ref{tab:2jet-trip-M00} and \ref{tab:2jet-trip-MM0}.

\begin{table}[!h]
  \centering
  \renewcommand{\arraystretch}{1.2}
  \begin{tabularx}{\textwidth}{|c|X|X|X||X|X|X|}
    \hline
    Dipoles &\multicolumn{3}{c||}{Gluons($C_A$ part)} &\multicolumn{3}{c|}{Quarks ($T_R n_f$ part)} \\ [0.5ex]
    \hline
    (M0) & \multicolumn{1}{c|}{$\tilde W_{I j}^{(2),{\rm inc}}$} & \multicolumn{1}{c|}{$\Delta W_{Ij}^{\rm dip}$} & \multicolumn{1}{c||}{$\tilde W_{Ij}^{(2)}$} & \multicolumn{1}{c|}{$\tilde W_{I j}^{(2),{\rm inc}}$} &  \multicolumn{1}{c|}{$\Delta W_{Ij}^{\rm dip}$} & \multicolumn{1}{c|}{$\tilde W_{Ij}^{(2)}$} \\
    \hline \hline
    \textbf{3}1 &  36.875(6) & -8.076(9)  & 28.80(1)   & -7.138(1) &  -4.6782(8) & -11.817(2)  \\
    \hline
    \textbf{3}2 & 15.439(2) & -6.605(3)  & 8.835(5)  &  -3.9967(5) & -2.2474(4) &  -6.2441(6) \\
    \hline
    \textbf{3}5 & 37.335(6) &  -8.332(8) & 29.003(9)  &  -6.693(1) & -4.7980(8) & -11.491(2) \\
    \hline
    \textbf{3}6  & 29.383(4) &  -8.745(9)  & 20.638(9)  & -4.9898(9) &  -3.7338(6) & -8.723(1) \\
    \hline
    \textbf{4}1 & 40.965(7)  &  -8.403(9) & 32.56(1)   & -8.041(1)    & -5.2895(9)  &  -13.331(2) \\
    \hline
    \textbf{4}2 & 12.617(2) & -6.505(4)  & 6.112(5)   & -3.3357(4) & -1.9966(3) &  -5.3323(5) \\
    \hline
    \textbf{4}5 & 43.973(9) &  -8.06(1) &  36.41(1)  &  -7.880(1)  & -5.748(1) & -13.628(2) \\
    \hline
    \textbf{4}6 & 19.964(3) & -8.198(6) &  11.766(7)  &  -3.5850(6) & -2.7683(4) &  -6.3533(6)\\
    \hline \hline
    (MM) & \multicolumn{1}{c|}{$\tilde W_{I J}^{(2),{\rm inc}}$} & \multicolumn{1}{c|}{$\Delta W_{IJ}^{\rm dip}$} & \multicolumn{1}{c||}{$\tilde W_{IJ}^{(2)}$} & \multicolumn{1}{c|}{$\tilde W_{I J}^{(2),{\rm inc}}$} &  \multicolumn{1}{c|}{$\Delta W_{IJ}^{\rm dip}$} & \multicolumn{1}{c|}{$\tilde W_{IJ}^{(2)}$} \\
    \hline \hline
    \textbf{34} & 11.228(1) & -0.4323(1) &  10.796(1) & -0.16252(8) &  -0.14659(2) & -0.30912(8) \\
    \hline
  \end{tabularx}
  \caption{Dipole coefficients for the two-jettiness case.}
  \label{tab:2jet-dip}
\end{table}

\begin{table}[ht]
  \centering
  \renewcommand{\arraystretch}{1.2}
  \begin{tabular}{|c|l|l|l|l|l|l|}
    \hline
    $U_{Ijk}$ &\multicolumn{6}{c|}{$jk$} \\ [0.5ex]
    \hline
    $I$ & \multicolumn{1}{c|}{12} & \multicolumn{1}{c|}{15} & \multicolumn{1}{c|}{16} & \multicolumn{1}{c|}{25} & \multicolumn{1}{c|}{26} & \multicolumn{1}{c|}{56} \\
    \hline\hline
    \textbf{3} & -30.228(8)  & 0.103(2)  & 0.625(3) & 1.217(3)  & 0.620(1)  & 0.746(2)\\
    \hline
    \textbf{4} & -42.051(7)  & 0.016(2)  & 0.378(4) & 1.643(4)  & 0.357(1)  & 0.299(3)\\
    \hline
  \end{tabular}
  \caption{Massive-massless-massless tripole coefficients for the two-jettiness case.}
  \label{tab:2jet-trip-M00}
\end{table}

\begin{table}[ht]
  \centering
  \renewcommand{\arraystretch}{1.2}
  \begin{tabular}{|c|l|l|l|l|}
    \hline
    $U_{IJk}$ &\multicolumn{4}{c|}{$k$} \\ [0.5ex]
    \hline
    $IJ$ & \multicolumn{1}{c|}{1} & \multicolumn{1}{c|}{2} & \multicolumn{1}{c|}{5} & \multicolumn{1}{c|}{6} \\
    \hline\hline
    \textbf{34} & 15.743(1)  & -6.397(1)  & 23.475(1)
                                                           & -22.533(1) \\
    \hline
  \end{tabular}
  \caption{Massive-massive-massless  tripole coefficients for the two-jettiness case.}
  \label{tab:2jet-trip-MM0}
\end{table}

\section{Conclusions}
\label{sec:conclusion}

We have presented the computation of the NNLO $N$-jettiness soft function for
the production of a heavy quark pair and an \emph{arbitrary} number of jets in
hadron collisions. With this result, all process-independent ingredients needed
for studying \emph{arbitrary} processes at colliders with up to two final-state
heavy quarks with the NNLO QCD accuracy using the $N$-jettiness slicing scheme
become available.

Our computation employs the recently proposed method \cite{Buonocore:2026yai}
that utilizes the so-called \emph{inclusive} $N$-jettiness variable to simplify
the calculation of the $N$-jettiness soft function. Using this method, we were
able to remove all divergencies of the soft function analytically, and express
the $N$-jettiness soft function through integrals that do not require an
infrared regulator.

The NNLO soft function that we studied in this paper contains quadrupole, dipole
and tripole color structures. Each of these structures receives contributions
from hard partons with different masses. The quadrupole contribution is always
given by an iteration of the (finite) NLO result, but dipole and tripole terms
receive genuine NNLO corrections.

We have shown that for the case with just one massive parton, both dipole and
tripole contributions can be written in a remarkably simple form. The case of
two massive partons is more involved and the complexity of intermediate
expressions for the finite remainders suggests that computing them numerically
is a preferable strategy. Since for a generic process with $N$ final-state jets,
one has to perform a complicated numerical integration of
$N$-jettiness-dependent contributions, additional numerical integrations do not
constitute a significant problem.

Recently, the calculation of the NNLO zero-jettiness soft function for top-quark
pair production in hadron collisions was reported in Ref.~\cite{Bell:2026pou}.
This calculation uses the method of Ref.~\cite{Bell:2018mkk} and relies on
numerical calculations, including the cancellation of the $1/\ep$ poles. We have
compared the results reported in Ref.~\cite{Bell:2026pou} with our results for
the zero-jettiness case and found excellent agreement for both dipole and
tripole contributions.

Finally, we have presented numerical results for the one- and two-jettiness soft
functions for processes when a pair of heavy quarks and the corresponding number
of jets are produced in hadron collisions. We have also provided analytic
formulas and ancillary files that should make coding up the results for the
$N$-jettiness soft function with heavy quarks and arbitrary number of jets a
straightforward endeavor.

\section*{Acknowledgments}

PA and KM are grateful to the Erwin-Schr\"odinger International Institute for
Mathematics and Physics at the University of Vienna for the partial support that
enabled their participation in the Program “New Paradigms for Harnessing Quantum
Field Theory at Colliders”, July 27 - August 28, 2026, during which parts of
this paper were completed. PA, KM and IP are partially supported by the Deutsche
Forschungsgemeinschaft (DFG, German Research Foundation) under grant 396021762 -
TRR 257. The work of AP is supported by the ERC Consolidator Grant 101169614.

\appendix
\section{Renormalization}
\label{sec:app-ren}
As discussed in Section~\ref{sec:definitions}, $Z$ is the renormalization matrix in color space. It satisfies the renormalization group equation
\begin{equation}
  \frac{d}{d \ln \mu} Z = -\frac{1}{2} Z \; \Gamma_S.
  \label{eq2.2}
\end{equation}
The anomalous dimension $\Gamma_S$ reads \cite{Becher:2009kw,Ferroglia:2009ep}
\begin{equation}
  \Gamma_S = \Gamma_S^{\rm dip} + \Gamma_S^{\rm trip},
\end{equation}
where $\Gamma_S^{\rm dip, trip}$ are the contributions of color-dipoles and
color-tripoles, respectively. We note that $\Gamma_S^{\rm trip}$ starts at
${\cal O}(\alpha_s^2)$ and is different from zero only if there are at least two
massive quarks in the process of interest. The dipole contribution reads
\begin{equation}
  \begin{split}
    \Gamma^{\rm dip}_S &= \sum_{(i j)} \frac{{\bf T}_i \cdot {\bf T}_j}{2} \; \Gamma_{\rm cusp}(a_s) \left[ L_{ij} + i \pi \lambda_{ij} \right] + \sum_{(i j)} {\bf T}_i \cdot {\bf T}_j \; \gamma_s(a_s) 
    \\
                       & + \sum_{(I J)} \frac{{\bf T}_I \cdot {\bf T}_J}{2} \; \Gamma_{\rm cusp}(a_s, v_{IJ})  + \sum_{(I J)} {\bf T}_I \cdot {\bf T}_J \; \gamma_S(a_s)
    \\
                       & + \sum_{I,j} {\bf T}_I \cdot {\bf T}_j \; \Gamma_{\rm cusp}(a_s) \left[ \frac{L_{Ij}}{2} + i \pi \lambda_{Ij} \right]
                         + \sum_{I,j} {\bf T}_I \cdot {\bf T}_j \left[ \gamma_S(a_s) + \gamma_s(a_s) \right].
                         \label{eq2.3}
  \end{split}
\end{equation}

\noindent We remind the reader that we use gothic letters $\gI,\gJ,$ etc. to
denote a generic color-charged particle, capital Latin letters $I,J$ to denote
massive particles, and lower-case Latin letters $i,j$ to refer to the massless
ones. In the sums we use the notation $(\gI \gJ)$ to indicate that they run over
all possible values of $\gI,\gJ$ with the constraint $\gI \ne \gJ$. The
quantities $\lambda_{\gI\gJ}$ are one if both particles $\gI \gJ$ are either in
the initial or in the final state, and zero otherwise. The quantities
$L_{ij}, L_{Ij}$ and $v_{IJ}$ that appear in the above equation are defined as
follows
\begin{equation}
  \begin{split}
    L_{ij} &= \ln \left( \frac{\bar u^2 \mu^2 \; p_i p_j}{2Q_i Q_j} \right), \quad
             L_{Ij} = \ln \left( \frac{\bar u^2 \; \mu^2 \; (p_I p_j)^2}{m_I^2\; Q_j^2} \right), \quad v_{IJ}= \sqrt{1- \frac{m_I^2 m_J^2}{(p_Ip_J)^2}}.
  \end{split}
  \label{eqA.4}
\end{equation}
Eq.~\eqref{eq2.3} contains various anomalous dimensions. They are parametrized
in the following way
\begin{equation}
  \begin{split}
    & \Gamma_{\rm cusp} = \sum_{n=0} \left( \frac{a_s}{2} \right)^{n+1} \; \Gamma_n,
      \;\;\;
      \Gamma_{\rm cusp}(a_s, v) = \sum_{n=0} \left( \frac{a_s}{2} \right)^{n+1} \; \Gamma_n (v),
    \\
    & \gamma_{s}(a_s) = \sum_{n=0} \left(\frac{a_s}{2}\right)^{n+1} \gamma_s^n,
      \;\;\;
      \gamma_{S}(a_s) = \sum_{n=0} \left(\frac{a_s}{2}\right)^{n+1} \gamma_S^n.
  \end{split}
\end{equation}
We will require the anomalous dimensions through ${\cal O}(a_s^2)$. The
coefficients for the cusp anomalous dimension read
\begin{equation}
  \begin{split}
    & \Gamma_0 = 4, \;\;\; \Gamma_1 =
      \left( \frac{268}{9} - \frac{4 \pi^2}{3} \right) C_A - \frac{80}{9} T_F n_f,
      \;\;\;
      \Gamma_0(v_{IJ}) =  \frac{\Gamma_0}{2 v_{IJ}} \ln \bigg[ \frac{1+v_{IJ}}{1-v_{IJ}}\bigg] + i \frac{\pi \Gamma_0 \lambda_{IJ}}{v_{IJ}},
      \;\;
    \\
    & \Gamma_1(v_{IJ}) =  \frac{\Gamma_1}{2 v_{IJ}} \ln \bigg[ \frac{1+v_{IJ}}{1-v_{IJ}}\bigg] + i \frac{\pi \Gamma_1 \lambda_{IJ}}{v_{IJ}}
      + 8 C_A \; g_{IJ}(v_{IJ}).
  \end{split}
\end{equation}
The various functions that appear in the above formula are 
\begin{equation}
  \begin{split}
    g_{IJ}(v) & =  \bigg\{ (1-v)\Bigg[-\frac{1}{24 v^2}\ln^3 ( x) - \frac{1}{4 v}\ln^2(x) - \left(\frac{1}{6v} + \frac{1}{12v^2}\ln x\right)\pi^2 \Bigg] + \left(1-\frac{1}{v^2}\right)\zeta_3  \\
              & \quad + \ln x \left(\frac{1}{v}\ln(1-x )  -\frac{1}{2v^2} {\rm Li}_2 (x) \right ) + \frac{1}{v} {\rm Li}_2(x) +  \frac{1}{v^2} {\rm Li}_3 (x)   \bigg\} \\
              & - \frac {\lambda_{IJ} \; \pi^2(1-v)}{2 v^2} \left( 2 v + \ln x \right) + i \pi \lambda_{IJ}\; g^{\operatorname{Im}}(v),
  \end{split}
\end{equation}
where we defined
\begin{equation}
  x = \frac{1-v}{1+v},
\end{equation}
and
\begin{equation}
  g^{\operatorname{Im}}(v) = - \frac{1}{v^2} \left[ \frac{\pi^2}{6} (1-v)+ 2 v \ln\left(1 - x \right)-\frac{1-v}{4} \ln^2 (x) + v (1+v) \ln (x) - {\rm Li}_2(x)\right].
\end{equation}
The soft anomalous dimension can be expressed as follows 
\begin{equation}
  2\gamma_s = - 2\gamma_H - \gamma_{B_1} - \gamma_{B_2}
\end{equation}
where $\gamma_H$ and $\gamma_{B}$ are the anomalous dimensions for the hard and
the beam functions. For the massless particles, they are known up to two loops
\cite{Stewart:2010qs}. We find
\begin{equation}
  \gamma_s^{(0)} = 0,
  \;\;\;\; \gamma_s^{(1)} =  \left( \frac{404}{27} - \frac{11 \pi^2}{18} - 14 \zeta_3 \right) C_A + \left( - \frac{112}{27} + \frac{2 \pi^2}{9} \right) T_F n_f,
\end{equation}
In the case of massive particles, $\gamma_B =0$ and $\gamma_H$ can be obtained
using the anomalous dimension of heavy-quark currents in HQET at two loops
\cite{Becher:2009kw}. Using these results, we obtain
\begin{equation}
  \gamma_S^{(0)} = -2,
  \;\;\;
  \gamma_S{(1)} = \left( -\frac{98}{9} + \frac{2 \pi^2}{3} - 4 \zeta_3 \right) C_A + \left( \frac{40}{9} \right) T_F n_f.
\end{equation}

We now turn to the tripole contribution. The anomalous dimension in this case can be fully retrieved from the results of Refs.~\cite{Becher:2009kw,Ferroglia:2009ep}
\begin{equation}
  \begin{split}
    \Gamma_S^{\rm trip} =
    &  \sum_{(I,J)} \sum_k i f^{a b c} {\bf T}^a_I {\bf T}^b_J {\bf T}^c_k \; f^*_2 \left(v_{IJ},\; \ln \left(\frac{m_I p_J p_k}{m_J p_I p_k} \right) \right) \\
    & + \sum_{(I,J,K)} i f^{a b c} {\bf T}^a_I {\bf T}^b_J {\bf T}^c_K \; f^*_1 \left(v_{IJ},v_{IK},v_{JK} \right).
  \end{split}
  \label{eq_Gammatri}
\end{equation}
For processes with up to  two heavy quarks, only the first term in Eq.~\eqref{eq_Gammatri} is needed. The function $f_2$ reads 
\begin{equation}
  f_2 \left(v_{IJ},\; \ln \left(\frac{m_I p_J p_k}{m_J p_I p_k} \right) \right) = - a_s^2 \; g^{\rm tri}(v_{IJ}) \; \ln \left(\frac{m_I p_J p_k}{m_J p_I p_k} \right),
\end{equation}
with
\begin{equation}
  \begin{split}
    g^{\rm tri}(v) =
    & \frac{1}{v} \bigg[ \frac{1-v}{4} \ln^2(x) - \ln (1-x) \ln (x) - {\rm Li}_2 (x)  +\frac{\pi^2}{6} (1-v) - \lambda_{IJ} \pi^2 (1-v) \\
    & - i \pi \lambda_{IJ}\left[ (1+v)\ln \left(x \right) + 2\ln \left( 1/x - 1\right) \right] \bigg].
  \end{split}
\end{equation}
Since the renormalization of the soft function involves the combination
$Z_{2,\rm tri} + Z_{2, \rm tri}^+$, only the real part of
Eq.~\eqref{eq_Gammatri} is relevant. We find
\begin{equation}
  Z_{2,\rm tri}^{\rm Re} = \sum_{(I,J),k} \;
  H_{IJk} \frac{\pi}{4 v_{IJ} \ep} \left[ (1+v_{IJ}) \ln \left( \frac{1-v_{IJ}}{1+v_{IJ}} \right) + 2 \ln \left( \frac{2v_{IJ}}{1-v_{IJ}} \right) \right]\ln \left(\frac{m_I p_J p_k}{m_J p_I p_k} \right),
\end{equation}
where the tripole color factor $H_{\gI \gJ \gK}$ is defined in
Eq.~\eqref{eq:tripHTTT-def}.
\\

Finally, solving the renormalization group equation, we write 
\begin{equation}
  Z = 1 + Z_1 + \left ( Z_{2,\rm dip} + \frac{1}{2}
    Z_1^2 \right ) + Z^{\rm Re}_{2,\rm tri}  + {\cal O}(a_s^3),
\end{equation}
where
\begin{equation}
  \begin{split}
    Z_1 & =a_s \sum_{(\gI \gJ) } {\bf T}_\gI \cdot {\bf T}_\gJ  \; G^{(1)}_{\gI \gJ},
          \;\;\;\;  Z_{2, \rm dip}  = a_s^2 \sum_{(\gI \gJ) } {\bf T}_\gI \cdot {\bf T}_\gJ \; G^{(2)}_{\gI \gJ},
  \end{split}
  \label{eq2.10}
\end{equation}
with
\begin{equation}
  \begin{split}
    G^{(1)}_{ij} &= \frac{1}{2 \ep^2} + \frac{ L_{ij}}{2 \ep} + i \left( \frac{\pi \lambda_{ij}}{2 \ep}\right), \\
    G^{(1)}_{IJ} &= \frac{ \Gamma^{\operatorname{Re}}_0 (v_{IJ}) -4}{8 \ep} + i \left( \frac{\pi \lambda_{IJ}}{ 2\ep v_{IJ}}\right), \\
    G^{(1)}_{Ij} &= \frac{1}{4 \ep^2} + \frac{ L_{Ij}-1}{4 \ep} + i \left( \frac{\pi \lambda_{Ij} }{ 2\ep}\right),
  \end{split}
  \label{eqA11}
\end{equation}
and
\begin{equation}
  \begin{split}
    \\
    G^{(2)}_{ij} &= -\frac{3 \beta_0}{8 \ep^3} + \frac{\Gamma_1-16 \beta_0 L_{ij}}{64 \ep^2} + \frac{\Gamma_1 L_{ij}+2 \gamma^1_s}{32 \ep} + i \pi \lambda_{ij}\left( -\frac{\beta_0 }{4 \ep^2} + \frac{ \Gamma_1}{32 \ep} \right),
    \\
    G^{(2)}_{IJ} &= -\frac{\beta_0 ( \Gamma^{\operatorname{Re}}_0 (v_{IJ}) + 2\gamma_S^0)}{16 \ep^2} + \frac{\Gamma_1^{\operatorname{Re}} (v_{IJ})+2 \gamma^1_S}{32 \ep} 
    \\ & + i \pi \lambda_{IJ} \left( \frac{-\beta_0 }{4 \ep^2 v_{IJ}} + \frac{\Gamma_1}{32 \ep v_{IJ}} + \frac{C_A \;g^{\operatorname{Im}}(v_{IJ})}{4 \ep} \right), \\
    G^{(2)}_{Ij} &= -\frac{3 \beta_0}{16 \ep^3} + \frac{\Gamma_1-16 \beta_0 ( L_{Ij}-1)}{128 \ep^2} + \frac{\Gamma_1 L_{Ij}+2\gamma_s^1 + 2\gamma_S^1 }{64 \ep} + i \pi \lambda_{Ij} \left( -\frac{\beta_0 }{4\ep^2} + \frac{ \Gamma_1}{32 \ep} \right).
  \end{split}
\end{equation}
In the above equations,
$\beta_0 = \frac{11}{6} \; C_A - \frac{2}{3} \; T_F n_f $, and the superscript
``$\operatorname{Re}$'' indicates that only the real part of the coefficients of
the massive cusp anomalous dimension should be taken. We note that throughout
this paper we assume that the two heavy quarks are in the final state, whereas
massless partons can be either in the initial or in the final state.

\section{Kinematic relations for the massive-massive case}
\label{sec:app-kin}

In this appendix, we present useful kinematic relations for the massive-massive
case. We consider partons $I,J$ with masses $m_{I,J}$ and momenta $p_{I,J}$.
These momenta are written as linear combinations of the corresponding light-cone
momenta $\tilde p_{I,J}$
\begin{equation}
  \begin{split}
    & p_I = \tilde p_I + \kappa_{IJ} \tilde p_J, \;\;\;\; p_J= \tilde p_J +
      \kappa_{JI} \tilde p_I,
  \end{split}
\end{equation}
where
\begin{equation}
  \kappa_{IJ} = \frac{m_I^2}{2 \tilde p_I \tilde p_J}.
\end{equation}
The scalar product of $p_{I}$ and $p_J$ evaluates to
\begin{equation}
  p_I p_J = \tilde p_I \tilde p_J (1 + \kappa_{IJ} \kappa_{JI}  ).
\end{equation}
We can solve this equation for $\tilde p_I \tilde p_J$, to find
\begin{equation}
  \tilde p_I \tilde p_J  p_I p_J \frac{(1+v_{IJ})}{2},
\end{equation}
where $v_{IJ}$ is defined in Eq.~\eqref{eqA.4}. Furthermore, it is easy to show
that the following equation holds
\begin{equation}
  \kappa_{IJ} \kappa_{JI} = \frac{1-v_{IJ}}{1+v_{IJ}}.
\end{equation}
Finally, momenta $\tilde p_{I,J}$ can be written as linear combination of the
original momenta $p_{I,J}$ using the following formulas
\begin{equation}
  \begin{split}
    & \tilde p_I = \frac{1+v_{IJ}}{2 v_{IJ}} \left (
      p_I - \kappa_{IJ} p_J
      \right ),
      \;\;\;\; \tilde p_J = \frac{1+v_{IJ}}{2 v_{IJ}} \left (
      p_J - \kappa_{JI} p_I
      \right ).
  \end{split}
\end{equation}

\section{Phase space angular integrals}
\label{sec:app-angular}
In this appendix, we define angular integrals that are used for calculations in
Section~\ref{sec:nnlo-dipole}. These integrals contain either massless propagators, a
massive and a massless propagator or two massive propagators. We denote these
propagators as
\begin{equation}
  \rho_{xy} = 1 - \vec n_x \cdot \vec n_y,
\end{equation}
with $\vec n_x^2 = 1$ for the massless and $\vec n_x^2 \ne 1$ for the massive
case. We use the notation $\langle \dots \rangle_x$ to indicate that integration
over directions of $\vec n_x$ in $(d-1)$ dimensional space is performed.
Finally, we remark that partons $m$ and $n$ are always massless.

We normalize the angular integrals in such a way that integrals without propagators are equal to one,
\begin{equation}
  \langle 1 \rangle_{\rm m} = \langle 1 \rangle_{\rm mn}=  1.
\end{equation}
To present the results for angular integrals, we start with the massless ones.
They can be expressed in terms of Gamma- and hypergeometric functions. For
$\rho_{xx}=\rho_{yy} =0$, we have,
\begin{equation}
  \bigg \langle \frac{1}{\rho_{mx}^\alpha} \bigg \rangle_m= \frac{\Gamma(1-\alpha-\ep) \Gamma(2-2\ep)}{2^\alpha \Gamma(1-\ep) \Gamma(2-\alpha -2\ep)},
  \label{app2}
\end{equation}
\begin{equation}
  \bigg \langle \frac{1}{\rho_{mx}^\alpha \rho_{my}^\beta } \bigg \rangle_m = \frac{\Gamma(2-2\ep)\Gamma(1-\ep-\alpha)\Gamma(1-\ep-\beta)}{2^{\alpha+\beta} \Gamma^2(1-\ep)\Gamma(2-\alpha-\beta-2\ep)} \;{}_2 F_1 \left(\alpha, \beta, 1-\epsilon, 1- \frac{\rho_{xy}}{2}\right).
  \label{app3}
\end{equation}
Angular integrals used in the calculations in Section~\ref{sec:nnlo-dipole} are defined as follows,
\begin{equation}
  I_\alpha^{(1)}[\rho_{xx}] =  \bigg \langle \frac{1}{\rho_{mx}^\alpha} \bigg \rangle_m \; \; ; \; \rho_{xx}\neq 0 ,
\end{equation}
\begin{equation}
  I_{\alpha,\beta}^{(1)}[\rho_{xy},\rho_{xx}, 0] =  \bigg \langle \frac{1}{\rho_{mx}^\alpha \rho_{my}^\beta } \bigg \rangle_m \; \; ; \; \rho_{xx}\neq 0 , \rho_{yy}= 0 ,
\end{equation}
\begin{equation}
  I_{\alpha,\beta}^{(2)}[\rho_{xy},\rho_{xx}, \rho_{yy}] =  \bigg \langle \frac{1}{\rho_{mx}^\alpha \rho_{my}^\beta } \bigg \rangle_m \; \; ; \; \rho_{xx}\neq 0 , \rho_{yy} \neq 0 .
\end{equation}
We note that, for generic indices $\alpha$ and $\beta$, some of these massive
angular integrals become quite complicated. However, for our purposes, integrals
with $\alpha=1$ and $\beta=1$ are relevant. Such integrals can be computed in a
closed form using the hypergeometric and Appell functions. We find
\begin{equation}
  I_1^{(1)}[\rho_{aa}] =  \frac{1}{1+ \sqrt{1-\rho_{aa}}} \;{}_2 F_1 \left(1, 1-\epsilon,2-2\ep; \frac{2 \sqrt{1-\rho_{aa}}}{1+ \sqrt{1-\rho_{aa}}}\right),
\end{equation}
\begin{equation}
\begin{split}
  I_{1,1}^{(1)}[\rho_{ab}, \rho_{aa},0 ] &=  \frac{2\ep-1}{2 \rho_{ab} \ep} \; F_1 \left(1,-\ep,-\ep,1-2\ep;1- \frac{1+\sqrt{1-\rho_{aa}}}{\rho_{ab}},1- \frac{1-\sqrt{1-\rho_{aa}}}{\rho_{ab}}\right).
\end{split}
\end{equation}

The double massive integral $I_{1,1}^{(2)}$ can be expressed in terms of $I_{1,1}^{(1)}$ using the following relation,
\begin{equation}
  I_{1,1}^{(2)}[\rho_{ab},\rho_{aa}, \rho_{bb}] = \frac{1}{\sqrt\chi} \left( \rho_{ak}\; I_{1,1}^{(1)}[\rho_{ak}, \rho_{aa},0 ] - \rho_{bk}\; I_{1,1}^{(1)}[\rho_{bk}, \rho_{bb},0 ] \; \right),
\end{equation}
where $\chi = \rho_{ab}^2 - \rho_{aa}\rho_{bb}$ and
\begin{equation}
  \rho_{ak} = \frac{\rho_{aa}\;(\rho_{bb} + \sqrt\chi) - \rho_{ab}\;(\rho_{ab}+ \sqrt\chi)}{\rho_{aa}+\rho_{bb}- 2\rho_{ab}},
\end{equation}
\begin{equation}
  \rho_{bk} = \frac{\rho_{bb}\;(\rho_{aa} - \sqrt\chi) - \rho_{ab}\;(\rho_{ab}- \sqrt\chi)}{\rho_{aa}+\rho_{bb}- 2\rho_{ab}}.
\end{equation}

\section{Simplifications of phase-space integrals for massive emitters}
\label{sec:app-ps-ints-simp}

As discussed in Section~\ref{sec:nnlo-dipole}, to determine the dipole contribution to the
$N$-jettiness soft function, we need to compute integrals of the following type
\begin{equation}
  \begin{split}
    & W_1  = \int \limits_{0}^{1}
      {\rm d} \xi \; \xi^{n \ep -1}
      \int \limits_{0}^{1} {\rm d}x \; x^{-\ep -1} (1-x)^{-\ep } f(\xi,x), \\
    & W_2(\xi)  =
      \int \limits_{0}^{1} {\rm d}x \; x^{-\ep -1} (1-x)^{-\ep } f(\xi,x),
  \end{split}
  \label{eqd.1}
\end{equation}
where $n$ is an integer. The function $W_2(\xi)$ is only needed for $\xi \ne 0$
and $f(\xi,x)$ can be either $\tilde F_{Ij}(1,\xi,x)$ or
$\tilde F_{Ij}(\xi,1,x)$, c.f. Eq.~\eqref{eq:FIJ-to-tld}. We also note that $\xi = 0$
and $x = 0$ are the only singular points of the integrands and that the function
$f(\xi,x)$ has at most logarithmic singularities at these points.

Hence, to compute $W_1$ and $W_2(\xi)$, we need to extract the asymptotic
behavior of the function $f(\xi,x)$ at $\xi \sim 0$ and $x \sim 0$ at fixed
$\ep$, and then use them to construct integrable quantities in
Eq.~\eqref{eqd.1}. To construct these asymptotics, we introduce an operator
$C_v$ which extracts the leading behavior of the function $f(\xi,x)$ in the
limit $ v \sim 0$. Then, to obtain a suitable integrand for the $W_1$, we write
the following identity
\begin{equation}
  f(\xi,x)
  = (I - C_\xi) (I - C_x)f(\xi,x)
  + C_\xi (I - C_x)f(\xi,x) + C_x (I - C_\xi)f(\xi,x) + C_x C_\xi  f (\xi,x),
  \label{eqc.2}
\end{equation}
whereas to compute the function $W_2(\xi)$, we only need
\begin{equation}
  f(\xi,x) = (I-C_x) f(\xi,x)
  +C_x f(\xi,x).
  \label{eqc.3}
\end{equation}
In the above expressions, when the operator $C_{v}$ acts on the function $f$,
the dependence of the integrand on the variable $v$ factorizes and integration
over $v$ at fixed $\ep$ can be performed. At the same time, whenever $(I- C_v)$
acts on $f$ or its remnant, the limit $v \to 0$ is not singular anymore and the
expansion of the integrand in $\ep$ can be performed.
\\

We will now discuss how to compute the relevant limits of the two functions
$\tilde F_{Ij}(1,\xi,x)$ or $\tilde F_{Ij}(\xi,1,x)$. We begin by noticing that
the extraction of the $\xi \to 0$ limit is only relevant for the function
$\tilde F_{Ij}(\xi,1,x)$ since the function $\tilde F_{Ij}(1,\xi,x)$ vanishes at
$\xi =0$. We will, therefore, discuss the function $\tilde F_{Ij}(\xi,1,x)$ but
we emphasize that the analysis of the $x \to 0$ limit for the two functions is
very similar.

To understand how to extract the $\xi \to 0$ and $x \to 0$ limits of the
function $\tilde F_{Ij}(\xi,1,x)$ defined in Eq.~\eqref{eq:tld-FIj-M0}, we note that the
many rational functions appearing there are \emph{smooth} functions in both
limits $ \xi \to 0$, $x\to 0$. One can verify this statement immediately using
the following equations
\begin{equation}
  q_I = p_{Ij}(1 + \xi \kappa_{Ij}),\;\;\;
  q_j = p_{Ij} \xi,
  \;\;\; q^2 = 2 p_{Ij} \xi x.
\end{equation}
Some care is only needed in the $q^2/q_j$ coefficient of the function
$I_{1,1}^{(1)}\left[\frac{p_{Ij} q^2}{q_I q_j}, \frac{m_I^2 q^2}{q_I^2}, 0\right]$
because this function itself is proportional to $q_j/q^2$, and the two factors
cancel each other exactly.

Before discussing the functions $I_{1,1}^{(1)}$, we consider the function
$I^{(1)}(m_I^2 q^2/q_I^2)$. For the case $\beta = 1$, $\alpha =\xi$, its
argument evaluates to
\begin{equation}
  \frac{m_I^2 q^2}{q_I^2}
  = \frac{4 \kappa_{Ij} \xi x}{( 1+
    \kappa_{Ij} \xi )^2}.
  \label{eqd.5}
\end{equation}
Thus,  the function $I_{1}^{(1)}$ reads
\begin{equation}
  I_{1}^{(1)}\left (
    \frac{m_I^2 q^2}{q_I^2}
  \right )
  = \frac{1}{1+\eta} \;
  F_{21}\left (1,1-\ep,2-2\ep,\frac{2 \eta}{1+\eta}
  \right ),
  \label{eqc.6a}
\end{equation}
where
\begin{equation}
  \eta = \sqrt{1 - \frac{4 \kappa_{Ij} x \xi }{(1+\kappa_{Ij} \xi )^2 }}.
\end{equation}
Since for $x = 0$ and/or $\xi = 0$, $\eta = 1$, the argument of the
hypergeometric function in Eq.~\eqref{eqc.6a} becomes 1. However, this is a
singular point of the hypergeometric function in Eq.~\eqref{eqc.6a}, and we need
to extract the asymptotic more carefully. This is easy to do using standard
identities that the hypergeometric functions satisfy
\cite{abramowitz1965handbook}. In this case, both $x \to 0$ and $\xi \to 0$
limits are described by the following asymptotic
\begin{equation}
  I_1^{(1)}
  \left ( \frac{ m_I^2 q^2}{q_I^2}
  \right )
  \sim   \frac{2\ep-1}{2\ep}
  \left (
    1 - \frac{2^{2\ep} \Gamma(1-2\ep)\Gamma(1+\ep)}{\Gamma(1-\ep)} \left ( \frac{m_I^2q^2}{q_I^2} \right )^{-\ep}
  \right ),\;\;\;{\rm at}\;\; q^2 \sim 0,
\end{equation}
where $m_I^2 q^2/q_I^2$ is given in Eq.~\eqref{eqd.5} for the $x \to 0$ limit or
equals to $4 \kappa_{Ij} \xi x$ for the $\xi \to 0$ limit.
\\

We continue with the investigation of the functions $I_{1,1}^{(1)}$ that appear
in Eq.~\eqref{eq:tld-FIj-M0}. In general, $I_{1,1}^{(1)}$ can be written in terms of an
Appell function
\begin{equation}
  I_{1,1}^{(1)}[\rho_{ab}, \rho_{aa},0 ] =  \frac{2\ep-1}{2 \rho_{ab} \ep} \; F_1 \left(1,-\ep,-\ep,1-2\ep;1- \frac{1+\sqrt{1-\rho_{aa}}}{\rho_{ab}},1- \frac{1-\sqrt{1-\rho_{aa}}}{\rho_{ab}}\right).
\end{equation}
To construct the expansion of the Appell function at small $x$ and/or $\xi$, we
employ the following integral representation
\begin{equation}
  \begin{split}
    & F_1 \left(1,-\ep,-\ep,1-2\ep;y_1,y_2 \right)
      = \frac{ \Gamma(1-2\ep)}{\Gamma(-2\ep)}
      \int \limits_{0}^{1}
      {\rm d} u \; (1-u)^{-2\ep-1} (1- u y_1)^\ep (1-u y_2)^{\ep}
    \\
    &
      = \frac{ \Gamma(1-2\ep)}{\Gamma(-2\ep)}
      (1-y_1)^{\ep} (1-y_2)^{\ep}
      \int \limits_{0}^{1}
      {\rm d} u \; u^{-2\ep-1} \left (1 +\frac{uy_1 }{1-y_1}  \right )^\ep \left (1+ \frac{uy_2}{1-y_2} \right )^{\ep} ,
  \end{split}
  \label{eqc.5}
\end{equation}
where
\begin{equation}
  y_1 = 1- \frac{1+\sqrt{1-\rho_{aa}}}{\rho_{ab}},
  \;\;\;\;
  y_2 = 1- \frac{1-\sqrt{1-\rho_{aa}}}{\rho_{ab}}.
  \label{eqc.6}
\end{equation}
For the generic values of $\rho_{ab}$, $\rho_{aa}$, it is straightforward to
derive the $\ep$-expansion using the above representation. For example, working
to first order in $\ep$, we obtain
\begin{equation}
  \begin{split}
    I_{1,1}^{(1)}(\rho_{ab},\rho_{aa},0)
    =& -\frac{1}{2\rho_{ab}}
       \Bigg \{ \frac{1}{\ep}
       - 2 +\ln \frac{\rho_{aa}}{\rho_{ab}^2}  + 2 \ep \bigg[ \left( \frac{1}{4} \ln \frac{\rho_{aa}}{\rho_{ab}^2} - 1\right) \ln \frac{\rho_{aa}}{\rho_{ab}^2} \\
     & + {\rm Li}_2 \left(1- \frac{ \rho_{ab}}{1+\sqrt{1-\rho_{aa}}} \right) + {\rm Li}_2 \left(1- \frac{ \rho_{ab}}{1-\sqrt{1-\rho_{aa}}} \right) \bigg] \Bigg\} + {\cal O}(\ep^2).
  \end{split}
\end{equation}

We continue with the discussion of the $x \to 0$ and $\xi \to 0$ limits of the
function $I^{(1)}_{1,1}(p_{Ij} q^2/q_I q_j, m_I^2 q^2/q_I^2,0)$ which, according
to Eq.~\eqref{eq:tld-FIj-M0} is multiplied by $q^2/q_j$. Since
\begin{equation}
  \rho_{ab} = \frac{p_{Ij} q^2}{q_I q_j} = \frac{2 x}{1+\kappa_{Ij} \xi},
  \;\;\;\; \rho_{aa} = \frac{4 \kappa_{Ij} \xi x}{(1+\kappa_{Ij} \xi)^2},
\end{equation}
we require this function in the following limits
\begin{equation}
  \begin{split}
    & x \to 0: \;\;\; \rho_{ab} \to 0, \;\;\;
      \rho_{aa} \to 0,  \;\;\;\frac{\rho_{ab}}{\rho_{aa}}=
      \frac{1+\kappa_{Ij}\xi}{2 \kappa_{Ij} \xi},
    \\
    & \xi \to 0 : \;\;\; \rho_{ab} = 2x ,  \;\; \rho_{aa} \to 0.
  \end{split}
\end{equation}

In case $x \to 0$, the quantities $y_{1,2}$, defined in Eq.~\eqref{eqc.6},
evaluate to
\begin{equation}
  y_1 \to - \frac{1+\kappa_{Ij} \xi}{x},
  \;\;\; y_2 \to \frac{1}{1+\kappa_{Ij} \xi}.
\end{equation}
Then, the following equations  hold
\begin{equation}
  x \to 0: \;\;\; \frac{y_1}{1-y_1}  \to -1,
  \;\;\;\;
  \frac{y_2}{1-y_2} \to  \frac{1}{\kappa_{Ij} \xi}.
\end{equation}
We use the above limits in Eq.~\eqref{eqc.5} and immediately find that at
$x \sim 0$
\begin{equation}
  \begin{split}
    \frac{q^2}{q_j }I_{1,1}^{(1)}
    \left (
    \frac{p_{Ij} q^2}{q_I q_j}, \frac{m_I^2 q^2}{q_I^2},0
    \right )
    & \sim
      -\frac{(1+\kappa_{Ij} \xi)}{2  \ep}
      \frac{\Gamma(2-2\ep) \Gamma(1+\ep)}{\Gamma(1-\ep)}
      \left (\frac{\kappa_{Ij} \xi}{x} \right )^\ep
    \\
    & \times F_{21}
      \left ( -\ep, -2\ep, 1-\ep, -\frac{1}{\kappa_{Ij} \xi}
      \right ).
  \end{split}
  \label{eqc.17}
\end{equation}
Eq.~\eqref{eqc.17} can be further used to determine the $x \sim 0$, $\xi \sim 0$
asymptotic of the $I_{1,1}^{(1)}$ function -- we only need to use the known
transformation rules for the hypergeometric function in that equation, to
express it through the hypergeometric function with the argument
$-\kappa_{Ij} \xi$. Once this is done, determining the $x \sim 0, \xi \sim 0$
asymptotic becomes straightforward.

Finally, we need to discuss the extraction of the $\xi \sim 0$ asymptotic at
fixed $x$. We find that in this case
\begin{equation}
  y_1
  \to 1 - \frac{1}{x},
  \;\;\; y_2 \to   1 -
  \kappa_{ij} \xi + {\cal O}(\xi^2).
\end{equation}
We note that we have to keep $y_2$ different from one because term $y_2 = 1$ is
the singular point of the Appell function in Eq.~\eqref{eqc.5}. We then find
that at $\xi \sim 0$
\begin{equation}
  \begin{split}
    \frac{q^2}{q_j }I_{1,1}^{(1)}
    \left (
    \frac{p_{Ij} q^2}{q_I q_j}, \frac{m_I^2 q^2}{q_I^2},0
    \right ) &  \sim
               (1-2 \ep)
               \left (\frac{\kappa_{Ij} \xi}{x} \right )^\ep
    \\
             & \times \int \limits_{0}^{1}
               {\rm d} u \; u^{-2\ep-1}
               \left ( 1 - (1-x) u \right )^{\ep} \left (1-u + \frac{u}{\kappa_{Ij} \xi}
               \right )^\ep \Bigg |_{\xi \sim 0}.
  \end{split}
\end{equation}
It is straightforward to extract the leading term of the expansion of the above
integral at small $\xi$; the simplest way to do so is the so-called ``strategy
of regions'' \cite{Beneke:1997zp}. We obtain
\begin{equation}
  \begin{split}
    \frac{q^2}{q_j }I_{1,1}^{(1)}
    \Bigg  (
    \frac{p_{Ij} q^2}{q_I q_j}, \frac{m_I^2 q^2}{q_I^2},
    & 0
      \Bigg  )  \sim
      (1-2 \ep) \;
      x^{-\ep} \\
    & \times
      \left (
      \frac{\Gamma(-2\ep) \Gamma(\ep)}{\Gamma(-\ep)} ( \kappa_{Ij} \xi)^{-\ep}
      -
      \frac{1}{\ep}
      F_{21}(-\ep,-\ep,1-\ep,1-x)
      \right ),\;\;\; \xi \sim 0.
  \end{split}
\end{equation}
The above results cover all the cases needed to construct formulas to describe
the contribution of the function
$I_{1,1}^{(1)}(p_{Ij} q^2/(q_I q_j),m_I^2 q^2/q_I^2,0)$ to the integrand in
Eqs~\eqref{eqc.2} and \eqref{eqc.3} that can be then integrated over $x$ and
$\xi$.

It remains to discuss the limiting behavior of the second function
$I_{1,1}^{(1)}(2 - p_{Ij}q^2/q_I q_j,m_I^2 q^2/q_I^2,0)$. The analysis proceeds
along similar lines so that we only present the required formulas without
additional comments. The $x \to 0$ limit in this case evaluates to
\begin{equation}
  x \to 0:\;\;\;  y_1 \to 0, \;\;\;
  y_2 \to 1 - \frac{\kappa_{Ij} \xi x }{(1+\kappa_{Ij} \xi)^2}.
\end{equation}
Using these results, we find that for  $x \sim 0$
\begin{equation}
  \begin{split}
    I_{1,1}^{(1)}\left (2
    - \frac{p_{Ij}q^2}{q_I q_j},\frac{m_I^2 q^2}{q_I^2},0 \right )
    & \sim  \frac{2\ep-1}{4 \ep}
      F_{21}\left( -\ep,1,1-2\ep, 1-
      \frac{\kappa_{Ij} \xi x }{(1+\kappa_{Ij} \xi)^2}.
      \right )
    \\
    & \to \frac{2\ep-1}{2\ep}
      \left (
      1 - \left (\frac{\kappa_{Ij} \xi x}{(1+\kappa_{Ij} \xi)^2} \right )^{-\ep}
      \frac{\Gamma(1-2\ep) \Gamma(1+\ep)}{2 \Gamma(1-\ep)}
      \right ).
  \end{split}
\end{equation}
The $x \sim 0$, $\xi \sim 0$ asymptotic follows straightforwardly from the above
equation.

In case $\xi \sim 0$ and $x \sim {\cal O}(1)$, we derive
\begin{equation}
  \xi \to 0:\;\;\;\; y_1 \to -\frac{x}{1-x},\;\;\; y_2 \to 1-\frac{\kappa_{Ij} \xi x}{1-x}.
\end{equation}
Using these expressions in Eq.~\eqref{eqc.5}, we obtain
\begin{equation}
  \begin{split}
    I_{1,1}^{(1)}\left (2
    - \frac{p_{Ij}q^2}{q_I q_j},\frac{m_I^2 q^2}{q_I^2},0 \right )
    \sim &\frac{1-2\ep}{2 (1-x)}
           (1-x)^{-\ep}\left(\frac{\kappa_{Ij} \xi x}{1-x} \right)^{\ep} \\
         & \times
           \int \limits_{0}^{1}
           {\rm d} u \; u^{-2 \ep -1}
           (1 - x u)^\ep \left ( 1 + \frac{1-x}{\kappa_{Ij} \xi x } u
           \right )^{\ep} \Bigg |_{
           \xi \sim 0}.
  \end{split}
\end{equation}
Extracting the $\xi \sim 0$ asymptotic, we find
\begin{equation}
  \begin{split}
    \lim_{\xi \to 0}
    I_{1,1}^{(1)}\left (2
    - \frac{p_{Ij}q^2}{q_I q_j},\frac{m_I^2 q^2}{q_I^2},0 \right )
    & \sim \frac{1-2\ep}{2 (1-x)}
    \\
    & \times \left (
      \frac{\Gamma(-2\ep) \Gamma(\ep)}{\Gamma(-\ep)} (\kappa_{Ij} \xi x)^{-\ep}
      -
      \frac{1}{\ep} F_{21}(1, -\ep, 1-\ep,
      x )
      \right ).
  \end{split}
\end{equation}

\section{Eikonal Functions}
\label{sec:app-eikonal-functions}

In this appendix, we provide the double real-emission eikonal functions that we
use to compute the relevant contributions to the $N$-jettiness soft function.
Following Ref.~\cite{Catani:1999ss}, we define
\begin{equation}
  \tilde S_{\gI \gJ }(k_1,k_2)
  = -\frac{C_A}{2} \tilde S^{gg}_{\gI \gJ }(k_1,k_2)
  + T_R n_f \tilde S^{qq}_{\gI \gJ }(k_1,k_2),
\end{equation}
where
\begin{equation}
  \begin{split}
    {\tilde S}_{\gI \gJ}^{gg}(k_1,k_2)
    &= \frac{1}{2} \Big[ 2 S_{\gI \gJ}^{gg}(k_1,k_2)  - S_{\gI\gI}^{gg}(k_1,k_2) - S_{\gJ\gJ}^{gg}(k_1,k_2) \Big],
    \\
    {\tilde S}_{\gI \gJ}^{q\bar q}(k_1,k_2)
    &= \frac{1}{2} \Big[ 2 S_{\gI \gJ}^{q\bar q}(k_1,k_2)  - S_{\gI\gI}^{q\bar q}(k_1,k_2) - S_{\gJ \gJ}^{q\bar q}(k_1,k_2) \Big].
  \end{split}
\end{equation}
The eikonal function for double gluon correlated emission
$S_{\gI \gJ}^{gg}(k_1,k_2)$ can be written as
\begin{equation}
  S_{\gI \gJ}^{gg}(k_1,k_2) 
  = S^{0}_{\gI \gJ}(k_1,k_2) 
  + [m_\gI^2 S_{\gI \gJ}^{m}(k_1,k_2) 
  +  m_\gJ^2 S_{\gJ \gI}^{m}(k_1,k_2) ] .
  \label{eqB2}
\end{equation}
The first term $S_{\gI \gJ}^{0}(k_1,k_2)$ is the same for massless and massive
emitters. It reads
\begin{equation}
  \begin{split}
    S^{0}_{\gI \gJ}(k_1,k_2)
    &
      =
      \frac{(1-\ep)}{(k_1 \cdot k_2)^2} \;
      \frac{(p_\gI \cdot k_1)(p_\gJ \cdot   k_2) + (p_\gI \cdot k_2)(p_\gJ \cdot  k_1)}{(p_\gI \cdot k_{12})(p_\gJ \cdot k_{12})}
    \\
    &
      - \frac{(p_\gI \cdot p_\gJ)^2}{2(p_\gI \cdot k_1) (p_\gJ \cdot k_2) (p_\gI \cdot k_2) (p_\gJ \cdot k_1)}
      \bigg[ 2 -
      \frac{(p_\gI \cdot k_1)(p_\gJ \cdot k_2) + (p_\gI \cdot k_2)(p_\gJ \cdot k_1)}{(p_\gI \cdot k_{12})(p_\gJ \cdot k_{12})}
      \bigg]
    \\
    & + \frac{(p_\gI \cdot p_\gJ)}{2(k_1 \cdot k_2)}\;\bigg[ \frac{2}{(p_\gI \cdot k_1)(p_\gJ \cdot k_2)}
      + \frac{2}{(p_\gJ \cdot k_1)(p_\gI \cdot k_2)}
      - \frac{2}{(p_\gI \cdot k_{12})(p_\gJ \cdot k_{12})}   \\
    &\times \left( 4 +
      \frac{\left[(p_\gI \cdot k_1)(p_\gJ \cdot k_2) + (p_\gI \cdot k_2)(p_\gJ \cdot k_1)\right]^2}{(p_\gI \cdot k_1)(p_\gJ \cdot k_2)(p_\gI \cdot k_2)(p_\gJ \cdot k_1)}
      \right)\bigg] ,
  \end{split}
\end{equation}
where $k_{12}= k_1 + k_2$. The other two contributions in Eq.~\eqref{eqB2}
become relevant for massive emitters. They are given by
\begin{equation}
  \begin{split}
    S_{\gI \gJ}^m (k_1,k_2)
    &= \frac{(p_\gI \cdot p_\gJ)(p_\gJ \cdot k_{12})}{2(p_\gI \cdot k_1)(p_\gJ \cdot k_2)(p_\gI \cdot k_2)(p_\gJ \cdot k_1)(p_\gI \cdot k_{12})} \\
    & - \frac{1}{2(k_1 \cdot k_2)(p_\gI \cdot k_{12})(p_\gJ \cdot k_{12}) } \left( \frac{(p_\gJ \cdot k_1)^2}{(p_\gI \cdot k_1)(p_\gJ \cdot k_2)} + \frac{(p_\gJ \cdot k_2)^2}{(p_\gI \cdot k_2)(p_\gJ  \cdot k_1)}\right) .
  \end{split}
\end{equation}
In case the soft partons are a quark and an anti-quark, we write
\begin{equation}
  S_{\gI \gJ}^{q\bar q}(k_1,k_2)
  = \frac{(p_\gI \cdot k_1)(p_\gJ \cdot k_2)+
    (p_\gI \cdot k_2)(p_\gJ \cdot k_1) - (p_\gI \cdot p_\gJ) (k_1 \cdot k_2)
  }{(k_1 \cdot k_2)^2( p_\gI \cdot k_{12})(p_\gJ \cdot k_{12} )}.
\end{equation}

\clearpage
\bibliographystyle{JHEP}
\bibliography{mNJNSF}
\end{document}